\documentclass[useAMS,usenatbib,twoside,a4paper,iop,numberedappendix,appendixfloats]{emulateapj} 
\slugcomment{{\sc Accepted for publication in ApJS on September 20, 2013 issue}}

\usepackage{graphicx}
\usepackage{epstopdf}
\usepackage{appendix}
\usepackage{color}

\newcommand{\Msol}{\mbox{$M_{\odot}$}}

\def\deg      {{\ifmmode^\circ\else$^\circ$\fi}}

\shorttitle{ZENS II. Galaxy Structural Measurements
and the Concentration of Satellites }

\shortauthors{Cibinel A. et al.}

\begin{document}

\title{The Zurich Environmental Study (\emph{ZENS}) \\ of Galaxies in Groups along the Cosmic Web. II. \\ Galaxy Structural Measurements and the Concentration of Morphologically-Classified Satellites in Diverse Environments\footnotemark[\dag]}

\author{A. Cibinel\altaffilmark{1,$\star$}, 
C.~M. Carollo\altaffilmark{1,$\star$}, 
S.~J. Lilly\altaffilmark{1}, 
F. Miniati\altaffilmark{1},
J.~D. Silverman\altaffilmark{2}, 
 J.~H. van Gorkom\altaffilmark{3}, 
E. Cameron\altaffilmark{1},  
A. Finoguenov\altaffilmark{4}, 
P. Norberg\altaffilmark{5},
 Y. Peng\altaffilmark{1},
 A. Pipino\altaffilmark{1}, 
C.~S. Rudick\altaffilmark{1}
}
 
\altaffiltext{$\star$}{E-mail: \texttt{cibinel@phys.ethz.ch}, \texttt{marcella@phys.ethz.ch}}

\altaffiltext{1}{Institute of Astronomy, ETH Zurich, CH-8093 Zurich, Switzerland}
\altaffiltext{2}{Kavli Institute for the Physics and Mathematics of the Universe (WPI), Todai Institutes for Advanced Study, The University of Tokyo, 5-1-5 Kashiwanoha, Kashiwa, 277-8583, Japan}
\altaffiltext{3}{Department of Astronomy, Columbia University, New York, NY 10027, USA}
\altaffiltext{4}{Max-Planck-Institut f\"ur extraterrestrische Physik, D-84571 Garching, Germany}
\altaffiltext{5}{Institute for Computational Cosmology, Department of Physics, Durham University, South Road, Durham DH1 3LE, UK}
\altaffiltext{$\dagger$}{Based on observations collected at the European Southern Observatory, La Silla Chile. Program ID 177.A-0680}

\begin{abstract}
We present structural measurements for the galaxies in the $0.05<z<0.0585$ groups of the \emph{Zurich Environmental Study}, aimed at establishing how galaxy properties depend on four environmental parameters:  group halo mass $M_{GROUP}$, group-centric distance $R/R_{200}$, ranking into central or satellite, and large-scale structure density $\delta_{LSS}$. Global galaxy structure is quantified both parametrically and non-parametrically. We correct all these measurements for observational biases due to PSF blurring and surface brightness effects as a function of galaxy size, magnitude, steepness of light profile and ellipticity. Structural parameters are derived also for bulges, disks and bars. We use the galaxy bulge-to-total ratios (B/T), together with the calibrated non-parametric structural estimators, to implement a quantitative morphological classification that maximizes purity in the resulting morphological samples. We investigate how the concentration $C$ of satellite galaxies depends on galaxy mass for each Hubble type, and on $M_{GROUP}$, $R/R_{200}$ and $\delta_{LSS}$.  At galaxy masses $M\ge10^{10}\Msol$, the concentration of disk satellites increases with increasing stellar mass, separately within each morphological bin of B/T. The known increase in concentration with stellar mass for disk satellites is thus due, at least in part, to an increase in galaxy central stellar density at constant B/T.  The correlation between concentration and galaxy stellar mass becomes progressively steeper for later morphological types. The concentration of disk satellites shows a barely significant dependence on $\delta_{LSS}$ or $R/R_{200}$. The strongest environmental effect is found with group mass for $>10^{10}\Msol$ disk-dominated satellites, which are $\sim10\%$ more concentrated in high mass groups than in lower mass groups.
 \end{abstract}

\keywords{surveys - galaxies: groups - galaxies: formation - galaxies: evolution - galaxies: structure}


\section{Introduction}
\setcounter{footnote}{0}
We  present the methodology used to derive structural measurements for the galaxies investigated in the \emph{Zurich Environmental Study (ZENS)} (\citealt{Carollo_et_al_2013}, hereafter Paper I).  The resulting  measurements are provided in the ZENS global catalog that we have published electronically with Paper I\footnote{The ZENS catalog is also downloadable from: http://www.astro.ethz.ch/research/Projects/ZENS.}.

ZENS is designed to address the  question of which specific environment  is most relevant for influencing the properties of different galaxy populations. 
Several definitions of  environment  have been commonly employed in the literature to study the relation between environment and galaxy evolution:
    the density of galaxies calculated out to a fixed 
  or an adaptive distance   \citep[e.g.][]{Dressler_1980,Hogg_et_al_2003, Cooper_et_al_2005,Baldry_et_al_2006}, the mass of the host group or cluster \citep{Weinmann_et_al_2006, Kimm_et_al_2009}, 
  the distance from the group/cluster center 
\citep{Whitmore_Gilmore_1991,Balogh_et_al_1997,Lewis_et_al_2002, DePropris_et_al_2003,Hansen_2009}
 or the location into larger structures such as cosmic filaments or superclusters \citep{Einasto_et_al_2007,Porter_et_al_2008}.
Recently, the ability to separate galaxies into centrals and satellites within their host group halos has produced mounting evidence  that the  environmental influence on the star-formation properties of galaxies may peak for  satellite  galaxies  \citep{VanDenBosch_et_al_2008,Peng_et_al_2010,Peng_et_al_2012,Knobel_et_al_2012}. There is cross-talk however between  different definitions of environment, which may also relate to one another from a physical perspective.  A key question is therefore to identify what is the  relative importance of different environmental conditions for well-defined  galaxy populations of different masses, star formation activity levels, and structural/morphological properties \citep[e.g.][]{Blanton_Berlind_2007,Wilman_et_al_2010,Peng_et_al_2012,Muldrew_et_al_2012,Woo_et_al_2012}. 

In  ZENS  we aim at helping clarifying which of the many environments that a galaxy experience has a larger impact on its evolution. We do so by using the same sample of suitably selected nearby galaxies to investigate the dependence of their properties, at fixed stellar mass, on four environment measurements: the host group halo mass,
 the radial segregation within the group, the large scale density field on which the group halos reside, and the galaxy rank within its group halo, i.e., whether it is the central or a satellite galaxy within the gravitational potential of its host group. 
In computing our  proxies for these different environments, we have attempted to minimize  cross-talks  between their   definitions, in order to better disentangle one from another of the physical conditions that galaxies experience (see Paper I).

This paper focuses on the quantification of robust galaxy  structural and morphological properties, which provide key information on the life histories of galaxies. The presence and properties of  massive disks and spheroids highlight the occurrence of relatively slow and dissipative gas accretion \citep{White_Rees_1978,Fall_Efstathiou_1980} or mergers \citep[e.g][]{Toomre_1977,Barnes_1988,Schweizer_1990,Naab_Burkert_2003},
respectively.   Inner cores or cusps \citep{Ferrarese_et_al_1994,Lauer_et_al_1995,Carollo_et_al_1997a,Carollo_et_al_1997b,Graham_Guzman_2003,Graham_et_al_2003,Truijillo_et_al_2004,Cote_et_al_2007, Kormendy_et_al_2009}  and  tidal debris \citep{Malin_Carter_1980,Malin_Carter_1983,Forbes_et_al_1992,vanDokkum_2005, Tal_et_al_2009,Janowiecki_et_al_2010} also trace the degree of dissipation involved in the evolution of galaxies (not surprisingly with some debate, \citealt[][]{Mihos_Hernquist_1994, Mihos_Hernquist_1996, Kawata_et_al_2006,Feldmann_et_al_2008,Hopkins_et_al_2009}).
Bars and pseudo-bulges are smoking guns for either secular evolution processes
 (\citealt{Kormendy_1979, Combes_et_al_1990,Courteau_et_al_1996,Norman_et_al_1996,Wyse_et_al_1997,Carollo_1999, Carollo_et_al_1997c,Carollo_et_al_1998,Carollo_et_al_2001,Balcells_et_al_2003,Debattista_et_al_2004,Debattista_et_al_2006,Kormendy_Kennicutt_2004,Fisher_Drory_2008}), or possibly for early bulge formation through instabilities in the proto-disks \citep{Immeli_et_al_2004,Carollo_et_al_2007,Dekel_et_al_2009a,Dekel_et_al_2009b}.
 A robust determination of galaxy morphology is hence essential for  understanding 
 whether this is linked to any of the environmental conditions above, how precisely it relates to the occurrence, enhancement or cessation of star formation activity, and thus for pinning down which physical processes  drive galaxy  evolution.

Determining galaxy structure is however notoriously not a trivial task.
Visual morphological classification is still a widely adopted method \citep[e.g.][]{Lintott_et_al_2008,Nair_Abraham_2010}, despite its subjectivity and failure to provide quantitative measurements for different components, which are necessary to trace galaxy  assembly over cosmic time.
For these reasons numerous publications have been devoted to the development of methods and 
software for the automated quantification of structure on large galaxy samples.  
There are a number of approaches to the problem which can be broadly divided into two categories:
 those which employ parametric descriptors for the galaxy morphology, namely a set of analytical profiles
  used to model the bulge, disk, or bar component \citep[e.g.][]{Simard_et_al_2002, Peng_et_al_2002,deSouza_et_al_2004}
and those which instead use the observed properties of the light distribution, such as the degree of asymmetry,
 isolation of bright pixels or decompositions into a set of  basis functions \citep[e.g.][]{Abraham_et_al_1996,Conselice_2003,Refregier_2003,Lotz_et_al_2004,Scarlata_et_al_2007}. The two methods have different strengths:  parametric
 decomposition is useful to obtain measurements of characteristic sizes and to have an estimate of the relative 
 importance of the bar, disk and bulge components; it also easily includes the effects of seeing. Non-parametric estimators well describe the inhomogeneities  in the light distributions of real galaxies, which typically display irregular, non axis-symmetric  features generated by recent star-formation, dust or galaxy interactions.

Both parametric and non-parametric measurements suffer however 
from a number of observational biases, which must be corrected for in order to perform comparisons between galaxies of different properties, and observed in different conditions.
 In particular, in ground-based surveys, the effect of atmospheric seeing is one of the major complications. 
 Several studies have shown the strong impact of the seeing on the photometric and structural properties of galaxies, 
 not only in the inner regions of galaxies, but also out to radii corresponding to several FWHM of the Point Spread Function (PSF)  \citep[e.g.][]{Schweizer_1979,Franx_et_al_1989, Saglia_et_al_1993,Trujillo_et_al_2001,Graham_2001}. 
    
 Another factor which affects the derivation of structural parameters is the inclination angle at which a galaxy is observed. The  overlap, in projection, of multiple subcomponents, as well as physical factors such as 
the non-uniform distribution of inter-stellar dust -- which causes a higher attenuation of  short-wavelength light in the central regions of edge-on galaxies  than in similar face-on galaxies \citep[e.g.][]{Driver_et_al_2007,Shao_et_al_2007} -- can substantially bias the measurements of sizes, bulge-to-disk ratios, concentration  and even stellar masses \citep[e.g.,][]{Maller_et_al_2009,Graham_Worley_2008, Bailin_Harris_2008}.

Finally, the background sky  makes the detection of faint components difficult,  a fact which introduces severe biases in  the measurements of the galaxy properties, especially magnitudes and sizes 
 \citep[e.g.][]{Disney_1976,Impey_Bothun_1997}. The strength of the bias depends on  galaxy size, inclination and stellar light profile.  Although a number of widely-used measurement techniques, such as the computation of Kron aperture fluxes \citep{Kron_1980} or the extrapolation of model galaxy surface brightness profiles \citep{Sersic_1968}, can help  recover light below the isophotal limit, significant systematic biases remain in the low surface brightness regimes specific to each survey \citep[e.g.][]{Graham_et_al_2005, Cameron_Driver_2007,Cameron_Driver_2009, Haussler_et_al_2007}. 
 
In ZENS we attempt {\it to correct}, when possible, all measurements of galaxy structure for systematic biases as a function of PSF-size, and also galaxy magnitude, size, axis ratio and radial shape of the light profile.  We also quantify the size of systematic biases in regimes of parameter space where the (statistical) recovery of the intrinsic information is not achievable, e.g., at small galaxy sizes and low surface brightnesses. We also stress that in ZENS each galaxy is handled individually, till self-consistent and both quantitatively- and  visually-checked  accurate measurements are achieved. This enables ZENS to tackle complementary questions regarding the galaxy-environments relationship relative to larger but less detailed galaxy samples.

The paper is organized as follows. In Section \ref{sec:survey} we briefly review the  specifications  and definitions for the four environments under scrutiny in ZENS.  We then devote the   first part of the paper to an overview of the structural measurements carried out  on the ZENS galaxy sample. These measurements  include isophotal analyses and bar detection/quantification (Section \ref{sec:IsophotalAnalysis}),  analytical surface brightness fits and bulge+disk decompositions (Section \ref{sec:GIM2Dmeasurements}), and derivation of non-parametric structural indices (concentration, Gini, asymmetry, M$_{20}$, smoothness;  Section \ref{sec:ZEST}). 
In Section \ref{sec:Simulations} we thoroughly investigate the sources of error in these measurements and derive a correction scheme  that recovers the intrinsic structural parameters.  In Section \ref{sec:MorphClass} we present the 
 morphological classification of the \textsc{ZENS} galaxies, based on a quantitative partition of the structural parameter space in regions that are associated with  elliptical, bulge-dominated disks, intermediate bulge-to-total ratio disks, late-type disks  and irregular galaxies. In the same Section we also discuss  the statistics of the structural properties for the
various morphological classes.
We describe in detail in Appendix \ref{app:DataReduction} the data reduction and photometric calibration of  the ESO $B$ and $I$ WFI/2.2m imaging data for the ZENS groups that are introduced in Paper I.
 Appendices \ref{sec:testGIM2D}-\ref{app:corrections_PSF} present additional details on the tests performed on the analytical surface-brightness fits and supplementary information for the derivation of the corrections for the structural parameters. Stamp images for galaxies in the different morphological classes are found in Appendix 
\ref{app:ClassDist}.  
In the final part of the paper we use the corrected structural measurements to study, at constant stellar mass,  the concentration of satellite galaxies as a function of Hubble type and environment (Section \ref{sec:Results}). We summarize the paper in Section \ref{sec:Conclusions}.

 This second  \textsc{ZENS} publication is complemented by a companion paper (\citealt{Cibinel_et_al_2013}, hereafter Paper III),  in which we present  the spectrophotometric properties of our galaxy sample and to which we refer for details on the derivation of, e.g.,  the galaxy spectrophotometric types, star formation rates and stellar masses that we use in this and other ZENS papers.
  
The following cosmological parameters are adopted in all the ZENS publications: $\Omega_m=0.3$, $\Omega_{\Lambda}=0.7$ 
and $h=0.7$. Unless otherwise specified,  magnitudes are in the AB system, and galaxy sizes are \emph{semi-major} axis measurements. All derived luminosities are corrected for Galactic extinction using the maps of  \citealt{Schlegel_et_al_1998}.


\section{A brief summary of ZENS}  \label{sec:survey}
    
\subsection{Data and sample}

\textsc{ZENS} is based on a sample of 1484 galaxies\footnote{Note that only 1455 galaxies are listed as members of the 141 2PIGGS groups; however, 29 of these single-entries are actually galaxy pairs/triplets, for which we measure the individual properties of both galaxies.}, members of 141 galaxy groups extracted from the 2-degrees 
Field Galaxy Redshift Survey (2dFGRS) (\citealt{Colles_et_al_2001,Colles_et_al_2003}), and specifically from the  
Percolation-Inferred Galaxy Group (2PIGG) catalogue (\citealt{Eke_et_al_2004a}).  
The 141 \textsc{ZENS}  groups are a random selection of the 2PIGG groups which are found in the very thin  redshift slice 0.05$<z<$0.0585 and have at least 5 confirmed
 members, down to a magnitude $b_J=19.45$. 
New $B-$ and $I-$band images were acquired for these groups with the WFI camera 
mounted at the Cassegrain focus of the MPG/ESO 2.2m Telescope at 
La Silla, over several observing runs between 2005 and 2009  (see Paper I for details). The data reduction and photometric calibration of  these WFI data is reported in Appendix \ref{app:DataReduction}. 
This paper describes in detail the structural analysis that we have performed on these WFI  $B$ and $I$ images. 
  
\subsection{The four environments investigated in ZENS}

For each \textsc{ZENS} galaxy, Paper I discusses and publishes four estimates of  environment: the group halo mass, the distance from the center of the group, the rank within the group (i.e., whether the galaxy is the central or a satellite) and the location on the large scale structure (LSS). In particular: 

$(1)$  Group masses  $M_{GROUP}$ are derived from the total group luminosities by assuming a mass-to-light ratio calibrated with mock catalogs (\citealt{Eke_et_al_2004b}).

$(2)$ The  ranking of galaxies in centrals and satellites factors in the errors on the galaxy stellar masses, and  includes a test of self-consistency requiring that a bona-fide central galaxy must, simultaneously,  be consistent with being the most massive galaxy of the group,   be located within a projected distance 
$<0.5 R_{200}$ from the mass weighted center of the group (with $R_{200}$ the characteristic size of the group, as defined in Paper I), and   have a relative velocity   within one standard deviation from the group velocity. 

$(3)$ The  centers of the groups,  on which the projected radial positions of satellite galaxies within the groups rely, are identified with the locations of the central galaxies. We operationally divide the  ZENS group sample into \emph{relaxed} and \emph{unrelaxed} groups, depending on whether a bona-fide central galaxy can be found, according to the prescription above. In Paper I we discuss   the observational biases which may hamper  the identification of the central galaxy in groups which are in fact dynamically relaxed. We therefore use our group classification scheme mostly to test that our results are not affected by the inclusion/exclusion of the unrelaxed groups from our studies, although we keep  an eye on the possibility that there may be a physical origin, related to the dynamical evolution of the host groups, for differences observed in otherwise similar galaxies that inhabit  the two classes of  groups.

 $(4)$ The LSS density at the \textsc{ZENS} group locations is defined using an Nth nearest-neighbor  analysis which adopts the groups (not the galaxies) as the tracers of the LSS density. Specifically, we define
 $\delta_{LSS}=\frac{\rho_{LSS}(z)-\rho_m}{\rho_m}\,$,
 with $\rho_{LSS}(z)$ the density of 2dFGRS groups in a projected circular area defined by the comoving distance of the 5th nearest-neighboring  group around the \textsc{ZENS} group,  and $\rho_m$ is the mean projected density calculated over the global 2dFGRS area at the given redshift. Our choice results in identical $\delta_{LSS}$ values for all galaxy members of any given group, and avoids the problems  that are associated with using Nth-nearest neighboring galaxy algorithms (see Paper I, and also \citealt{Peng_et_al_2012,Woo_et_al_2012}).

\subsection{Galaxy stellar mass completeness limits for ZENS}

In Paper III we discuss the mass completeness limits of ZENS which is determined by the 2dFGRS apparent magnitude selection. The strongest constraints are set by  passively evolving galaxies, which have the larger mass-to-light ratio, for which the 85$\%$ completeness  is reached above $10^{10}\Msol$.  This is the mass completeness threshold that we adopt for passive elliptical and S0 galaxies. Progressively more actively star forming galaxies have progressively lower mass completeness thresholds, with strongly  star forming galaxies mass complete at $10^{9.2}\Msol$  (see Paper III). Given the statistical mix of spectrophotometric types in the different morphological bins, the   mass completeness thresholds of $10^{9.93}\Msol$, $10^{9.78}\Msol$ and $10^{9.55}\Msol$ are statistically adequate  respectively for the   morphological classes of  bulge-dominated spirals, intermediate bulge-to-total disk galaxies, and late-type disk (or irregular) galaxies.

\begin{figure}
\begin{center}
\includegraphics[width=90mm]{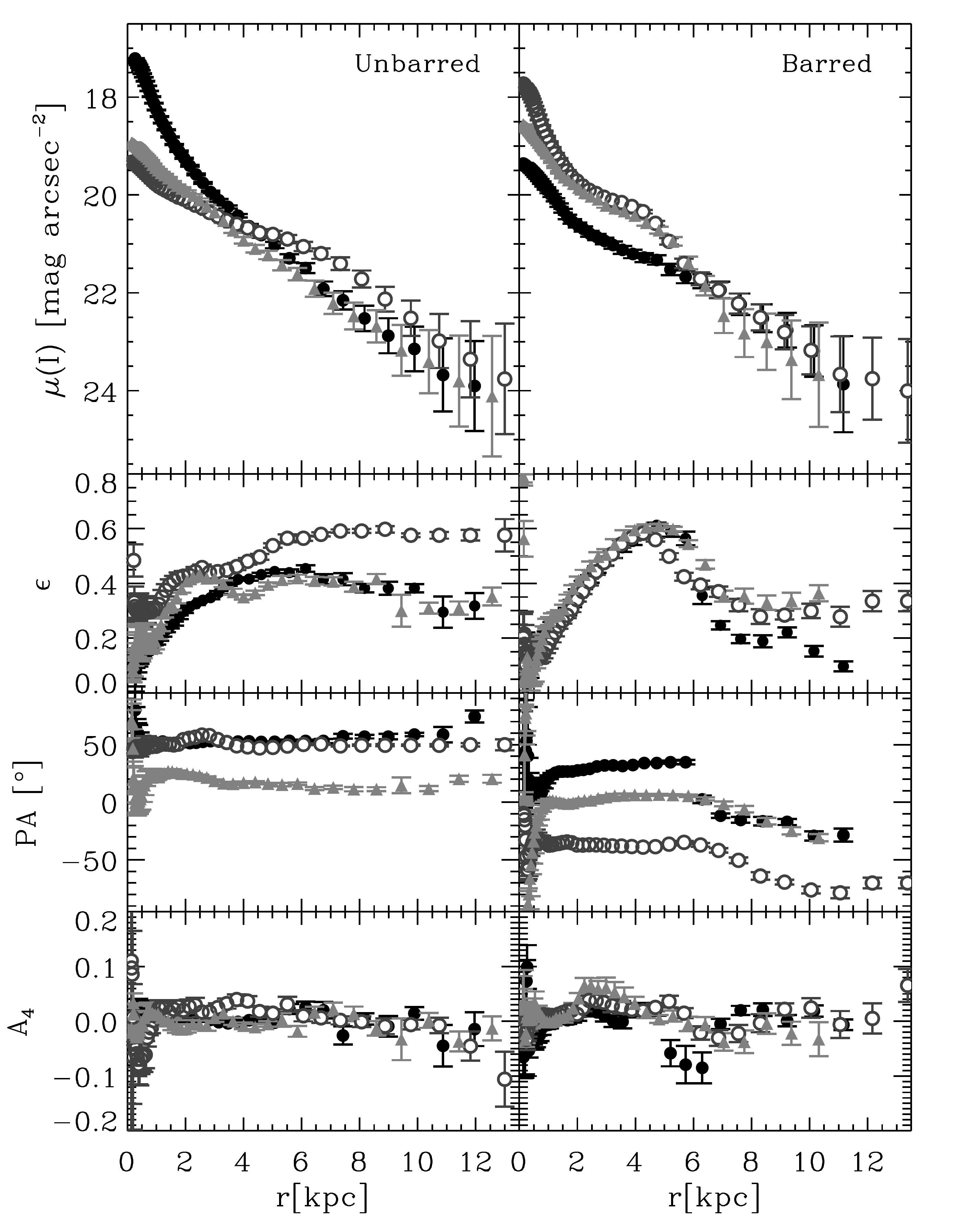}
\end{center}
\caption{\label{fig:SB_profiles} From top to bottom, the panels show the radial profiles of $I-$band surface brightness, ellipticity, position angle and the high order Fourier coefficient $A_4$ for three example of ZENS disk galaxies classified as non-barred (left) and barred (right). Different symbols and gray shades are used to distinguish the profiles for the three individual galaxies.}
\end{figure}


\section{Quantification of galaxy structure. I. An isophotal analysis} \label{sec:IsophotalAnalysis}

Surface brightness profiles and isophotal parameters were obtained  with the  IRAF {\ttfamily{ELLIPSE}} routine.
This well-tested algorithm fits the intensity at a given radius with the Fourier series
$I(\theta)=I_0+\sum(A_n\cos n \theta+B_n \sin n \theta)$ \citep{Jedrzejewski_1987}.
Here $I_0$  is the mean intensity within the isophote,  and $A_n$, $B_n$ the high harmonic 
coefficients which quantify isophotal deviations from perfect ellipticity. We truncated the series after the fourth order, so that,
for each isophote, the fits return the mean intensity, position angle (PA), ellipticity ($\epsilon$), and 
the amplitude of the fourth-order coefficients $A_4$ and $B_4$.  
The coefficient $A_4$ has been extensively used in the past to measure `boxiness'  ($A_4<0$) 
and   `diskyness'  ($A_4>0$)  of the isophotes (e.g., \citealt{Bender_et_al_1988}, \citealt{Franx_et_al_1989}, 
\citealt{Naab_et_1999}). 

 In running {\ttfamily{ELLIPSE}}, we allowed  the position angle and ellipticity to vary freely with radius, and limited wanderings of the isophotal 
 center to within 3 pixels from the center of the innermost isophotes.
 The PA and ellipticity determined by the \textsc{SExtractor} algorithm (\citealt{Bertin_et_Arnouts_1996})  served as initial guesses for the {\ttfamily{ELLIPSE}} algorithm.
The semi-major axis of adjacent isophotes was increased in logarithmic radial steps of 0.1, 
in order to increase the signal-to-noise ratio when measuring the external isophotes.   
We  allowed the code to perform a two iteration 3-sigma clipping of the discrepant pixels during the fits, and terminated the fitting procedure when 
$>50\%$ of the pixels in a given step of the calculation were flagged as discrepant.

The procedure was applied independently to the $I$ and $B$ images, and returned independent  surface brightness profiles in each of the
 two passbands. We also derived surface brightness profiles in the $B$ filter using the isophotal parameters derived from the $I$ images; 
 this returned more reliable surface brightness profiles  also at the shorter wavelength for those galaxies whose $B$ light distributions
  were too irregular (because of dust absorption and star formation knots) for a reliable measurement directly on the $B$ images.

\begin{figure*}
\begin{center}
\includegraphics[width=90mm, angle=90]{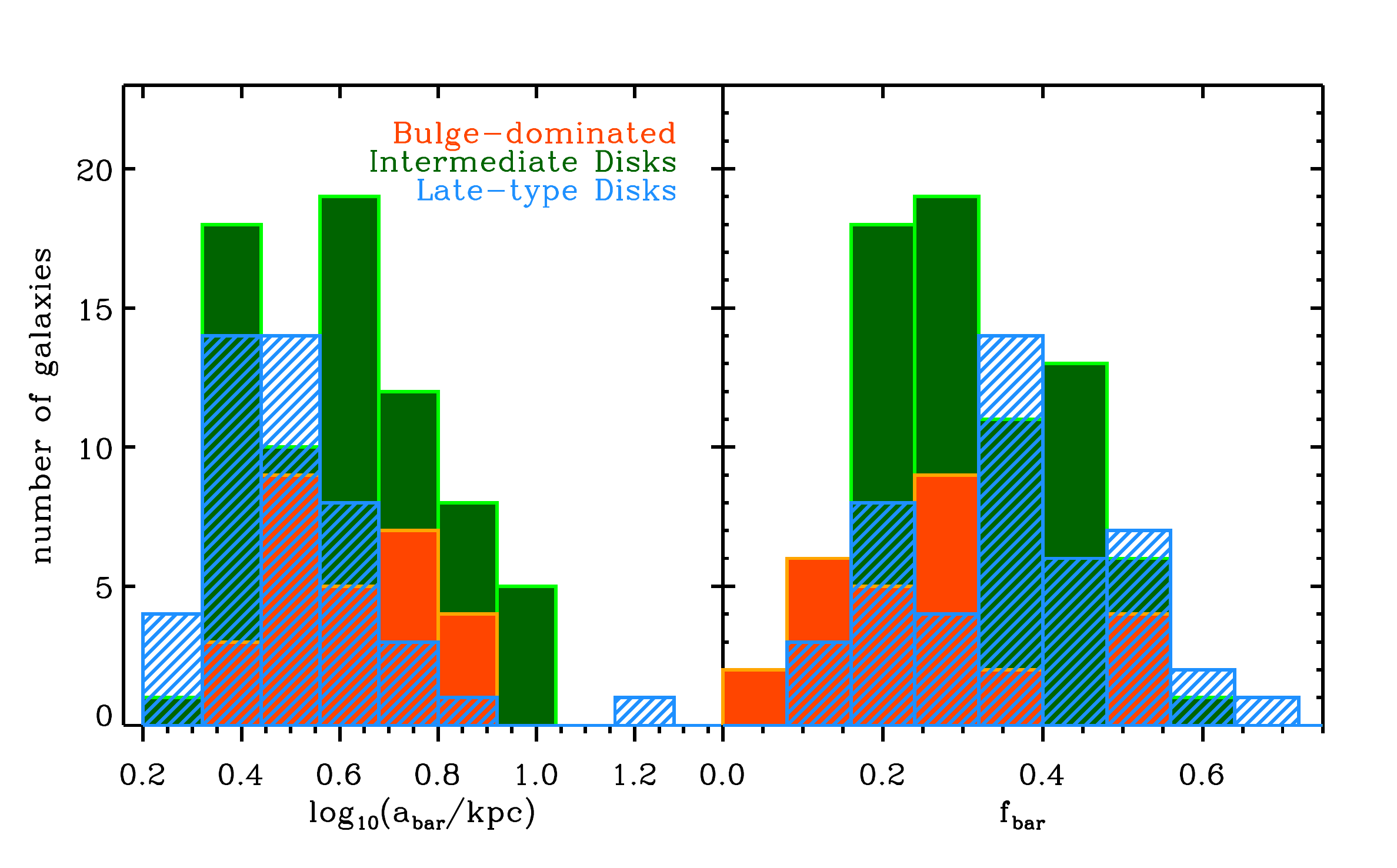}
\end{center}
\caption{\label{fig:barred_hist}Distribution of bar semi-major axes (left) and strengths (right) for the \textsc{ZENS} galaxies which are classified as barred. 
Histograms are colored differently according to the disk galaxy type hosting the bar: dark grey (orange in the online version)=S0/bulge-dominated spirals; grey (green)=intermediate-type disks; black (blue)=late-type disks.
(A color version of this figure is available in the online journal.)}
\end{figure*}

\subsection{Bar identification and quantification of bar strength} \label{sec:bars}

Changes in  the radial profiles of PA and ellipticity were used as a first diagnostics to identify bars in the (disk) galaxies. 
The presence of a bar produces a characteristic signature on the isophotes shapes: 
within the bar region the ellipticity increases smoothly to a maximum value with almost constant PA,
 and then drops abruptly at the end of the bar, where also the PA changes substantially. 

We inspected the $I$-band ellipticity and position angle radial 
profiles of the disk galaxies in our sample (S0 and later, see Section \ref{sec:MorphClass} for the definition of the morphological types).
We classified a disk galaxy as barred if its ellipticity grows to an absolute maximum greater than 0.2, and shows a variation in ellipticity 
and PA greater than 0.1 and 10$\deg$, respectively. In the bar region, we furthermore request that the PA 
profile is flat within $\pm 20 \deg$.  
The size of the bar,  $a_{bar}$, was defined to be the semi-major axis associated with the maximum value 
of the ellipticity profile,  subject to the condition that it had to be at least twice the FWHM of the PSF, which is typically $\sim 1^{\prime \prime}$. Note that galaxies with bar sizes smaller than this threshold are not considered as barred systems. Therefore, our final sample of barred disks provides a conservative lower limit to the total number of such galaxies in the entire sample.  Illustrative examples of the isophotal parameters profiles for a random selection of ZENS galaxies classified as barred and non-barred are shown in Figure \ref{fig:SB_profiles}.

The above mentioned criteria are widely used in the search for bars (e.g. \citealt{Knapen_et_al_2000}, \citealt{Menendez_Delmestre_et_al_2007} 
or \citealt{Sheth_et_al_2008}) and are shown to provide robust results, 
but they can fail in identifying a bar if, e.g., the bar has a similar PA to that of the disk (see e.g., \citealt{Menendez_Delmestre_et_al_2007}). 
For this reason, we also visually inspected all the disk galaxies and those which clearly showed a bar structure -- and had an ellipticity 
profile consistent with it, but no drop in PA -- were included in the barred sample as well. 
Combining the {\ttfamily{ELLIPSE}}  fitting method with the visual validation we identified a   clear bar signature  in 148  disk galaxies.

To quantify the bar strength we used the definition of \citealt{Abraham_and_Merrifield_2000}, i.e.,
\begin{equation}
f_{bar}=\frac{2}{\pi}\left[ \arctan \left( \frac{b}{a} \right)_{bar}^{-\frac{1}{2}}-   \arctan \left( \frac{b}{a} \right)_{bar}^{
+\frac{1}{2}}\right] \, ,
\end{equation} 
where $\left( \frac{b}{a} \right)_{bar}$ is the intrinsic axial ratio of the bar. The bar strength ranges from 
zero to unity for unbarred galaxies and infinitely strong bars, respectively.
The intrinsic axial ratio was derived from the observed one using the transformation  of \cite{Abraham_et_al_1999}, i.e.,
\begin{equation}
\left( \frac{b}{a}\right)^2_{bar}=\frac{1}{2}(X-\sqrt{X^2-4})\, ,
\end{equation}

 with  
\begin{eqnarray*}
X=\sec^2i [2\cos^2\phi \sin^2 \phi \sin^4i + \\
(b/a)^2_{inner}(1-\sin^2\phi \sin^2 i)^2 + \\
(b/a)^{-2}_{inner}(1-\cos^2\phi \sin^2 i)^2 ] \, .
\end{eqnarray*}

Here $i$ is the inclination angle obtained from the minor and major axis of the galaxy considered 
as whole ($i=\arccos (b/a)_{out}$), $(b/a)_{inner}$ is the observed axis ratio of the bar and $\phi$ is the twist angle
 between the bar and galaxy semi-major axis.
For our sample we computed $(b/a)_{inner}$ at the bar semi-major axis $a_{bar}$ and we used the values of 
ellipticity and PA obtained from \textsc{SExtractor} \citep{Bertin_et_Arnouts_1996} to calculate $(b/a)_{out}$ and $\phi$.

Figure \ref{fig:barred_hist} shows the distribution 
of bar strength and bar sizes measured in the \textsc{ZENS} samples of bulge-dominated galaxies (including both S0 and bulge-dominated spirals), intermediate-type and late-type disk galaxies. 
The median bar strength mildly increases with Hubble type,
changing from a value of  $<\!\!f_{bar}\!\!> = 0.24^{+0.01}_{-0.02}$ for S0 or bulge-dominated spirals, to
$<\!\!f_{bar}\!\!> = 0.31 ^{+0.02}_{-0.01}$ for intermediate-type disks and  $<\!\!f_{bar}\!\!> = 0.36 ^{+0.02}_{-0.02}$ for late-type disk galaxies.
 We also find a decrease in median bar  size from late-type to early-type disks:
  $<\!\!a_{bar}\!\!>=3.95^{+0.44}_{-0.20}$ kpc for S0 and bulge-dominated spirals, $<\!\!a_{bar}\!\!>=3.78 ^{+0.31}_{-0.14}$ kpc for  intermediate-type disks and  $<\!\!a_{bar}\!\!>=2.95^{+0.25}_{-0.07}$  for late-type disks. For both the bar size and strength a Kolmogorov-Smirnov test rejects a common parent distribution, between bulge-dominated galaxies and late-type disks, with a probability of 99\%. 
  The correlation  between bar properties and Hubble types (which could be a reflection of   a correlation  with galaxy stellar mass,  \citealt[e.g.][]{Sheth_et_al_2008,Cameron_et_al_2010,Nair_Abraham_2010b}) has been investigated in several works in the literature, whose results are consistent with our measurements. For example, 
a weakening of bars in early type galaxies  is reported  by   \citet{Laurikainen_et_al_2007,Buta_el_al_2005,Barazza_et_al_2008,Aguerri_et_al_2009}, and an increase in bar size in early- relative to late-type disks is found by \citet{Erwin_2005}.

 
\section{Quantification of galaxy structure. II.  Parametric characterization}\label{sec:GIM2Dmeasurements}

Two dimensional fits to the surface-brightness distributions of all ZENS  galaxy images were carried out 
with the Galaxy IMage 2d (\textsc{GIM2D})  software  package (\citealt{Marleau_Simard_1998}, \citealt{Simard_et_al_2002}).  
We used a single S\'ersic  profile to describe the total galaxy light distribution, i.e.:
$
\Sigma_b = \Sigma_e \exp \{-k_n[(r/r_{1/2})^{1/n}-1]\}
$
where $\Sigma_e$ is the surface intensity at the half-light radius $r_{1/2}$. 
The value of the parameter $k_n$ is such to ensure that the flux within $r_{1/2}$ is half of the total flux and it is approximated  to $k_n=1.9992n-0.3271$ \citep[e.g.][]{Caon_et_al_1993,Graham_Driver_2005}.
 
For the ZENS galaxies which are not classified as ellipticals (in Section \ref{sec:MorphClass}), we also performed two-components, bulge+disk decompositions of the two-dimensional galaxy light distributions.  
For this purpose we assumed a S\'ersic profile for the bulge and a perfectly exponential  disk  represented by $
\Sigma_d = \Sigma_0 \exp(-r/h)$, with  $\Sigma_0$  the central surface intensity and $h$ the disk scale length.
Pure exponential models, with no bulge component, were furthermore generated for galaxies classified as late-type disks (see again Section \ref{sec:MorphClass}).
Table \ref{tab:GIM2D_parameters} summarizes  the range of values between which the model parameters were allowed to vary.

\begin{deluxetable*}{rrrrrrrrrr}
\tabletypesize{\small}
\tablewidth{0pt}
\tablecaption{Range of allowed values for the parameters of the GIM2D fits.}
\tablehead{
\colhead{}  & \multicolumn{3}{c}{Double Component Fits}  &  \multicolumn{3}{c}{Single S\'ersic Fits}  &  \multicolumn{3}{c}{Pure Exponential Fits} \\ 
\cline{2-10} \\ 
\colhead{Parameter}  & \colhead{initial guess} & \colhead{min} & \colhead{max}  & \colhead{initial guess}   & \colhead{min} & \colhead{max} & \colhead{initial guess}   & \colhead{min} & \colhead{max}}
\startdata
$m_{T,B}$ / $m_{T,I}$  (mag)                  & 18/16.5   &  21.5/20   & 13/12      &  18/16.5 & 21.5/20  & 13/12 &  18/16.5 & 21.5/20  & 13/12 \\        
B/T                       	            &    0.5         &  0              &  1             &  1             & 1             & 1          &  0           & 0               & 0  \\                                                                     
$R_{e,bulge}$ (kpc)        	   &    2.3        &  0.5           &  20           &  2.3         &  0.5         &  30      &  -             & -            & - \\ 
$\epsilon_{bulge}$      	   &  0.5          &   0             &  0.7          &  0.5         & 0              & 1.0      &  -             &-            & - \\                                                                                                                                        
$\phi_{bulge}$ ($\deg$)  	   &  45           &  -180        &  180         &  45          &  -180      &  180    &  -             & -            & -\\                                                                     
$h $  (kpc)                  &   2.3         &  0.5           &  20           &  -             & -            & -         &   2.3         &  0.5           &  20\\                                                                     
$i_{disk}$ ($\deg$)               &    45         &  0              &  90           & -              & -            & -          &    45         &  0              &  90  \\                                                                     
$\phi_{disk}$ ($\deg$)         &     45        & -180         &  180        & -               &             & -        &     45        & -180         &  180 \\                                                                     
$\Delta x $ (pixels)               &      0          &  -2            &  2             & 0              & -2           & 2         &  0              & -2           & 2\\                                                                     
$\Delta y $ (pixels)               &      0          &  -2            & 2              &  0             & -2           & 2          &  0              & -2           & 2\\                                                                     
$n$                                         &     2.1        &  0.2           &  10          &  2.1         & 0.2          & 10       &   -             &  -           &-      \\                                                             
\enddata
\tablecomments{Permitted ranges for the variation of model parameters used in the \textsc{GIM2D} fits. From top to bottom: total galaxy magnitude in $B$ and $I$ bands, bulge-to-total ratio ($B/T$), bulge half-light radius, bulge ellipticity, bulge position angle, disk scale length, disk inclination angle, disk position angle, x and y galaxy center offset from the \textsc{SExtractor} center (further refined by using the IRAF {\ttfamily imcntr} task), and S\'ersic index. \label{tab:GIM2D_parameters}}
\end{deluxetable*}

GIM2D convolves the theoretical models with the PSF before fitting them to the galaxy images.  
 We modeled the PSF of each of the \textsc{ZENS} group with a two-dimensional gaussian having FWHM equal
 to the mean of the full-widths measured on several unsaturated stars over the \textsc{ZENS} fields.
 A single PSF was used for a whole ZENS field as the WFI PSF varies only slightly across the field of view of the camera, showing in particular a mild $\sim10\%$  increase of the PSF FWHM very close to the edges of the CCDs (as tested on  \textsc{SExtractor} sources with stellarity class $>0.9$ in all ZENS fields).
Due to statistical scatter and signal-to-noise variations, errors  introduced by applying a correction for this effect are comparable to or even marginally larger than those resulting from the adoption of a single PSF for each field. Larger PSF variations in ZENS are instead observed  between different fields (i.e. pointings); in this case,  the PSF can change by as much as 0.6-0.7 arcseconds  (see Figure \ref{fig:skyConditions}).
 This effect is taken into account not only during the model fitting, but also when applying the correction scheme that we develop in Section \ref{sec:Simulations}.

 During the fitting procedure, the code was allowed to re-compute the initial parameters from the image moments and 
 to fit the sky background level in each individual postage stamp galaxy image; the postage stamps were sized proportionally to the  \textsc{SExtractor} Petrosian radius of the galaxies, and set to be equal to 3 times the Petrosian radius. 
 On a set of simulations calibrated on the GEMS survey \citep{Rix_et_al_2004}, fixing the background to 
a locally defined value during the GIM2D fit was shown to improve the performances for low surface brightness galaxies  \citep{Haussler_et_al_2007}.  
 Our alternative approach was  to enable the local sky subtraction and to quantify any systematic bias in the GIM2D fits using a large set of artificial galaxy images which are specifically tailored to the ZENS observations (Section \ref{sec:Simulations}).  
 The sky pixels were identified as those pixels in the galaxy postage stamps that were located outside an aperture equal to 1.5 times the \textsc{SExtractor}  Petrosian radii of the galaxies.
 Bright star-forming clumps were  masked during the fitting procedure. 
 Finally, although \textsc{GIM2D} offers an option to correct for the effects of the disk optical thickness -- which consists in adding to the disk total magnitude a geometric factor $2.5\times \log (a/b)$ where $a/b$ is the axis ratio -- we decided not to use this feature, and disks were assumed to be optically thin in order to fit the actually observed light distribution without a prior assumption about the dust distribution.

\subsection{Details of the variable-$n$ single S\'ersic fits and $n=1$ pure exponential fits}\label{sec:SersicFits}

To be able to detect color gradients, for the variable-$n$ single S\'ersic fits as well as for the $n=1$ pure exponential models, the $B-$ and $I-$band images were fitted independently without imposing the structural parameters of one filter to the other band.
The position angles and ellipticity derived from the $B-$ and $I-$ band images agree well, with differences being limited to 15$\deg$ and 0.15, respectively.
In only $2\%$ of the galaxies we find differences which are larger than this, with a maximum change in ellipticity $\sim$0.2 and a maximum change in position angle $\sim$30$\deg$.
Generally speaking, we thus conclude that the single component fits are  robust and strong twists between the $B-$ and $I-$band isophotes are not a problematic issue when performing independent fits for the two passbands. 

 To verify the validity of the GIM2D single-component fits, we ran on all the \textsc{GIM2D} models the task {\ttfamily{ELLIPSE}}, keeping the isophotes  fixed at the radii, position angle and ellipticity of the {\ttfamily{ELLIPSE}} fits to the real galaxy images described in Section \ref{sec:IsophotalAnalysis}. The comparison between the total magnitudes and the {\ttfamily{ELLIPSE}}  profiles derived for the \textsc{GIM2D} models and those derived for the real galaxies, as well as the inspection of the   residual images between the \textsc{GIM2D} models and the real galaxies, enabled us to reject unphysical \textsc{GIM2D} models.

 Overall, we could obtain reliable S\'ersic fits for $96\%$ of the $B-$band and $I-$band images.  A similar fraction
of successful fits is obtained for the pure exponential models of late-type disks ($95\%$). 
The distribution of stellar masses (from Paper III), 
sizes and morphological types for the $4\%$ of  ZENS galaxies with no single S\'ersic fits  is shown in Figure 
 \ref{fig:failedGIM2D} and Table \ref{tab:GIM2DRejected} 
 of Appendix \ref{sec:testGIM2D}.
  In this  Appendix  we also present,  for the subset of   ZENS galaxies for which the information is publicly available from the NYU-VAGC \citep{Blanton_et_al_2005}, a comparison of our estimates of the S\'ersic parameters with those published in this other catalogue; the comparison shows a good agreement between  our {\it uncorrected} measurements and  the previously published data.
    We show below however that, in some regimes of parameter space,  these measurements require further corrections to be cleaned by residual observational biases, indicating that also measurements in the quoted and other public catalogues necessitate of similar attention. 
 
  \subsubsection{Comparison between single-component  $B-$ and $I-$band models}
 
\begin{figure*}
\begin{center}
\includegraphics[width=110mm,angle=90]{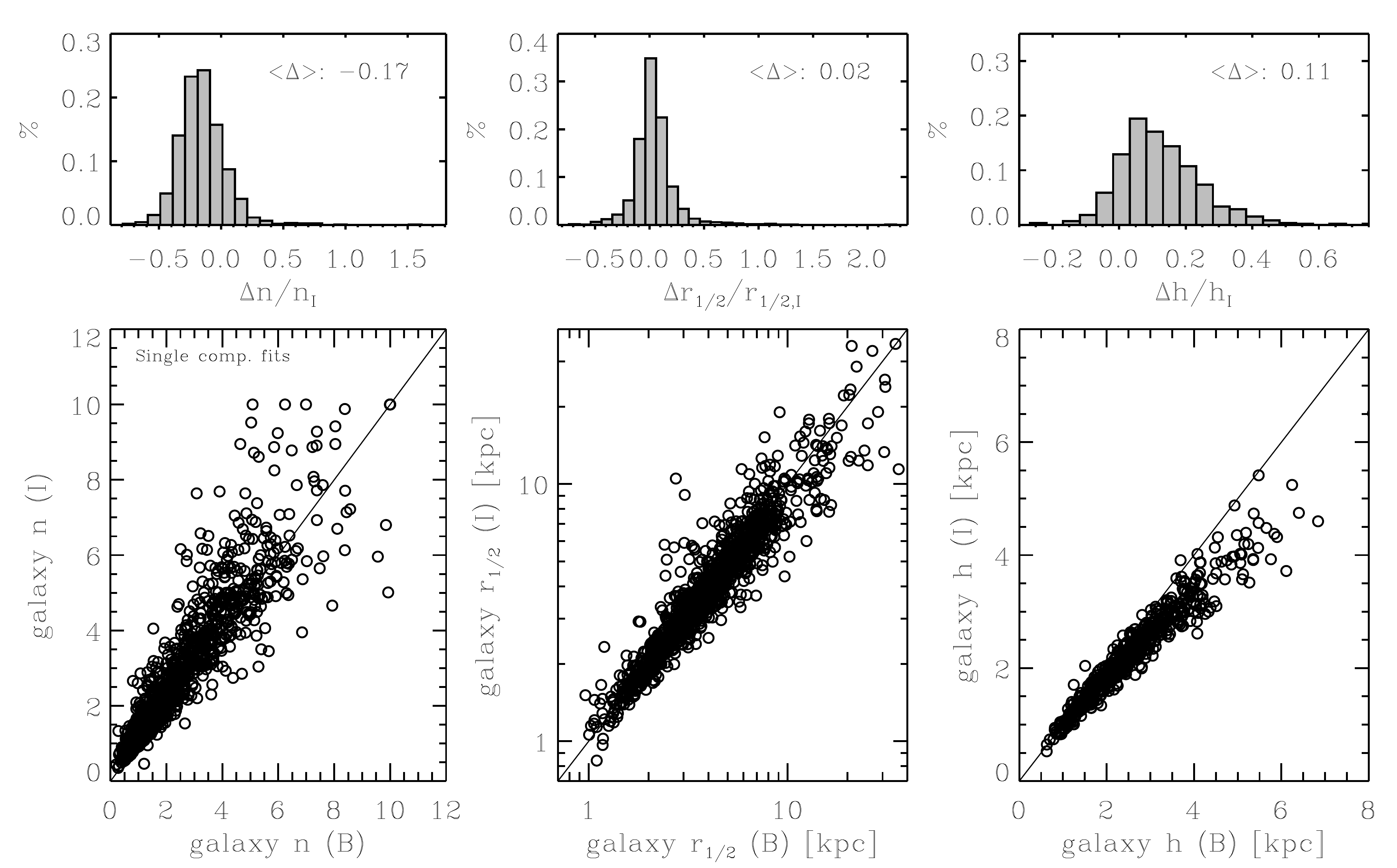}
\end{center}
\caption{\label{fig:GIM2D_comp1}Comparison  between half-light radii (left) and S\'ersic indices (center) obtained from the {\ttfamily GIM2D} single S\'ersic fits to the $I-$ and $B-$band data. The upper histograms show the distributions of the parameter differences in the two filters ($B-$band minus $I-$band), normalized to the $I$-band measurements.
The median value of the structural variation between the $I-$ and $B-$band is given inside these upper panels.   Values are presented for the \emph{uncorrected} GIM2D measurements, i.e., before applying  
our correction scheme for systematic biases described in Section \ref{sec:correctionMaps}. The rightmost panels show 
the comparison between the disk scalelengths obtained from the pure exponential  fits  in the two pass-bands
 to galaxies classified as late-type disks. In the $I$-band, variable-$n$  S\'ersic fits result in  $\sim17\%$ larger $n$  values than those measured for the same galaxies  in the $B$-band. Furthermore, 
 the $I-$band scalelengths  of $n=1$ pure exponential fits to relatively large late-type disks are  $\sim10-20\%$ smaller  than those derived from similar fits to the $B-$band images.}
\end{figure*}

 The  comparison between the structural parameters obtained from the single  S\'ersic fits with variable-$n$ in the two available passbands is shown in the left and central plots of Figure \ref{fig:GIM2D_comp1}. 
The single component fits in the two different pass-bands generally provide consistent measurements with little scatter. In particular, half-light radii measured in the two pass-bands are   in very good agreement with each other.
The $I$-band fits, however, result  in slightly steeper profiles: at this wavelength galaxies have $n$ indices which are bigger by   $17\%$ with respect to those measured in the $B$-band. This is not unexpected in an `inside-out' galaxy formation process, as younger stellar populations at large radii would lead to this effect.

Another   difference  is   observed between  $I-$ and $B-$band  disk scalelengths of large late-type galaxies, as derived from the pure exponential fits (right-hand side plot in\ Figure \ref{fig:GIM2D_comp1}):  disk scalelengths  $h$ derived from $n=1$ fits to  relatively large late-type disk galaxies are  smaller at the longer wavelength by $\sim10-20\%$.   We have checked that the $I$ magnitude distribution of these late-type disks is not biased toward faint values that would raise concerns on surface brightness detections.   As we show in detail in Section \ref{sec:correctionMaps},  the $I-$band \textsc{GIM2D} sizes that we derive for relatively large, $n\sim1$ galaxies are not severely affected by observational biases. We thus interpret this result again as mostly due to a genuine color gradient in large late-type disks. 
Note that the average difference in the $I$ and $B$ disk scalelengths in the pure exponential fits of late-type disk galaxies  is consistent with a similar difference reported for a sample of local late-type galaxies by \citet{deJong_1997} and \citet{Barden_et_al_2005}.

Summarizing, the larger $h$ values from $n=1$ fits  and smaller S\'ersic indices from variable-$n$ fits of late-type disks in the $B$-band relative to the $I-$band are consistent with being the joint result of the segregation of young stars in the galaxy outskirts and, also, of light absorption from the center of the galaxies  by interstellar dust, 
as discussed in a number of observational and theoretical works \citep[e.g.][]{Byun_et_al_1994, Beckman_et_al_1996,Cunow_2001,Moellenhoff_et_al_2006}.

The  more pronounced variation of disk scalelength in the pure exponential fits with respect to the half-light radius in the  S\'ersic fits for the late-type galaxies can be readily understood: whereas in the variable-$n$ single S\'ersic fits a lower central concentration of light can be modeled with  a smaller value of the S\'ersic index $n$ and  a relatively small change in the effective radius, fixing the index   to $n=1$   in the  pure exponential fits forces an increase of the characteristic scalelength
in order to obtain  a milder radial  decline in  surface brightness in the $B$ band. This is  consistent with  the fact that the largest variation in $n$ values between the $B$- and $I$-band  variable-$n$  S\'ersic fits to late-type disks are observed for those galaxies which also have the largest variations in scalelength $h$ when fitted with an $n=1$ profile. 

Finally we note that a self-consistent correction for dust effects would require radiative transfer simulations of the light scattered and re-emitted by the dust grains \citep[e.g.][]{Byun_et_al_1994,Cunow_2001,Tuffs_et_al_2004}; this is beyond the scope of this paper.  Moreover it remains difficult to disentangle dust effects from genuine radial segregation in the stellar populations.
 For these reasons, instead of attempting a correction for dust absorption, we choose the empirical approach of  employing the less dust-sensitive $I$-band data as the fiducial reference for
 our structural measurements,  and to discuss separately possible dust-reddening effects, when relevant, e.g., in studying  color profiles and star formation rates (see Paper III).

\subsection{Details of the bulge+disk(+bar) decompositions}\label{sec:GIM2D_decomp}

Double component, bulge+disk decompositions were also performed on both the $B$- and $I$-band images of all galaxies
which are not classified as ellipticals or irregulars (Section \ref{sec:MorphClass}).
For the $I$-band, our reference filter for structural measurements, no a priori constraints on the bulge and disk parameters were imposed during the fitting procedure, except for the wide limits listed in Table \ref{tab:GIM2D_parameters}.
In the $B$ filter,  each galaxy was instead fitted in four different ways: $(i)$ by performing a separate decomposition to the $B$-band letting the structural parameters completely unconstrained (unconstrained model fit,  hereafter $UF$), $(ii)$ by fixing the disk and bulge position angles, ellipticity and inclination to those of the $I$-band (constrained model fit number one, hereafter $C1$), $(iii)$ by also fixing the bulge half-light radius and S\'ersic index $n$ (constrained model $C2$) and $(iv)$ by keeping all the parameters tied to those of  $I$-band except for the $B/T$ and total flux (constrained model $C3$).
In all the models we allowed a maximum wandering of 2 pixels for the disk and bulge components from the  \textsc{SExtractor} center.

There are clear  advantages and disadvantages in performing either independent or constrained fits to the two bands:  by fixing the $B$-band structural parameters to the $I$-band, one ensures that bulge and disk colors are measured consistently over the same regions; this however prevents the detection of structural differences and color gradients.  
For this reason we decided to adopt a mixed approach to determine the bulge and disk parameters to the $B$-band.

First, not all  models returned by the GIM2D fits are physically-meaningful bulge+disk decompositions. We hence adopted a filtering scheme to reject unreliable or unphysical models to all \textsc{GIM2D} models, i.e., the fiducial $I-$band models for each galaxy, and the  four versions of the $B-$band models.  Our filtering scheme is  described in Appendix \ref{sec:testGIM2D}.  

We then followed a quantitative procedure to select, amongst the physically-valid alternatives for the $B$ 
bulge+disk fits for each galaxy,  our fiducial (i.e., in our judgement, the most reliable) bulge+disk $B-$band decomposition.
In brief, we required that all $B-$band disk and bulge fits  always have bulge and disk position angles, disk inclinations and bulge axis ratios consistent, within a sensible range\footnote{The  allowed ranges of  variations for $B$ and $I$ bulge and disk position angles, disk inclination angles and bulge ellipticities were respectively $15\deg$, $15\deg$ and 0.15.}, to those of the $I-$band fits. When this was achieved with unconstrained fits to the $B$ band images, these unconstrained fits were retained as a fair description of the $B$ bulge+disk  decompositions. This   choice maximizes the detection of possible wavelength-dependent structural differences and color gradients. For galaxies in which such a consistency requirement was not achieved with the unconstrained $B$ fits,  we adopted as fiducial  $B-$band fits those which satisfied such requirement  with the minimum number of $B$ parameters tied to the $I$-band fit parameters (i.e., in order of priority, the $C1$, $C2$ and $C3$ fits).  

\begin{figure*}
\begin{center}
\includegraphics[width=120mm]{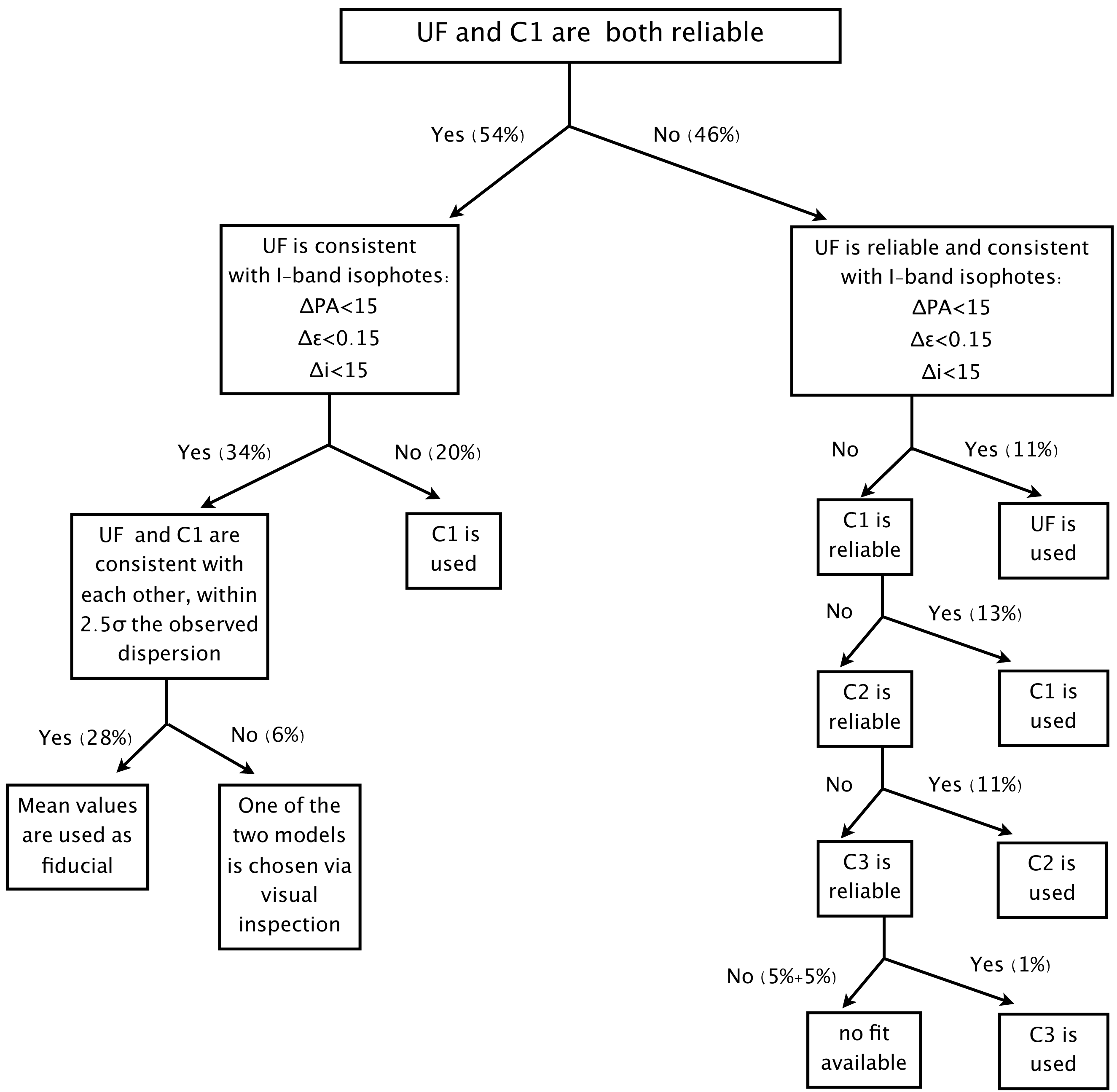}
\end{center}
\caption{\label{fig:BTchart}Flow chart  describing our  scheme for selecting  the fiducial \textsc{GIM2D} bulge+disk $B-$band 
models. $UF$, $C1$, $C2$ and $C3$ are the four fits that we performed on the B images, respectively by keeping the $B$ bulge and disk parameters completely unconstrained, fixing their position angles and ellipticities  to those derived from the $I$-band fits, adding to the $I-$band-fixed parameters the bulge half-light radius and S\'ersic index $n$, and finally, by tying   to the $I-$band values also the $B$ disk scalelengths.  Number in parenthesis give the fraction of $B$-band bulge+disk decompositions which fall in each category; note that sub-branches sum up to the fractions listed on the previous level.
The fractions refer to  disk galaxies with a detected bulge component in the \textsc{ZENS} sample, i.e., to S0, bulge-dominated spirals and intermediate-type disks.
The two numbers for the models with no bulge+disk decomposition in the $B$-band give the fraction of galaxies for which either an $I$-band
fit is available but no good $B$-band fit can be achieved ($5\%$), or the fraction of galaxies for which neither an $I$ nor a $B$ bulge+disk decomposition can be achieved ($5\%$).}
\end{figure*}

In cases where both the  $C1$  and $UF$ models were in principle both potentially good representations of the bulge+disk properties of considered galaxy, we applied the following decision scheme: if the two models gave disk and bulge size, $B/T$ and S\'ersic indices within $2.5\sigma$ of the dispersion measured around the identity from all the model falling in this latter category, then both models were validated as reliable, and we adopted as our fiducial bulge+disk parameter estimates the 
mean of the structural parameters returned by these two fits.
If these two fits returned discrepant values, both were inspected and a judgement was made on which model to use, on the basis of the residual images and the difference in total magnitude between the galaxy and the model.
Only $6\%$ of the bulge+disk $B$ fits needed this further visual validation. This entire procedure is schematized in  Figure \ref{fig:BTchart}.

\begin{figure*}
\begin{center}
\includegraphics[width=110mm,angle=90]{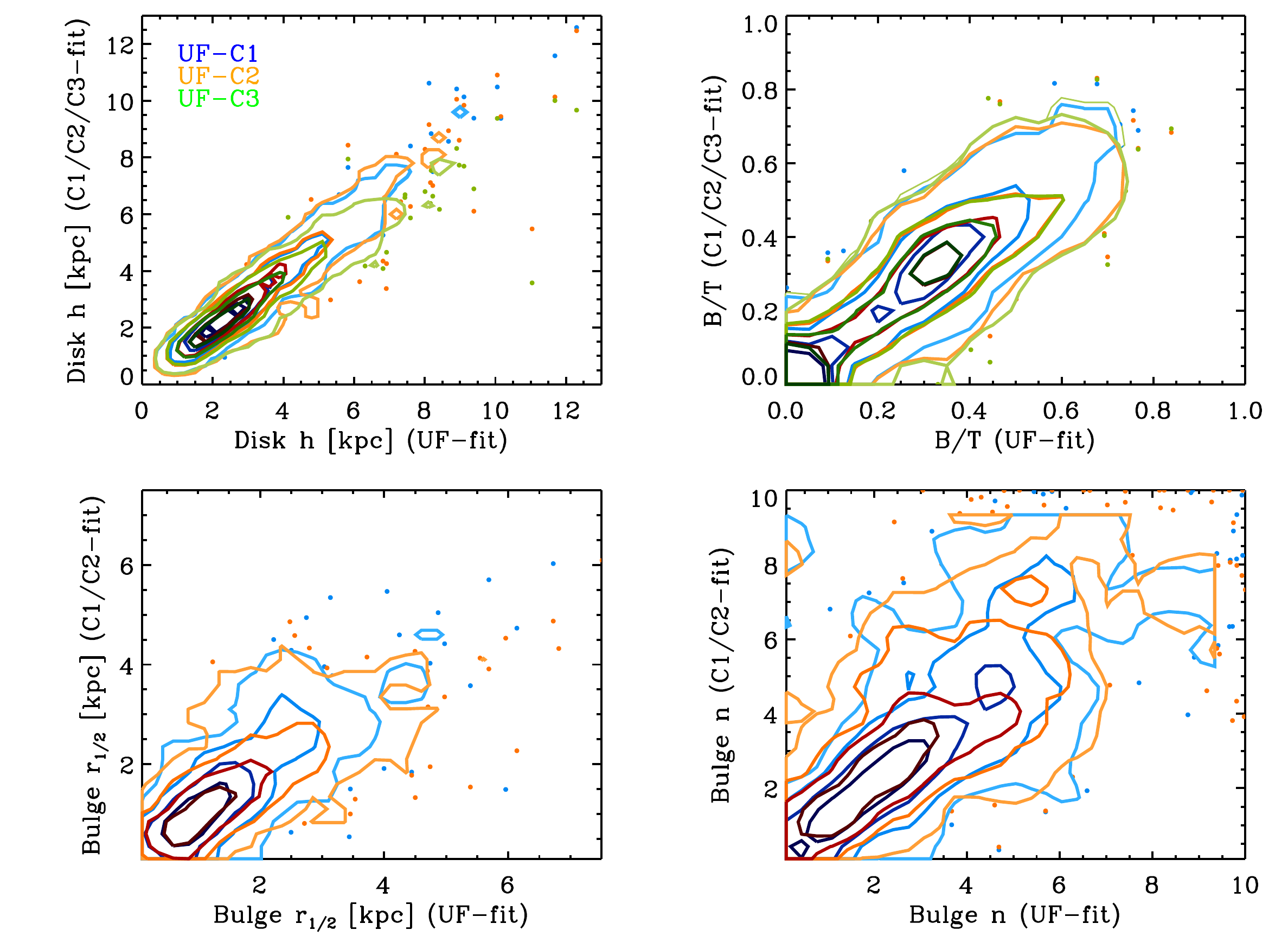}
\end{center}
\caption{\label{fig:GIM2D_BRuns}  Comparison  between the bulge and disk structural parameters obtained with {\ttfamily GIM2D} in the four different $B$ fits described in section \ref{sec:GIM2D_decomp}. These four $B$ fits vary between being totally unconstrained ($UF$), and fully tied too the $I-$band best fit parameters ($C3$).
From top to bottom and left to right we show the disk scale-length, bulge-to-total ratio, bulge effective radius and bulge S\'ersic index. The contours identify isodensity regions containing 25$\%$, 50$\%$, $75\%$ and $95\%$ of the data. The points show the models which are not encompassed by the contours.
The three different colors highlight the comparison between the unconstrained   $B$ fits ($UF$) with the $B$ models in which the position angles, ellipticity and inclination angle are tied to the $I$-band ($C1$, blue contours), with those in which also the bulge half-light radius and S\'ersic index are kept fixed ($C2$, orange contours) and with the models which are fully tied to the $I$-band structural parameters ($C3$,  green contours). 
The green curves  are shown only for  disk scalelengths and  bulge-to-total ratios, as model $C2$ and $C3$ have the same bulge parameters.}
\end{figure*}

It is important to notice that, while disk scalelength and bulge-to-total ratios are quite consistently returned by all four fits to the $B$ images,  bulge sizes and S\'ersic indices show much larger variations from one fit to the other.
This is illustrated in Figure \ref{fig:GIM2D_BRuns}, where we compare the different measurements obtained for the $B$-band in the four cases $UF$, $C1$, $C2$, $C3$.  The robust dispersion around the identity line in the four fits is  $\sim 0.2$ kpc and 8$\%$ for $h$ and $B/T$,   and $\sim$0.6 kpc and 1.6 for the bulge half-light  radii and S\'ersic indices $n$, respectively.
 Especially for the bulge $n$ values, models with the isophote's position constrained and unconstrained can indeed provide very different results. Similar conclusions on the reliability of the bulge and disk parameters are drawn from
 tests on simulated galaxies which we discuss in Section \ref{sec:BulgeDiskBias}.

 Not surprisingly, the success rate for the bulge+disk decomposition is lower than the one for the
 single component. The single component fits return robust measurements  for  $\sim95\%$ of  S0, bulge-dominated spirals and intermediate-type disk galaxies in the both the $B$- and $I$-bands (see Appendix \ref{sec:testGIM2D} for a detailed summary for the individual morphological classes).  Bulge+disk decompositions in both  $B$- and $I$-bands are available for $\sim80\%$ of these galaxies.

\begin{figure*}
\begin{center}
\includegraphics[height=14cm,width=10cm,angle=90]{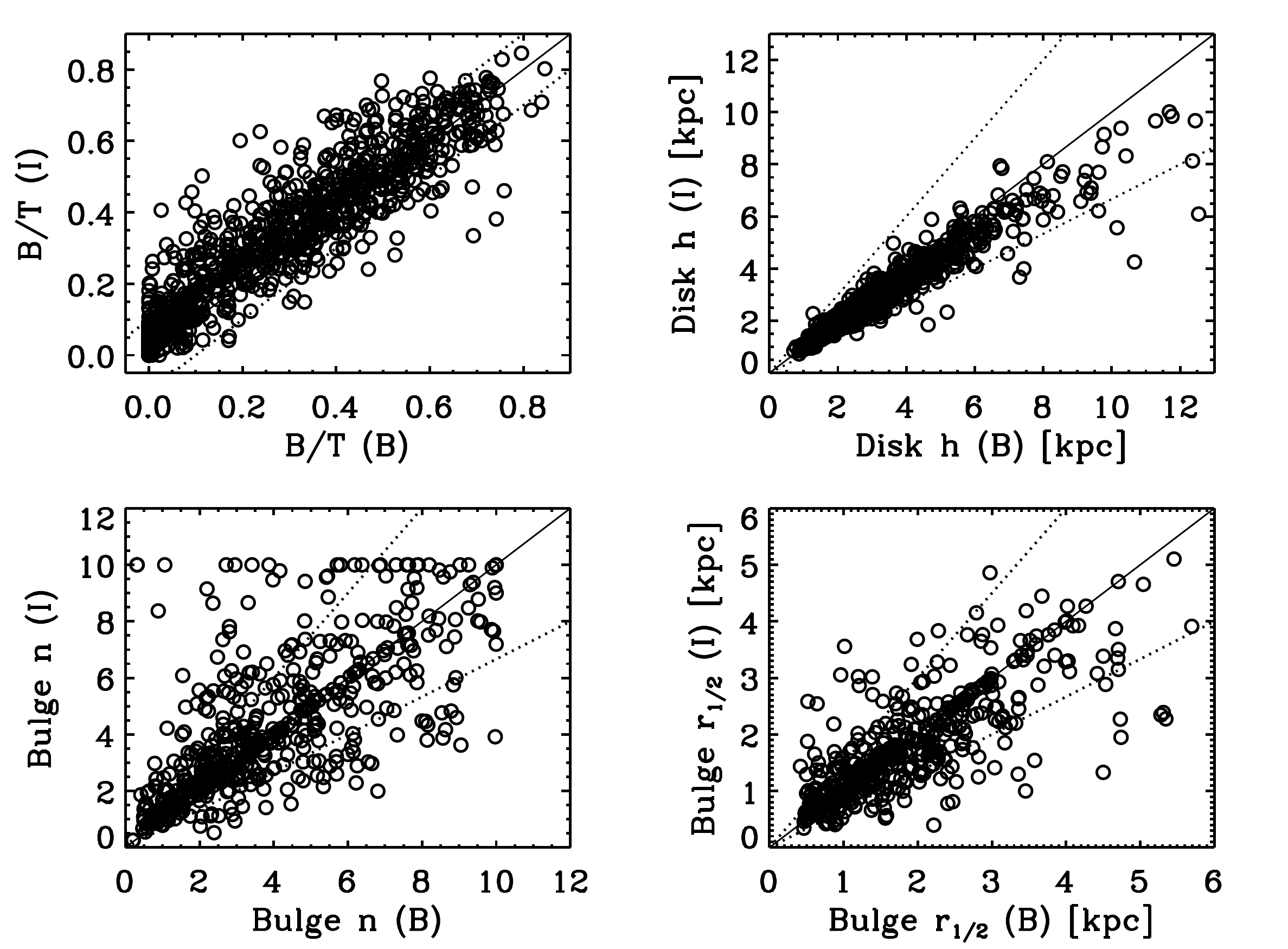}
\end{center}
\caption{\label{fig:GIM2D_comp2} Comparison between the  {\ttfamily GIM2D} double-component $I$-band fit parameters, and the  corresponding parameters in the finally-adopted, fiducial   $B$-band fits. Specifically, the upper panels show the comparison between the  bulge-to-total ratios and disk scale lengths in the two filters. The lower panels show the comparison between the bulge half-light radii and bulge S\'ersic indices. In the upper left panel, values are plotted only for  galaxies which have detected bulge and disk component  in both bands. In the other panels,  galaxies are further constrained to have $B/T>0.1$ when considering bulge parameters. The dashed lines in the upper left
panel indicate a positive and negative variation of $0.1$ in the bulge-to-total ratio with respect to the identity line; in the other three panels, they indicate a relative change of a factor of 1.5.}
\end{figure*}

For galaxies with a reliable decompositions in both bands, Figure \ref{fig:GIM2D_comp2} presents the parameters obtained in the two pass-bands. Partly by construction, the scatter around the identity line for the bulge-to-total ratio is less than $0.1$ in the vast majority of the cases, as shown on the top left panel of the Figure.
Given that our morphological classification is based on the $I-$band bulge-to-total ratios, this means that our classification would not change substantially  if we had used the $B$-band measurements instead.
The disk sizes obtained in the two filters agree well with each other, although disks are more extended in the $B$-band than in the $I$-band, as also discussed above.
As anticipated, the bulge radii and indices have instead a broader scatter. 
Note that, despite the scatter, in the vast majority of  cases the S\'ersic indices for the bulge are consistent with either a `concentrated' or a `diffuse' central bulge component in both filters; truly discrepant results are obtained  in a small number of cases ($\sim 5\%$).  

Finally, in  Appendix \ref{sec:testGIM2D} we show  the comparison of our bulge+disk GIM2D fits with similar  obtained with the package GALFIT \citep{Peng_et_al_2002}, and also with bulge+disk+bar fits; the latter provide an estimate for the impact of a bar component on the bulge (and disk) parameters for disks which host such a third component.

 \subsubsection{Comparison between galaxy half-light radii estimated from single S\'ersic fits and from bulge+disk fits}\label{sec:LargeRadiiDiff}

\begin{figure*}
\begin{center}
\includegraphics[height=10cm,angle=90]{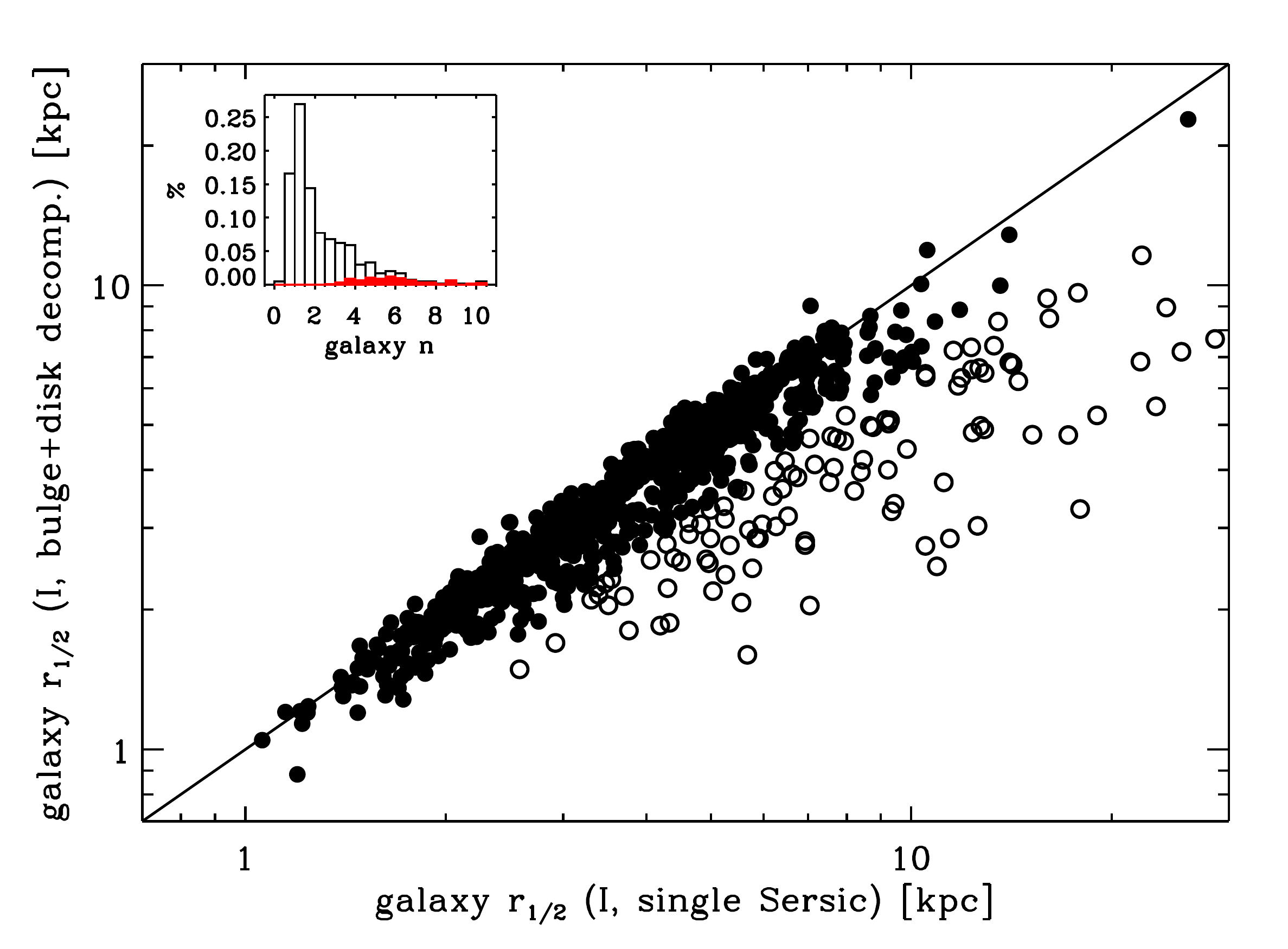}
\end{center}
\caption{\label{fig:Re_SersicBD} Comparison between $I$-band galaxy half-light radii
 obtained from single S\'ersic fits and bulge+disk decompositions.
 The empty circles highlight those galaxies for which the two size estimates differ by more than a factor 1.5.
 The distribution of galaxy S\'ersic indices for such discrepant cases are shown in grey (red in the online version) in the inset on the top left corner of the plot,  in comparison with the indices for the rest of the sample, shown in black.
 (A color version of this figure is available in the online journal.)}
\end{figure*}

As  discussed extensively in Section \ref{sec:correctionMaps}, the calculation of a galaxy size
radius is made difficult by a number of observational biases which can lead to substantial uncertainties.
Here we additionally show how the estimation of the galaxy half light radius depends on the choice of the specific  model to the galaxy light,  by comparing the global galaxies half-light radii derived from the single S\'ersic fits with those obtained from the bulge+disk decomposition.
This is shown in Figure \ref{fig:Re_SersicBD} for the $I$-band, but similar results are obtained in the $B$-band.
It is clearly seen that for about $10\%$ of  the galaxies with formally `reliable' bulge+disk decomposition and single component fits, 
half-light radii from the single S\'ersic models are larger by more than a factor 1.5 than the sizes inferred from the bulge+disk decompositions. 
 As illustrated in the inset in Figure \ref{fig:Re_SersicBD}, the majority of the discrepant galaxies have steep light
 profiles with Sersic index $n>2$.
We note that for all these galaxies both the single and double component fits were inspected to confirm their formal reliability;
 furthermore, the magnitude difference between the model galaxy and the real ZENS image within an aperture equal to 1.5 times the Petrosian radius is $<0.3$ mag 
 for both the single and double component fits.

 For these  galaxies we decided to keep both radii estimates in our ZENS catalogue (published with Paper I) but, to keep memory of the discrepancy,  to add to the formal errors an uncertainty  equal to half the difference between the two radii estimates.
As an indication that these difference in the measured radii are not peculiar to the ZENS sample only or to the GIM2D software, we remark that other recent publications have found evidence of an overestimation of the half-light radius when a single S\'ersic fit is performed on a galaxies which have intrinsically a bulge and a disk component, for simulations reproducing SDSS noise properties and also using different fitting algorithms \citep{Meert_et_al_2012,Mosleh_et_al_2013}.
 
 \section{Quantification of galaxy structure. III. Non-parametric analysis} \label{sec:ZEST}
 
We further used the \emph{ Zurich Estimator of Structural Types Plus} (\textsc{ZEST+}), 
 an upgrade of the \textsc{ZEST} approach published in \citealt{Scarlata_et_al_2007}, to derive non-parametric measurements of structure for the ZENS galaxies. Such estimates add further information on the structural properties of the galaxies, are useful in cases where no parametric fit could be performed, and enable comparisons with other published samples.

\textsc{ZEST+} is a {\ttfamily{C+}}-based code for the study of galaxy structure and it is designed for automated morphological classification of galaxies through either a principal component analysis or
a support vector machines technique; it features  several improvements relative to the  \textsc{ZEST} algorithm and implementation of Scarlata et al.\ (2007; see for example \citealt{Cameron_et_al_2010} for a first  application to the COSMOS field; \citealt{Scoville_et_al_2007}). The classification scheme uses both user supplied parameters and/or  a set of non-parametric morphological coefficients computed by \textsc{ZEST+} itself. 
For the \textsc{ZENS} galaxies we employed \textsc{ZEST+}  only to derive the structural coefficients, rather than using the morphological classification option, since, given the limited size of our sample, we were able to  perform  visual and quantitative checks   on the reliability of the classification for each galaxy individually, which we mainly based on the available bulge-to-disk ratios (see Section \ref{sec:MorphClass}).

The measurements provided by  \textsc{ZEST+} are galaxy concentration $(C)$,  asymmetry ($A$),  Gini and M$_{20}$ coefficients, and smoothness ($S$). 
These  parameters  are widely used in the literature to study galaxy structure and morphology (e.g. \citealt{Conselice_2003}, \citealt{Lotz_et_al_2004}, \citealt{Scarlata_et_al_2007}, \citealt{Zamojski_et_al_2007}).
 For sake of brevity in the following we will refer to the $C$, $A$, Gini, M$_{20}$ and $S$ set of measurements as to the ``$CASGM$" parameters.
 
The concentration $C = 5 \log (r_{80}/r_{20})$ was defined as the ratio of elliptical radii containing $80\%$ and $20\%$ of the total flux of the galaxy, which is provided as input by the user. For  \textsc{ZENS}, the \textsc{SExtractor} Kron flux ({\ttfamily{FLUX$\_$AUTO}}) was employed.

The asymmetry index $A$ gives information on the degree of rotational symmetry of the galaxy light. This is parametrized through the difference between the original galaxy image and a 180$\deg$-rotated version of it. To account for the effect of background noise, \textsc{ZEST+} uses the procedure introduced by \citet{Zamojski_et_al_2007}, which involves the calculation of the asymmetry for both the original image $A_0=\frac{1}{2}\frac{\sum_{i,j} |I (i,j) -I_{180}(i,j)| } {\sum_{i,j}|I(i,j)|}$ and for a smoothed version of it, $A_{0,S}=\frac{1}{2}\frac{\sum_{i,j} |I^S (i,j) -I^S_{180}(i,j)| } {\sum_{i,j}|I^S(i,j)|}$, where $I$ is the intensity of the image on the pixel $(i,j)$ and $I_{180}$ the intensity of the rotated image.  The final asymmetry value is given by $A=A_0-\frac{A_0-A_{0,S}}{1-1/\sqrt{5}}$,
where the last term corresponds to the background correction factor.

The Gini coefficient $G$ introduced by \citet{Abraham_et_al_2003} contains information on how uniformly the light is distributed within the galaxy: if the flux is equally distributed among all pixels, then $G$ is equal to zero, whereas if all the light is concentrated in just one pixel, the coefficient $G$ is equal to unity. We  defined it as in \citealt{Lotz_et_al_2004}: $G =  \frac{1} {\bar{I}n(n-1)}\sum_i^n(2i-n-1)\bar{I}_{i}$,
where $\bar{I}$ is the mean of the flux of the galaxy pixels, sorted in increasing order.

The parameter $M_{20}$ is the normalized second-order moment of the brightest $20\%$ of the galaxy pixels. It describes the spatial distribution of bright substructures within the galaxy, such as spiral arms, bars or bright nuclei.
The computation of $M_{20}$ involves the following steps: $(i)$ galaxy pixels  are ordered by flux, $(ii)$  for the $20\%$ brightest pixels, the sum of moments $\Psi= \sum_i^{n_{20}}I_{i}[(x_i-x_c)^2+(y_i-y_c)^2]$ with respect to the light-center in the Petrosian ellipse $x_c$, $y_c$
is computed, and $(iii)$  the latter is normalized by the total sum of moments to give $M_{20} = \log_{10} \left (\frac{\Psi}{M_{tot}} \right)$, with $M_{tot}=\sum_i^{n_{tot}}M_i$.

Finally, the smoothness $S$ is a measure of the degree of clumpiness of the galaxy light distribution, and is thus useful to trace patches in the light profile such as star-forming regions. 
To calculate $S$, a smoothed version of the original image, obtained by convolving it with a gaussian filter of FWHM equal to 0.25 times the petrosian radius $R_p$, was subtracted from the image itself. 
Clumpy regions were then identified from the residual image  as those pixels for which the intensity $I_{res}$ is $k$ times higher than the background standard deviation in the residual image  $\sigma_{bkg}$.
We used a default threshold factor of $k=2.5$. 
The pixels so identified were then used to calculate $S=\left(\frac{\sum_{i,j} I_{res}(i,j)}{\sum_{i,j}|I(i,j)|}\right)_{I_{res}(i,j)>2.5\sigma_{bkg}}.$
A region of radius $0.25*R_p$ from the center was masked out during the calculation to avoid including the highly concentrated centers of the galaxies  which will boost the final value of the smoothness. 
The Gini, M20, $A$ and $S$ indices were all calculated within one petrosian ellipse.

Along with the structural indices, \textsc{ZEST+} also gives an estimate for the ``elliptical aperture" galaxy half-light radius,  based on the user-provided total flux. These are the measurements we refer to when discussing \textsc{ZEST+}  half-light radii for our \textsc{ZENS} galaxy sample.



\section{Corrections for systematic biases in the structural parameters} \label{sec:Simulations}

The careful inspection and filtering of the measurements thus far presented, although necessary, 
does not provide a quantification of  systematic  errors
 in the derived structural parameters.  
In ZENS, and more generally in ground-based imaging galaxy surveys,
 the major contributors to such errors can be identified in the noise of the night sky, and
 in the  PSF width. In ZENS, the median $I$-band PSF FWHM  is $1"$, which is about 30\% of the median half-light radius of the galaxy sample. 
Both analytical surface brightness fits and non parametric algorithms are affected 
by these observational limitation and none is completely free from pitfalls:   
the \textsc{ZEST+} measurements are not PSF-deconvolved and, to obtain a consistent measure of structure on the
different ZENS fields, an homogenization to a common resolution of the 
$CASGM$ parameters  is necessary. In contrast,
\textsc{GIM2D} uses the provided PSF to derive, in principle, seeing-corrected quantities, and indeed
biases related to PSF-blurring are less severe in GIM2D-based measurements;
 however, systematic uncertainties remain, especially
in low surface brightness regimes, where both aperture photometry and two-dimensional fits are well known to underestimate
 galaxy sizes and fluxes \citep[e.g.][]{Bernstein_et_al_2002a, Bernstein_et_al_2002b,Benitez_et_al_2004,Haussler_et_al_2007,Cameron_Driver_2007}.    
 
We  asses here the impact of these effects specifically for our ZENS measurements through tests on artificial galaxy images. In particular, we derive recipes to correct  the observed
galaxy structural parameters (e.g., sizes, concentrations, ellipticities, etc.)
for  biases in the observations. In areas of parameter space where we cannot recover the true values of the given parameters, we  provide
 an estimate for the systematic uncertainty that affects the measurements.
 
 We perform this analysis on the $I$-band, which     provides a  view of the intrinsic galaxy structure less affected by dust or young stars thus is used as our fiducial band to classify galaxies structurally and morphologically.  The corrections in the $B$-band are then derived from those obtained for the $I$ measurements, suitably rescaled to the $B$-band luminosity and PSF.

\subsection{Methodology}

The derivation of the corrections is done by applying  the entire process of object extraction, followed by parametric and non-parametric photometric/structural measurements with \textsc{SExtractor/ZEST+}   and \textsc{GIM2D}, in exactly the same way as on the real data, to a set of artificial galaxy images  for which the intrinsic structural/photometric properties are know precisely by construction. The difference between the model input and output parameters provides an estimate of the uncertainties in the  measurements. 

A self-consistent study of the errors in the quantification of the galaxy structural properties must be a
function of five parameters:  the galaxy luminosity (magnitude, $mag$), the galaxy size (half-light radius $r_{1/2}$), the inclination (ellipticity $\epsilon$),  the  steepness of the
light profile (S\'ersic index $n$ or concentration $C$) and  the PSF under which the galaxy was observed. 
The effects of each of such parameters on the quantification of structure, including galaxy sizes, are tightly interconnected and hence need to be considered simultaneously.
It is immediately clear that, as an example, the impact of the PSF is stronger --  with all other parameters fixed -- for highly concentrated galaxies than for those with shallow light profiles, and for galaxies with sizes comparable to the seeing than for more extended ones.

We thus adopt a sampling approach in which we construct many thousand artificial galaxy models to fully explore the observed parameter space of galaxies in the ZENS data set, and such to have, for each combination of the five parameters $mag-r_{1/2}-\epsilon-n$(or $C)-PSF$, a sufficient number of models on which to test the measurements.
This approach is also used in \citealt{Carollo_et_al_2013b} on a higher redshift sample of galaxies extracted from the COSMOS survey.

Two cautionary remarks need to be made: first, our artificial galaxies are generated to populate \emph{uniformly} a broad grid in the five-dimensional $mag-r_{1/2}-\epsilon-n$(or $C)-PSF$ space. Real galaxies are not   uniformly distributed in this parameter space, and this is not taken into account in our corrections.
Second, we deliberately ignore dust or stellar population segregation effects, as our models are created smooth and neglect dust attenuation. 
In the following we will test  the robustness of our corrections towards the specific design of the simulations.
 
\subsection{Generation of  the artificial galaxy images}\label{sec:simulations_design}

To derive our correction functions for the structural parameters of \textsc{ZENS} galaxies  we created both S\'ersic models and bulge+disk artificial galaxies  on which we 
tested the corresponding \textsc{GIM2D} fits. 
In both cases, models were constructed on a grid of points
 in the ellipticity, magnitude, size and S\'ersic index (or B/T ratio) parameter space.
 Specifically, in the single component case galaxies were simulated around the following regions: $r_{1/2}=[0.4,0.7,0.9,1.3,1.5,2,3,8,20]$ kpc, 
 $I_{AB}=[13,14,15,16,17,18,19,20]$, $\epsilon=[0.1,0.3,0.6,0.9]$ and $n=[0.5,1,1.5,2.5,3.5,4.5,7.5]$.
For each grid node 30 models were generated randomly to have: radii and S\'ersic indices within $\pm$ 30$\%$ of the nominal radius and index $n$ at the grid point,  magnitude within $\pm$ 0.25 magnitudes and a difference in ellipticity equal to $\pm$ 0.05. 
This makes a total of $\sim 60'000$ single-component model galaxies.
 
 For the double component model galaxies, which span a wider range of combination of  bulge and disk parameters and are also computationally more expensive, we used a coarser sampling of the parameter space, creating models in the following way: given a value of the magnitude for the \emph{entire} galaxy randomly generated around the points $I_{AB}=[13.7, 15.4, 16.8, 18.2, 19.6]$ and a disk scale length $h$ similarly chosen within $\pm30\%$ from the positions $[0.5, 1, 1.5, 3, 6, 15]$ kpc, artificial galaxies were constructed in three bins of bulge-to-total ratios centered at values of  $B/T=0.15, 0.4, 0.65$ and of width $\Delta(B/T)=0.15$, hence allowing a maximum $B/T=0.8$.  The bulge half-light radius was selected on the same grid used for the disk scale length, but imposing that  $R_{e}<1.678\times h$. 
 The bulge S\'ersic index was allowed to have values  $n=0.5, 1, 2.5, 4, 8$ with a scatter of $\pm 30\%$ and three bins of disk ellipticity (i.e galaxy inclination) $\epsilon_{disk}=0.1, 0.4, 0.7 \pm 0.1$ were employed.
 We furthermore assumed that bulges cannot be very elongated and hence explored only two  values of the ellipticity for the bulge component, namely $\epsilon=0.1$ and $\epsilon=0.6$, so to bracket any other value in between.

 All model galaxies described above were then convolved with three
 PSF sizes, for a total of 180'000 single component models and $\sim 90'000$ double-component models. The three PSF sizes were taken to reproduce the best, medium and worst ZENS seeing in the $I-$band (respectively  0.7$^{\prime\prime}$, 1$^{\prime\prime}$, and 1.5$^{\prime\prime}$). The PSF-convolved models were inserted with Poisson sampling into sky-subtracted  empty regions extracted from the real ZENS fields.
To mimic noise in the reconstruction of the PSF in the real data, when performing the \textsc{GIM2D} fits on the simulated galaxies we provided as input to the code a  rotated  version of the PSF originally used to convolve the artificial image.
 
For brevity, we focus in the following on the results obtained for the simulations convolved with the median \textsc{ZENS}  PSF, which are hence representative of the bulk of the ZENS observations. In Appendix \ref{app:corrections_PSF} a full account of the results derived with the best and worst PSF can be found
(see Figures \ref{fig:GIM2DCorrections_best}, \ref{fig:GIM2DCorrections_worst}, \ref{fig:ZESTCorrections_best} and \ref{fig:ZESTCorrections_worst}), which shows correction matrices that are consistent with those here presented. Note that the corrections to each \textsc{ZENS} galaxy were obtained  through linear interpolation, at  the PSF size value relevant for any given galaxy, of the correction matrices  describing PSF sizes that bracketed the PSF in question.

\subsection{Implementing the derived corrections to galaxy sizes and magnitudes}\label{sec:correctionMaps}

We start the description of the resulting correction functions by focusing on those parameters which our approach can self-consistently correct without recurring to additional information, namely galaxy sizes and magnitudes. We show in Section \ref{sec:CorrDegeneracy} how for other structural properties, such as the concentration coefficient, the recovery of the ``true" values is instead less straightforward, and depends on the intrinsic  distribution of parameters in the simulated sample (and requires therefore a different approach). 

Our fiducial measurements of galaxy sizes in \textsc{ZENS}  are those that we derive from the GIM2D analytic fits, which are less prone to systematic biases then those inferred with \textsc{ZEST+}, as we show below.
It is nonetheless worthy to discuss here also the results for the \textsc{ZEST+} measurements, given that  these measurements are used for the few ZENS galaxies for which no reliable \textsc{GIM2D} fit could be achieved.   
 
\begin{figure*}
\begin{center}
\includegraphics[width=140mm]{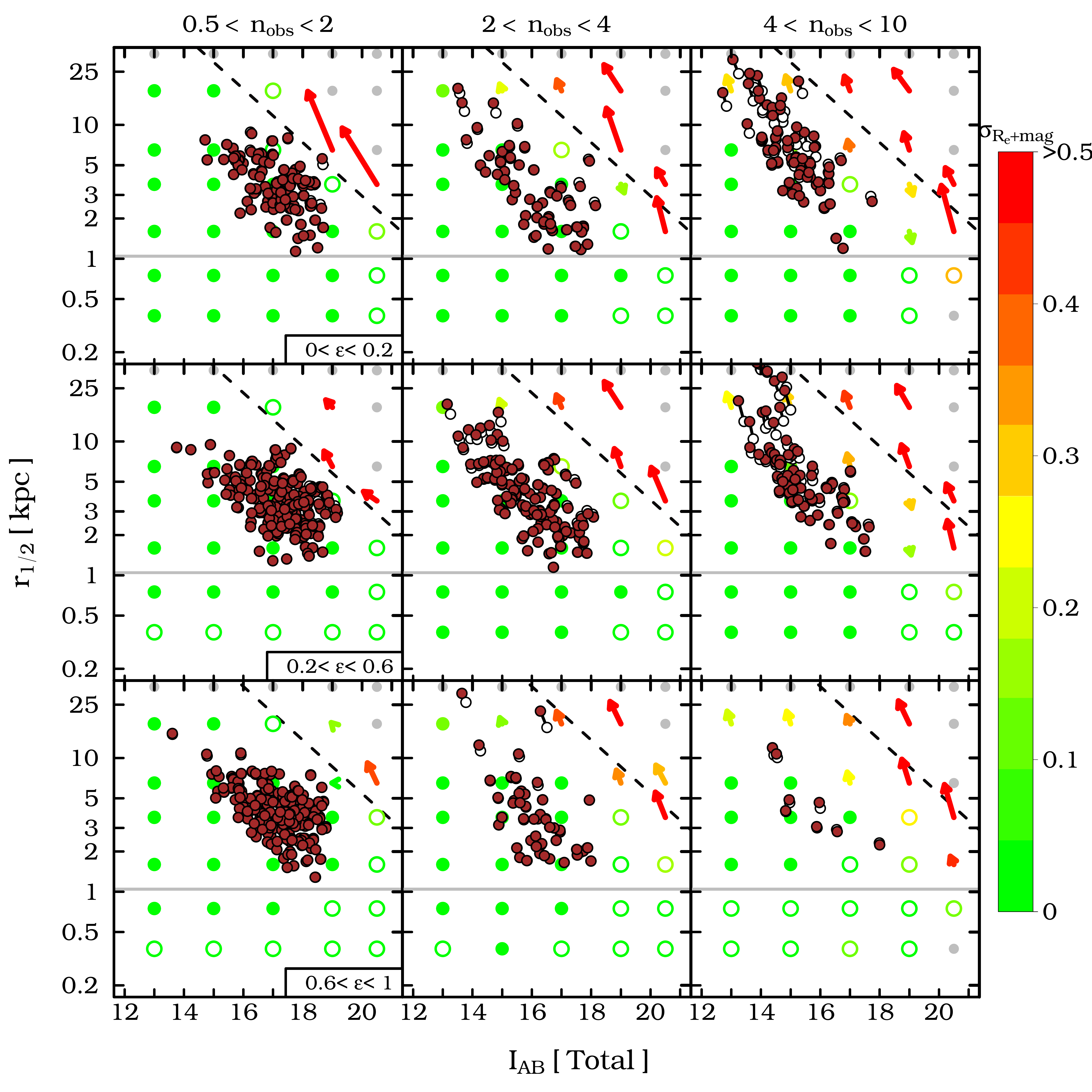}
\end{center}
\caption{\label{fig:GIM2DCorrections} The size-magnitude plane with arrows  illustrating the strength of the correction vector that must be applied,  at each grid point, to recover intrinsic total magnitudes and sizes from the observed magnitude and sizes (as measured by \textsc{GIM2D} in the ZENS $I$-band imaging). The corrections  are based on single S\'ersic fits to single S\'ersic galaxy models; they are shown binned in three different panels of observed S\'ersic index (from left to right) and three panels  of observed galaxy ellipticity (from top to bottom).  Arrows represent the direction and strength of the corrections at each grid point; they are obtained as the median difference in magnitude ($\bar{\Delta m}$)  and radius ($ \bar{\Delta r_{1/2}}$) between the input models and the measured parameters at the given grid point.
Colored circles show positions in the plotted parameter space in which no corrections to magnitudes and sizes are required; these are defined as grid points in which 80$\%$ of the models have a correction in radius and magnitude which are, respectively, $<20\%$  and $<0.3$ mag.
The colors of the arrows and circles indicate the amount of  scatter in the individual contributing models relative to the shown median correction. The scatter increases from green to red and is defined as the quadratic sum of the median absolute deviations of $\frac{\Delta R_i}{R_i}$ and 
$\frac{\Delta F_i}{F_i}$, with  $\Delta R_i$ and  $\Delta F_i$  the size and flux  differences between each individual model at the corresponding median value. The precise values of the scatter around the medians at any color is given in the color chart on the right-hand side  of the figure.
Empty colored circles represent  regions in the $mag-r_{1/2}$ plane where the recovery of the models' ellipticity and S\'ersic index is subject to large uncertainties; precisely, empty colored circles show those grid points in which at least 25$\%$ of the models have a difference between input and output $\epsilon$ or $n$ which is larger than 0.08 and 15$\%$, respectively.
The gray horizontal line marks  the value of the typical PSF FWHM for the ZENS observations; the dashed black line highlights the surface brightness limit 
of our images. Gray dots highlight regions of incompleteness in the observed space, i.e., which are populated by less than 10 model galaxies.
Corrected radii and magnitudes for the real ZENS galaxies that were observed with a seeing $\sim 1^{\prime\prime}$  are plotted as small  brown circles; for galaxies with magnitude-radii measurements falling on grid points with a correction arrow, we show as empty circles their pre-correction, raw measurements; corrections up to $\sim 40\%$ had to be applied, in particular to galaxies with steep light profiles. }
\end{figure*}

\begin{figure*}
\begin{center}
\includegraphics[width=140mm]{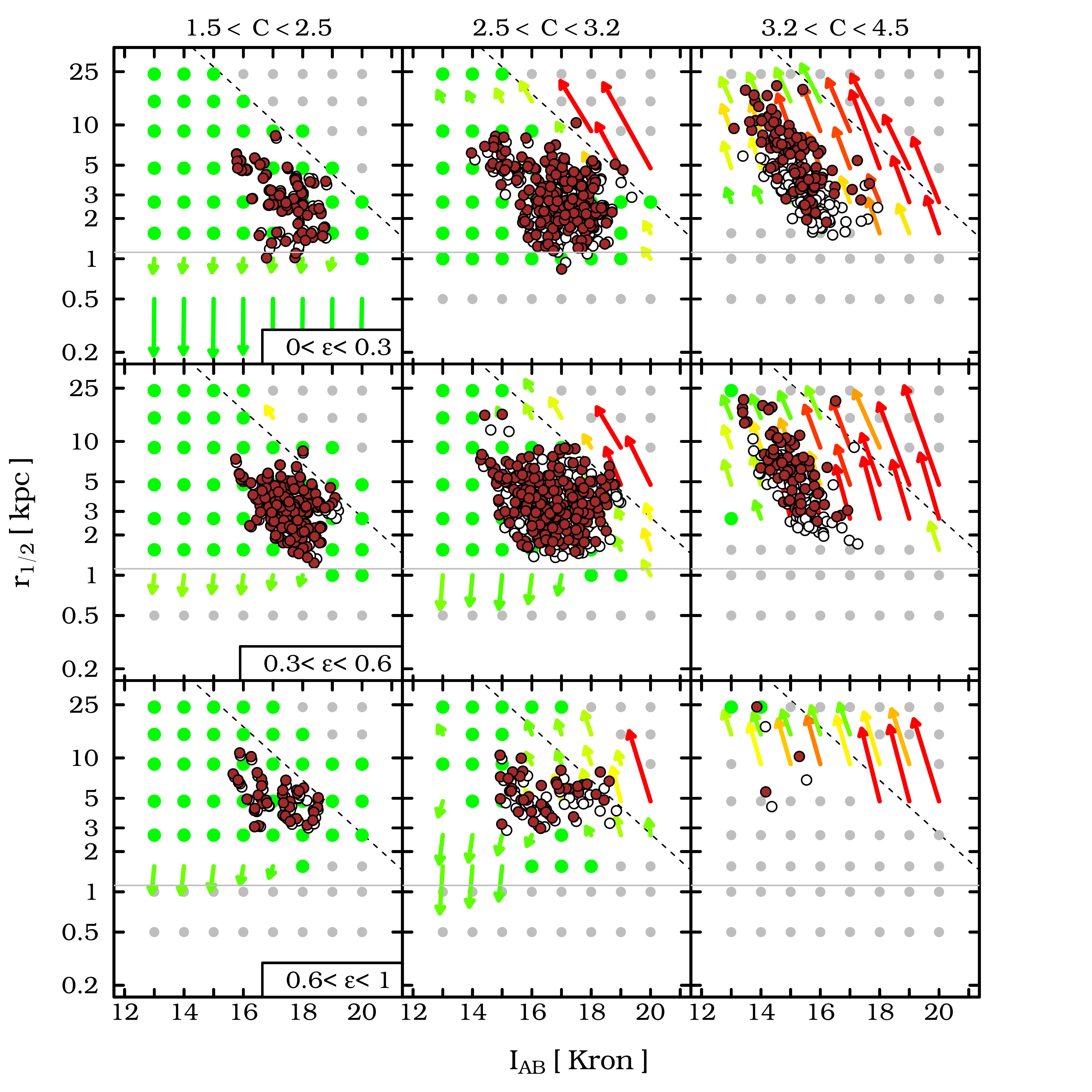}
\end{center}
\caption{\label{fig:ZESTCorrections} Same as  Figure \ref{fig:GIM2DCorrections}, but for measurements obtained with ZEST+. In this case, models and observed galaxies are binned in three panels according to the measured concentration index $C$.
 Colors and symbols are as in Figure \ref{fig:GIM2DCorrections}. Note that the corrections are, as expected, more substantial than those shown in Figure \ref{fig:GIM2DCorrections} for the GIM2D-based measurements.}
\end{figure*}

After the extraction/fitting process on the artificial galaxies was completed, we produced ``calibration maps" for sizes and magnitudes as shown in Figures \ref{fig:GIM2DCorrections} and \ref{fig:ZESTCorrections} (see also Carollo et al.\ 2013b for further details on this calibration approach). 
The arrows in the maps show, at any point of the  \emph{observed} $r_{1/2}-mag-\epsilon-n($or $C)$ plane, the direction and strength of the correction which is needed to recover the ``intrinsic" galaxy size and magnitudes from the observed values; these correction arrows are obtained by taking the median difference between the model nominal input parameters and the parameters measured with \textsc{ZEST+/GIM2D} of all artificial galaxies in the given point in the grid.  The correction matrices are binned in  three separate panels of concentration/S\'ersic index and three separate panels of ellipticity.
The colors of the arrows gives the amount of scatter shown by the individual models around the median correction: in green are  grid points where all corrections are coherent in strength, in red are shown those which have a high scatter and hence our correction is representative 
on average but not for the single models.  
We use these maps to derive corrections for observed magnitude and sizes in the real ZENS galaxy sample.  
The choice of a discretized but dense grid allows us to pinpoint the correction/uncertainty maps at well localized positions on the considered planes. As indicated above, the corrections for any given real ZENS galaxy are obtained by interpolation at the position of the ZENS galaxy in the \emph{observed} $\epsilon-mag-r_{1/2}-n(C)-PSF$ space.

To discuss Figures \ref{fig:GIM2DCorrections} and \ref{fig:ZESTCorrections}  it is useful to identify three regions in the $mag-r_{1/2}$ plane:
$(1)$ the region populated by models which are close to the surface brightness limit of the ZENS study (indicated with the dashed black lines in the plot), $(2)$ the region of well-resolved, high-signal to noise ratio measurements, and $(3)$ the region  close or  below the size of the PSF (highlighted with a gray horizontal line). 

Above the detection and resolution limit, and at low ($C<2.5$, $n<1.5$) to intermediate ($C\simeq 3$, $n\simeq 2.5$) concentrations, both GIM2D and ZEST+ perform fairly well, and only small corrections are needed for both $r_{1/2}$ and magnitude.

Not surprisingly, below the sky noise surface brightness, basically no galaxy can be recovered by both \textsc{ZEST+} and GIM2D, as models  are a priori not detected during the \textsc{SExtractor} source extraction  (necessary to define the total galaxy flux and the initial guess for the size needed as input by \textsc{ZEST+} and GIM2D). 
For galaxies with low concentration/S\'ersic index, the detection rate falls rapidly to zero when the surface brightness limit is reached (no model with $I_{AB}\sim 19$ and $r_{1/2}\sim10$kpc is recovered); conversely, the centrally peaked light distribution in galaxies with higher concentration pushes the detection limit to slightly fainter surface brightnesses.
Close to the surface brightness limit, magnitude and sizes are severely underestimated for both GIM2D and  \textsc{ZEST+}  fits:  sizes are typically smaller by more than a factor of two (three) and magnitudes are dimmer by  half (one) magnitude for GIM2D (ZEST+) measurements. 

 At $C>3$, the \textsc{ZEST+} ``aperture" measurements suffer from  a strong underestimation of sizes and magnitudes at any signal-to-noise. This is a consequence of the well known  tendency to miss a substantial 
fraction of the flux from the faint wings in steep light profiles when performing aperture photometry measurements. 
Integration to total light mitigates this effect in the GIM2D models, which nonetheless results in sizes which are about $30\%$ smaller than the intrinsic ones for  extended galaxies with steep light profiles ($r_{1/2}\gtrsim10$ kpc).

When moving close to the resolution limit of the survey, 
we notice further differences between the performance of ZEST+ and GIM2D. Thanks to the PSF deconvolution,
 the analytical fits are able to reliable recover the sizes also for models with sizes below the PSF FWHM.  
 Systematic effects become visible only at the worst observing condition (see Figure \ref{fig:GIM2DCorrections_worst} in Appendix \ref{app:corrections_PSF}).
   ZEST+ suffers from much stronger biases in this regime, causing an
 artificial increase of the size of models with $r_{1/2}<PSF$ for any value of the seeing. 
We note however that our approach, being based on idealized, regular galaxy light distributions produced with \textsc{GIM2D}, may return in general an optimistically good performance of \textsc{GIM2D} than when recovering the parameters for real galaxies with irregular, clumpy light distributions. Our corrections are thus to be considered as the ``minimal" correction functions that must be applied to the data to put the structural measurements on a comparable grid.

 \subsection{The impact of the PSF on concentration and ellipticity measurements}\label{sec:CorrDegeneracy}

 Another clear effect observed for the \textsc{ZEST+}-based  parameters is a lack of recovered parameters in the galaxy half-light radius $r_{1/2}$ vs.\, magnitude plane at high concentrations and small radii (see right most panels of Figure \ref{fig:ZESTCorrections}); this becomes increasingly more severe at higher ellipticities and larger PSF FWHM (see also Figures \ref{fig:ZESTCorrections_best}-\ref{fig:ZESTCorrections_worst}). A similar trend is observed for the GIM2D fits in the worst seeing conditions and  at  small radii (Figure \ref{fig:GIM2DCorrections_worst}).  
   Note that, consistently,  the distribution of the real ZENS galaxy measurements based on \textsc{ZEST+}  in Figure \ref{fig:ZESTCorrections}  presents the same bias.
 
\begin{figure*}
\begin{center}
\includegraphics[width=140mm]{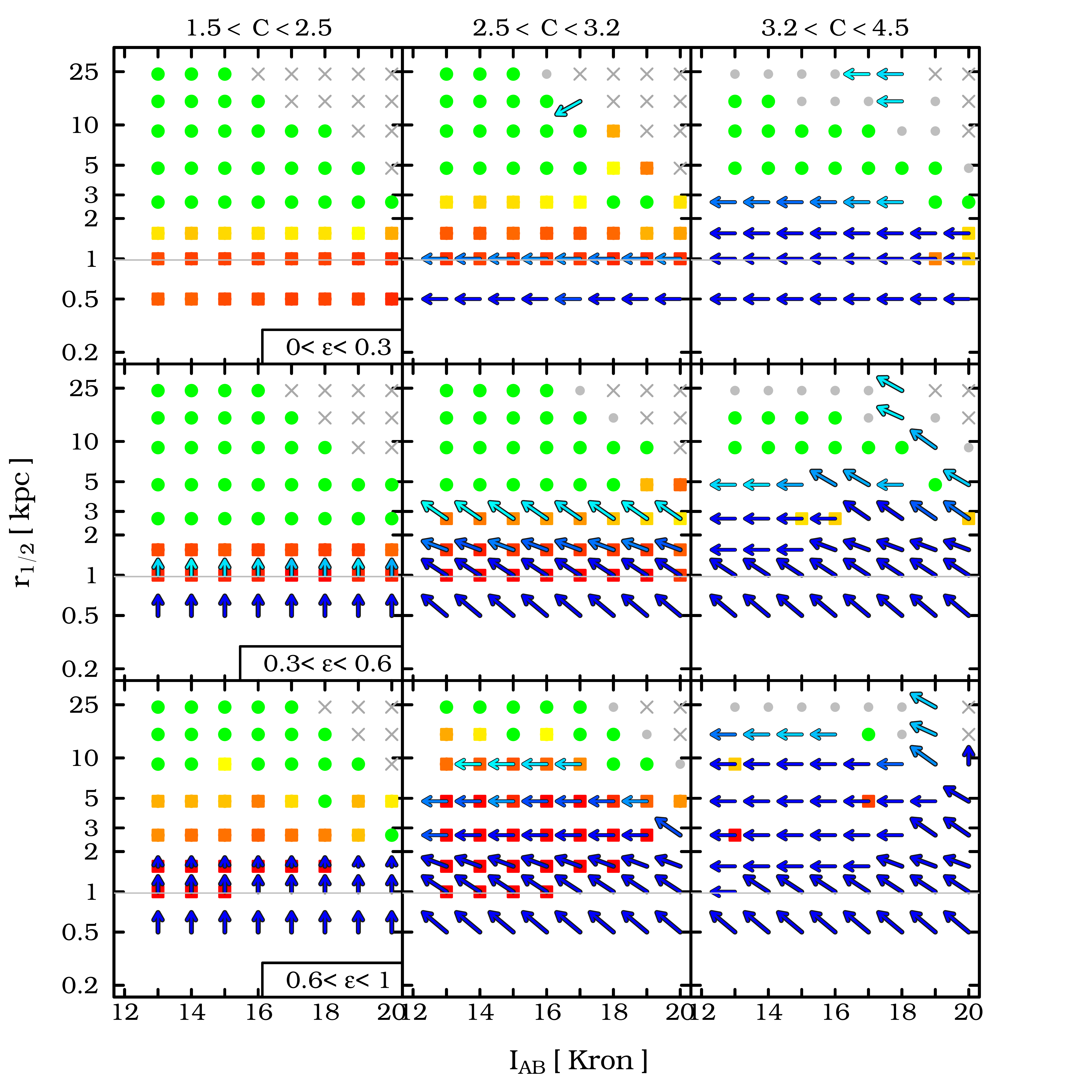}
\end{center}
\caption{\label{fig:ZEST_Scatter}The size vs.\, $I$-band magnitude plane on which we show the effect of measurement biases on the recovery of galaxy concentration and ellipticity calculated with ZEST+. Colored arrows and squares indicate, respectively, the scattering of model galaxies out of their  intrinsic $C$ and $\epsilon$ panels, and into a different panel of measured $C$ and $\epsilon$: e.g., an upward-pointing arrow indicates that  model galaxies with intrinsic parameters in that grid point are observed in a higher $b/a$ ratio panel, while a left-pointing arrow indicates that  the model galaxies  are observed in a panel of lower concentration. An arrow is drawn on a grid point  if at least $50\%$ of the models that are generated at that given $r_{1/2}-mag-\epsilon-C$ point are scattered out of it, into a different concentration or ellipticity panel; to draw the horizontal (vertical) component of the arrow, we further require that at least 25$\%$ of the scattered models change, when observed,  panel of concentration (ellipticity). Blue colors become darker with increasingly larger fractions of objects which are scattered out of a given panel.   The length of the arrows is arbitrary set for plotting purposes and has no specific meaning. With yellow square we show the points of the $r_{1/2}-mag$  grid in which at least $50\%$ of  the \emph{observed} model galaxies were generated in a different panel of intrinsic concentration or ellipticity (up to 100$\%$,  shown with full red squares).
Green points indicate grid nodes in the observed $C$ or $\epsilon$ parameter space which do not suffer  from  strong scattering of intrinsic model galaxies either into it or out of it.  Gray crosses and dots  respectively highlight the grid points which are below the surface brightness limit of the ZENS study, and  in which  galaxy  models with intrinsic $C$ and $\epsilon$ in those grid points are recovered by \textsc{ZEST+} at a different 
radius and/or magnitude, but within the same concentration/ellipticity panel (i.e., at a different grid point within the same panel).}
\end{figure*}

 The origin of such  effect is investigated in Figure \ref{fig:ZEST_Scatter}, which illustrates where the artificial galaxies, created in a given region of the parameter space, are
 placed in the \emph{observed} space. In this map, the arrows indicate  the direction in which models generated with a certain $\epsilon$ and $C$ are displaced relative to their intrinsic concentration and ellipticity: 
  an upward-pointing arrow indicates that models are observed on average as less elongated, and a left-pointing arrow indicated that  they are observed at lower concentrations.
     Intrinsic concentrations are calculated analytically from the original models S\'ersic indices, and refer to the ratio of the radii containing $80\%$ and $20\%$
  of the \emph{Petrosian} flux (see e.g. \citealt{Graham_et_al_2005}).
  In the same figure,  red-to-yellow squares indicate points in the grid in which at least $50\%$ (up to  100\% for full red squares) of the artificial galaxies that are \emph{observed} at that location originate from a different intrinsic  concentration or 
  ellipticity grid point.
  Green points in the figure indicate grid nodes  which are not affected by either scattering of  galaxies into the grid point from different intrinsic concentration/ellipticity panels, or by
  scattering of galaxies out of the grid point into   a  different panel of observed ellipticity and/or concentration.

  It is clearly seen in Figure \ref{fig:ZEST_Scatter} that  model galaxies which were originally generated at high concentrations and ellipticities have a  high probability to be scattered  into lower  $\epsilon$ and $C$ bins by the aperture-based \textsc{ZEST} measurements;  this bias is exacerbated at sizes $\lesssim 2-3$ kpc.
 This is caused by the effect of the PSF convolution
  which artificially lowers a galaxy concentration and circularizes their axis ratio.  
Consequently, this \textsc{ZEST+}  region of the $mag-r_{1/2}$ plane out to $\sim2$ PSF radii 
 is highly degenerate, being populated by both galaxies which have intrinsic parameters in that location of  parameter space, and by
galaxies which are  scattered into it from higher ellipticity and/or concentration regions.
Given this degeneracy, it is important to verify to which extent  the corrections for radii and magnitude that we discussed above depends on the precise way in which the simulation grid is populated. 
If model galaxies which are ``scattered in" and those which have intrinsic parameters in that location of  parameter space required different corrections, then a precise modeling of the relative fractions of such galaxies in the given grid point would be needed. 
To test this, we created magnitude and sizes correction maps  using either only artificial galaxies which were scattered into the given $\epsilon-C$ bin, or only galaxies born in situ. We show the outcome of such an experiment  for the median PSF in Figure \ref{fig:ZESTCorrections_withnoContaminers} in Appendix \ref{app:corrections_PSF}.
 Both realizations resulted in very similar corrections for size and magnitudes, indicating that these are robust independent of the intrinsic concentration/ellipticity of the galaxies.

On the other hand, models which are scattered into lower $C$ or $\epsilon$  bins require by definition  stronger corrections in $\epsilon$ and $C$ themselves than those which were generated within the bin. While for ellipticity it is reasonable to assume that also real (disk) galaxies would have an uniform distribution, and we hence can consider our corrections to be representative, the correction for concentration depends on the relative fraction of truly low concentration galaxies with respect to galaxies with a high intrinsic concentration. 
Given that we have a no priori knowledge of the true distribution of concentrations for the real galaxies, and to avoid introducing biases associated to the choice of the simulation grid, we choose to follow a different method to correct the non-parametric structural estimators, which uses the available information on the S\'ersic index.
For data sets which do not have an as comprehensive set of measurements as  ZENS,  this approach could not be applied, and a statistical modeling of the underlying distribution of galaxies will be needed.
These results should be taken as a cautionary note in using solely the concentration index as a morphological discriminant for galaxy types, especially if galaxies are close to the resolution limit of the given survey and if no correction to this parameter is attempted. 

Conversely to the ZEST+  measurements, an equivalent map as in Figure \ref{fig:ZEST_Scatter}, but for S\'ersic indices and ellipticities measured by GIM2D (not shown)  demonstrates instead only a marginal  contamination/scattering of galaxy models across the broad ellipticity and $n$ bins, thank to the PSF-deconvolution performed by the GIM2D fitting algorithm. 
 For this reason,  we apply to the  ellipticities and S\'ersic indices  the corrections obtained  by interpolating between the relevant grid points of the previously discussed vector maps, which we regard as statistically representative of the average correction.
 We stress however that systematic effects are observed  for model galaxies convolved with the worst PSF of \textsc{ZENS}:
 in this case,  models with high S\'ersic indices and sizes smaller than $\sim1-1.5\times$ the PSF size are  scattered  to lower $n$.  Interestingly, GIM2D tends to overestimate the ellipticity of such galaxies as opposed to what observed for the measurements performed with ZEST+.  Furthermore, although under typical observing conditions the error on ellipticity and S\'ersic index measured with GIM2D are small enough to keep model galaxies within the same bin in these parameters, moving closer to the resolution or detection limit of the \textsc{ZENS} WFI  images increases the randomicity in the measurements; inferred ellipticities and S\'ersic indices have a typical scatter of  $\sim 0.1$ and $20-30\%$, respectively. We highlight these problematic regions with empty symbols in Figure \ref{fig:GIM2DCorrections}. 
 We finally note that a PSF correction is often included -- typically as a convolution-kernel in surface brightness fitting algorithms --  in structural studies based on other ground-based datasets, including the SDSS   (whose typical PSF width is similar to the worst ZENS PSF; see e.g., \citealt[]{Blanton_et_al_2003,Lackner_et_al_2012}). As we show above, it is possible however that further residual  (size-, magnitude-, ellipticity- and concentration-dependent) corrections to the structural parameters may be needed, depending on the specific PSF, signal-to-noise and redshift of the studied samples.

 \subsection{Corrections of non-parametric structural indices}\label{sec:GiniCorr}

To overcome  potential biases associated with calculating a correction for the $CASGM$ parameters which may depend on the distribution of  intrinsic structural properties, we  use   the S\'ersic indices  of the galaxies from the GIM2D fits as a prior.  S\'ersic indices are robustly determined in the vast majority of the models and are less prone to systematic biases, as discussed in the previous section.   
 
The corrections for concentration, Gini and $M_{20}$ indices are thus derived by splitting the artificial and real galaxies in similar bins of magnitude, radius, ellipticity and PSF as those employed for the
size corrections (see  Figure \ref{fig:ZESTCorrections}). In this case however we characterize the galaxy structure according to the observed S\'ersic index rather than observed concentration, dividing the samples in three broad bins of $0.2<n_{obs}<1.5$, $1.5<n_{obs}<3.5$ and $3.5<n_{obs}<10$, respectively. 

Following the same approach as for sizes and magnitudes, the corrections for non-parametric structural estimators are then defined as the median difference between 
the models' intrinsic indices and those calculated with ZEST+ at the given position in the observed $mag-r_{1/2}-\epsilon-n-PSF$ parameter space. 
The intrinsic concentration is computed analytically from the input S\'ersic index, as specified above, while for the Gini and $M_{20}$ indices we use the measurements performed on ``pure" models that are neither  PSF-convolved nor degraded with the ZENS typical noise.   
Given that our analysis is based on intrinsically smooth and axisymmetric models, we do not attempt to correct the asymmetry and smoothness  index. These quantities will nonetheless be affected (the smoothness possibly by the largest amount), and this caveat should hence be kept in mind.   
For the few ZENS galaxies for which no GIM2D fit is available, we applied an average correction obtained as the median of the correction for the
ZENS galaxies having GIM2D fits and similar observed concentrations.

\subsection{Robustness of the bulge-to-disk decompositions} \label{sec:BulgeDiskBias}

A similar approach as the one employed for the single S\'ersic fits was used to test the reliability of  the \textsc{GIM2D} bulge+disk (B+D) decompositions. 
All the B+D model galaxies were processed in  exactly the same way as the real galaxies: i.e., after the decomposition, the output models were filtered to reject the unphysical fits and to identify acceptable B+D decompositions (see Appendix \ref{app:FilterBD}).
Consistently with the approach used for the real galaxies,  only these were used for deriving the correction functions for the B+D parameters.

\begin{figure*}[htpb]
\begin{center}
\includegraphics[width=160mm]{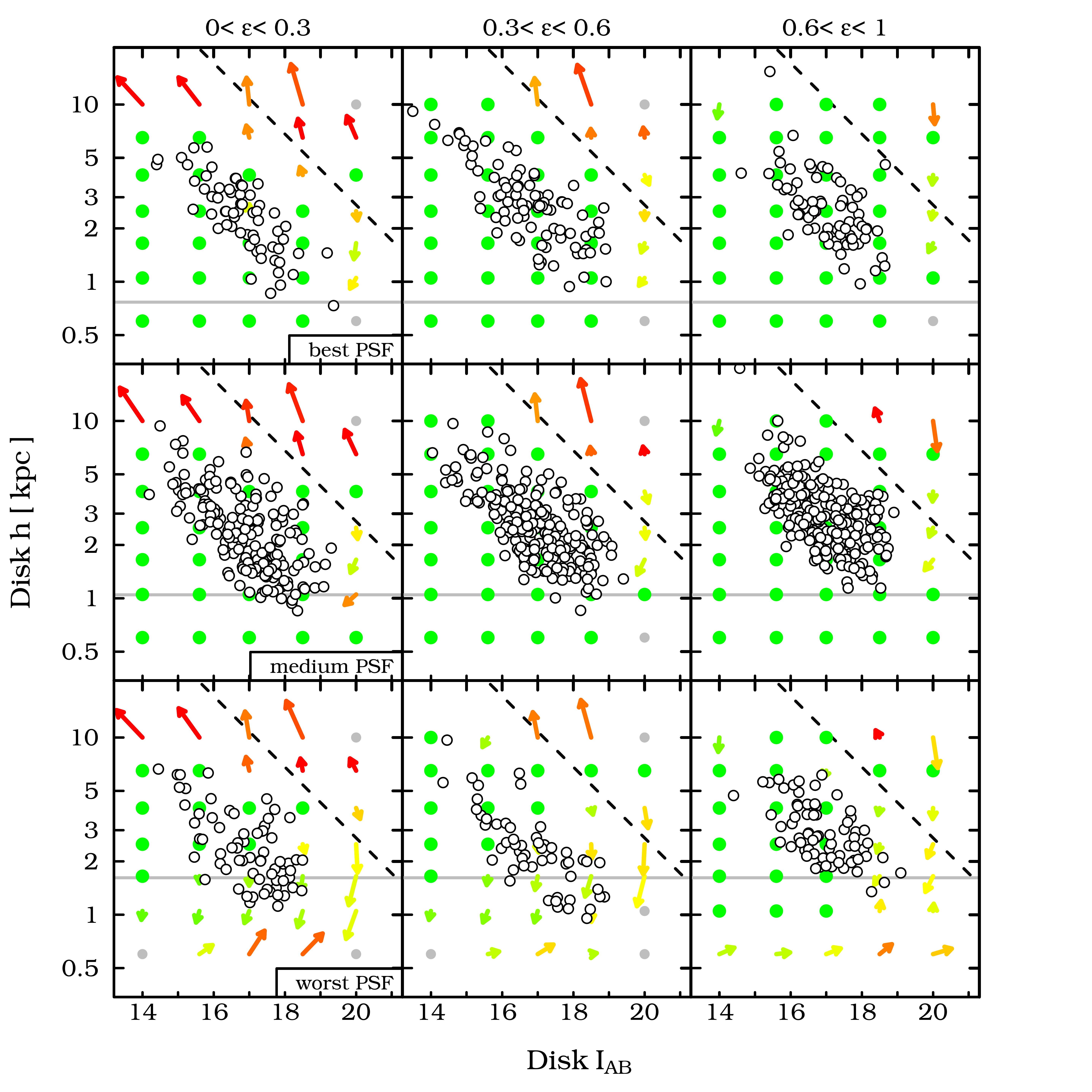} 
\end{center}
\caption{\label{fig:DoubleCorrDisks}
Disk scale length $h$ vs.\, disk $I-$band magnitude plane, with highlighted the error vectors illustrating the systematic errors on the disk scalelength and $I$ magnitude recovered by GIM2D relative to the intrinsic values. These vector maps are the results of our extensive simulations described in Section \ref{sec:Simulations}.  Only models with measured bulge-to-disk ratio  $B/T<0.80$ are considered in this figure. From top to bottom, the panels refer to the best, median and worst ZENS PSF. Results are presented, from left to right, in three bins of  disk ellipticity. Colors and symbols are as in Figure \ref{fig:GIM2DCorrections}. Small black empty circles indicate the real measurements for the disk components of ZENS galaxies.}
\end{figure*}

\begin{figure*}[htpb]
\begin{center}
\includegraphics[width=160mm]{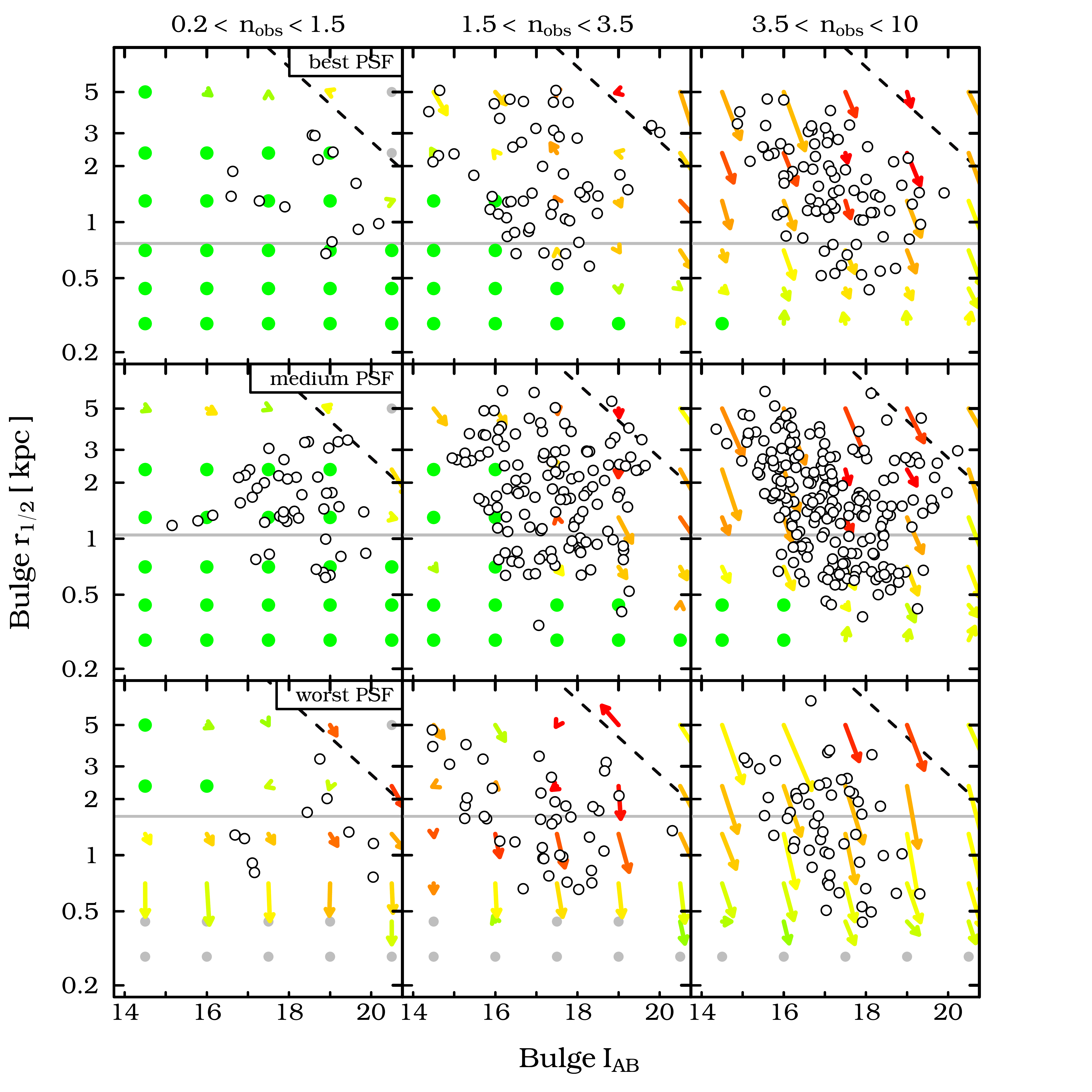}
 \end{center}
\caption{\label{fig:DoubleCorrBulges}  Bulge half-light radius $r_{1/2}$ vs.\, bulge $I-$band magnitude plane, with highlighted the error vectors illustrating the systematic errors on the bulge radii and magnitudes recovered by GIM2D relative to the intrinsic values. Only models with measured bulge-to-disk ratio  $0.1<B/T<0.80$  are considered in this figure. From top to bottom, the panels refer to the best, median and worst ZENS PSF. Results are presented, from left to right, in three bins of S\'ersic indices. Colors and symbols are as in Figure \ref{fig:GIM2DCorrections}. Small black empty circles indicate the real measurements for the disk components of ZENS galaxies.}
\end{figure*}

The correction maps for the B+D parameters are shown in Figure \ref{fig:DoubleCorrDisks} and \ref{fig:DoubleCorrBulges}, for the three PSFs. 
For bulges, we present the results for all  disk inclinations together, as no strong dependence on the model axis ratio was detected.

Both scale lengths and magnitudes of disks are very well recovered by the GIM2D code for most of the cases.
Systematic deviations from the input parameters are observed only for very small disks (i.e., $h\sim0.5\times$PSF),  especially for the largest PSF size,  and in the low surface brightness limit, where disk scale-lengths are underestimated  by a factor of about two, and the measured magnitudes are systematically fainter than the input ones, as already discussed above.
Globally, in only $\sim10\%$ of the cases the normalized difference between input and output disk sizes is larger than $\sim50\%$, and predominantly in the low surface brightness regime.

The bulge parameter are instead subject to much larger uncertainties, as illustrated in Figure \ref{fig:DoubleCorrBulges}.  Although in many regions of the magnitude-size relation there are no systematic corrections, the scatter of measured bulge sizes around the input ones is generally large, typically $\sim30-40\%$ for the measured bulge half-light radii. The strongest differences between input and output model parameters are measured for bulges with high S\'ersic indices  ($n\gtrsim4$) in which case a substantial over-estimation of bulge effective radii is observed.
Low S\'ersic index bulges (leftmost panels in the Figure)  suffer from smaller measurement errors.
Below the resolution limit and for the worst PSF, sizes are over-estimated also for bulges with $n<4$ whereas the effect is less evident for the best or intermediate PSF.

\begin{figure}[htpb]
\begin{center}
\includegraphics[width=70mm]{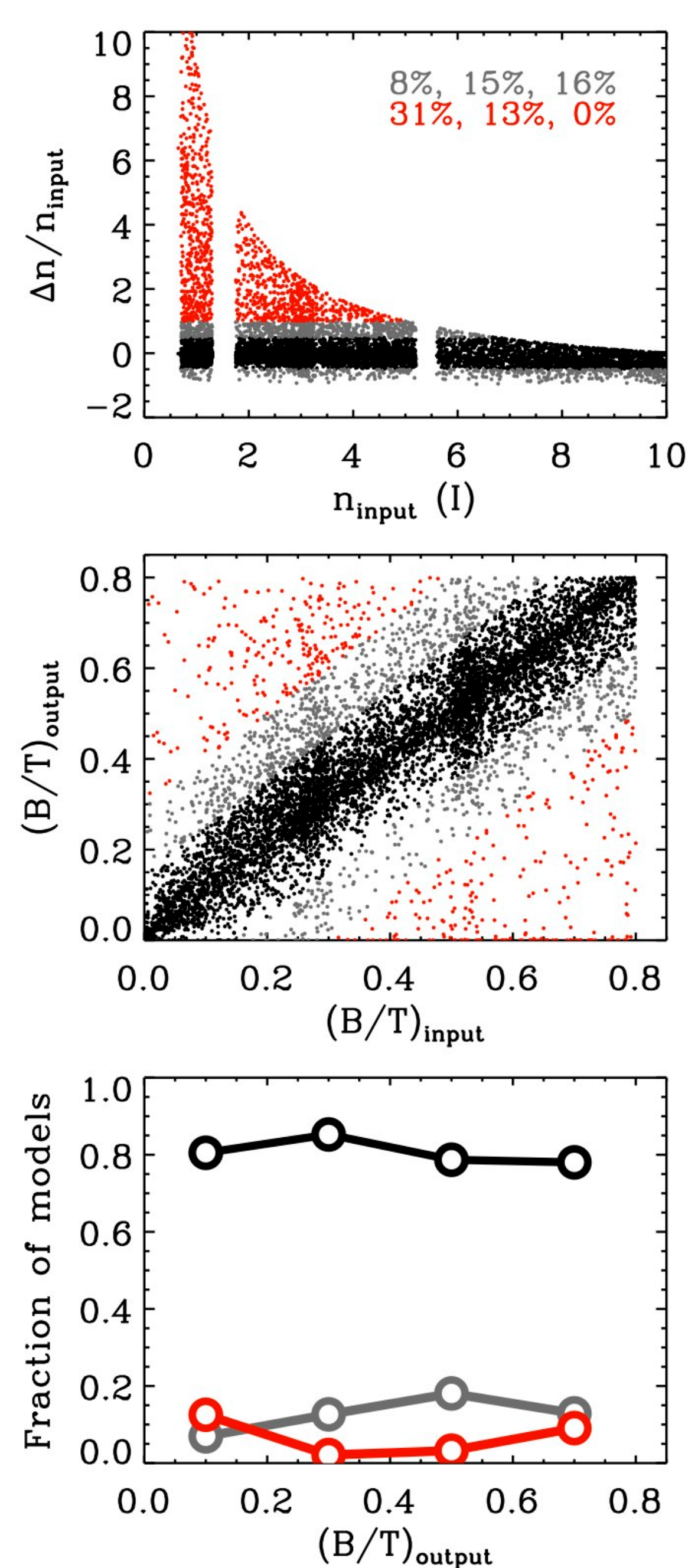}
 \end{center}
\caption{\label{fig:BTscatter}In the top panel we show the comparison between the intrinsic bulge S\'ersic indices for the set of simulated galaxies used to test the reliability of GIM2D bulge+disk decompositions, and those recovered by GIM2D. The variation $\Delta$ corresponds to the difference between the intrinsic and measured values. 
With black points we highlight models for which the structural parameters are well recovered, i.e., have relative error  $<50\%$. In dark gray are the points which have a fractional error between $50\%$ and $100\%$, and in gray (red in the online version) are the extreme outliers with errors which are $>100\%$. 
The reported  fractions are the fraction of discrepant models relative to the total  in the three bins of S\'ersic index $n$ shown in the Figure.
In the  central panel we show the comparison of intrinsic and measured bulge-to-total ratios (B/T) for the model galaxies. Black shows models for which $|\Delta (B/T)|<0.15$,  dark gray those with  $0.15<|\Delta (B/T)|<0.3$ and gray (red in the online version) is for models for which the measured B/T deviates more than 0.3 from the intrinsic value.  
The fraction of models within each of the three types, as a function of the \emph{observed} B/T, is shown in the bottom panel.
(A color version of this figure is available in the online journal.)}
\end{figure}

These uncertainties in the bulges sizes reflects in (or are possibly generated by) a general difficulty in recovering the correct bulge S\'ersic index. As an illustration of this fact, we plot on the top Panel of Figure \ref{fig:BTscatter} the comparison between input and measured bulge S\'ersic indices  for the median PSF. 
It can be noticed how the measured S\'ersic index, especially below $n\simeq2$,  can deviate substantially from the model intrinsic values. This particularly severe for bulges with half-light radii which are close to the PSF size: small bulges with low index $n$ can be misclassified as  bulges with larger half-light radii and steeper light profiles.

Although the structural properties of the bulges are subject to relatively large uncertainties,  the fractional contribution of the bulge to the total light is generally well measured by the GIM2D decompositions, as shown on the middle panel of Figure \ref{fig:BTscatter}. In the bottom panel we plot, for a given value of the observed bulge-to-total ratio, the fraction of input models which where originated with a $B/T$ which differed less than 0.15 from the measured one, between 0.15 and 0.3 and more than 0.3 (black, gray and red lines respectively). In $\sim$80-85$\%$ of the cases the bulge-to-total ratio is recovered within a scatter of 0.15 and really catastrophic failures ($\Delta (B/T) >0.3$) happen in $\lesssim 10\%$ of the models.
It is worth  to notice that the fraction of galaxies for which the $B/T$ is robustly recovered is largely independent of the value of $B/T$ itself, and the typical scatter around the input bulge-to-total ratio is of order $\sim0.1$.
Problematic fits which results in large differences in $B/T$ are again mostly associated with   flat  ($n<2$) and small ($r_{1/2}<1$kpc$\simeq$PSF) bulges, for which is  difficult to
disentangle the disk component from the small, disk-like bulge. 

In the light of these results we decided \emph{not} to apply any corrections to the bulge and disks structural parameters: on the one hand disk sizes are well recovered by the GIM2D fits, hence no substantial correction is needed; on the other hand, the large scatter in the recovered bulge half-light radii and S\'ersic indices make
the derived corrections noisy and dependent on the specific sampling of parameter space with our simulations.
For this reason, we prefer to use the direct output from GIM2D for the bulge half-light radii, but with associated  a typical uncertainty of 30$\%$, as estimated from our tests.

\subsection{Testing the robustness of the applied corrections} \label{sec:ReliabilityCorrections}

We present a number of diagnostic tests that we performed to verify the reliability of the corrections for the structural parameters that are described in Section 
\ref{sec:correctionMaps} and \ref{sec:GiniCorr}.

\begin{figure*}[hbtp]
\begin{center}
\includegraphics[width=120mm,angle=90]{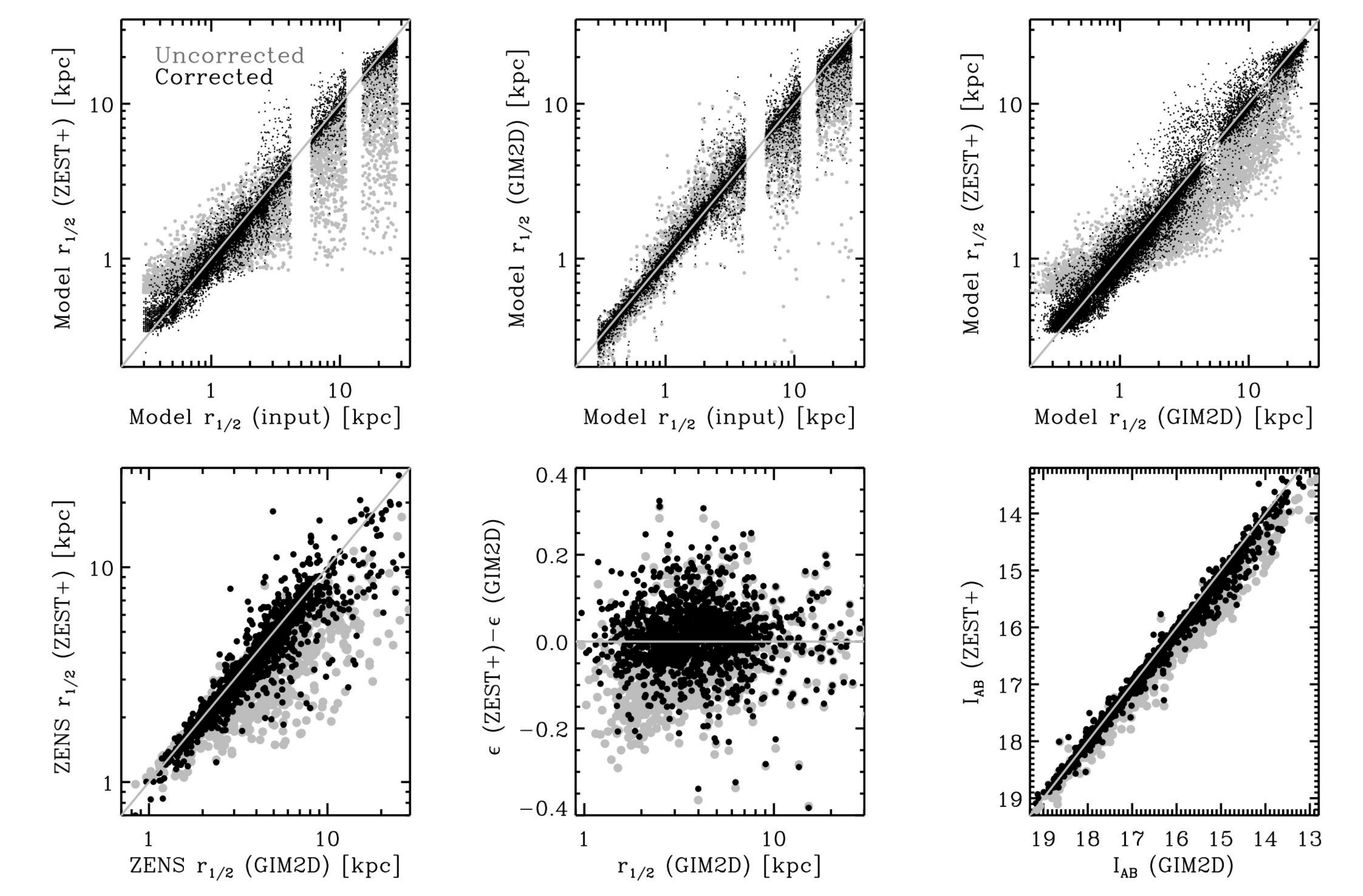}
\end{center}
\caption{\label{fig:ZEST_GIM2DCorectionComp} Tests of the correction scheme outlined in Section \ref{sec:Simulations} both for the GIM2D and ZEST+ measurements  on simulated (top panels) and real \textsc{ZENS} data (bottom panels). In all panels, black symbols are the corrected measurements and  gray the raw, uncorrected  measurements. The top panels specifically show the comparison between intrinsic (input) half-light radii of simulated galaxies versus half-light radii recovered with ZEST+ (left) and GIM2D (center), and the comparison between ZEST+ and GIM2D half-light radii (right).    
The identity lines (shown in gray) are well matched by the corrected sizes, with systematic shifts largely cured by our correction scheme. 
The bottom panels show the application of our corrections to the real ZENS data. Specifically, we show the comparison between ZEST+ and GIM2D half-light radii measurements, the ellipticity differences between ZEST+ and GIM2D as a function of (corrected) GIM2D half-light radius, and the comparison between  $I-$band magnitudes recovered by ZEST+ and GIM2D. Also in these case, large discrepancies observed in the raw measurements vanish once we apply our correction scheme. }%
\end{figure*}

As a first sanity check we run our correction scheme, in exactly the same fashion as for the real galaxies, on the measured properties for the  models  themselves; the outcome of this exercise is illustrated in the upper panels of Figure \ref{fig:ZEST_GIM2DCorectionComp} for sizes obtained with GIM2D and  ZEST+,   for models convolved with the median PSF. The  intrinsic (input) and corrected radii for the model galaxies agree very well, and systematic biases that are present in the uncorrected sizes are largely cured by our correction scheme. Furthermore, the large discrepancies between the ZEST+ and GIM2D raw measurements disappear in the corrected data.   

The bottom panels in Figure \ref{fig:ZEST_GIM2DCorectionComp} show the  comparisons of ZEST+ and GIM2D half-light radii, ellipticities and $I-$band magnitudes, before and after the application of our correction schemes for the real ZENS galaxies. The raw  sizes derived by ZEST+ are systematically smaller than those obtained by the GIM2D fits, by up to a factor of  $\sim 2-3$ in the worst cases. 
Our correction scheme nicely brings the two estimate into a very good agreement. 
As a consequence of the PSF convolution, galaxies are also measured to be rounder in ZEST+ than in GIM2D, especially at low radii, a bias which is largely cured by the implementation of the correction scheme.  The latter also mitigates the underestimation of the total fluxes in ZEST+. 

\begin{figure*}[hbtp]
\begin{center}
\includegraphics[width=90mm,angle=90]{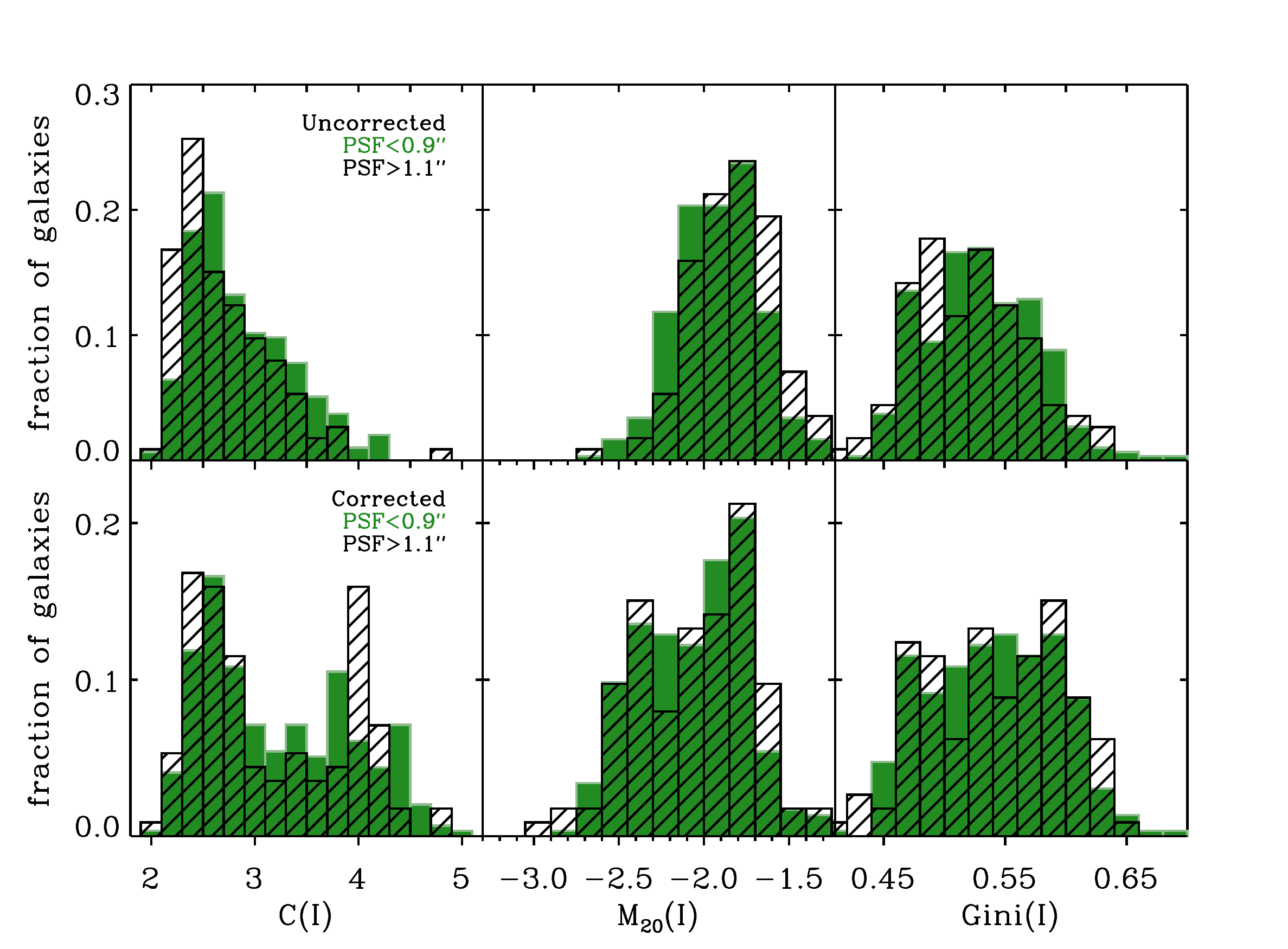}
\end{center}
\caption{\label{fig:ZEST+_corrected} Comparisons of the distributions of  ZEST+ concentration (left), $M_{20}$ (center) and Gini (right) $I-$band parameters before (top row) and after (bottom  row)  the application of our correction scheme to eliminate PSF-induced biases.   The distributions are shown for two separate bins of PSF FWHM, i.e.,  FWHM $<0.9^{\prime\prime}$  (gray histograms, green in the online version) and FWHM $>1.1^{\prime\prime}$ (black histograms). Galaxies observed with smaller PSF appear more concentrated than those observed with a larger seeing,  a consequence of the blurring of the structural features caused by the PSF size. Also, the effect of PSF-blurring  is to move high-concentration, highly-negative $M_{20}$ galaxies to lower concentration, less negative $M_{20}$ regions of parameter space. Both biases are eliminated by our correction scheme.
(A color version of this figure is available in the online journal.)}
\end{figure*}

The comparison between uncorrected and corrected $I$-band 
$C$, Gini and $M_{20}$ indices  is shown in Figure \ref{fig:ZEST+_corrected}. Specifically, 
to test whether our corrections deal properly with PSF-induced biases, we compare, for corrected and uncorrected measurements,  the distributions of these parameters for two bins of PSF size, i.e.,   PSF $FWHM<0.9^{\prime\prime}$ and  $>1.1^{\prime\prime}$, respectively.
The uncorrected indices show different distributions for galaxies observed with small or large PSF's FWHM. The effect  is mostly evident for the concentration parameter, for which a two-sided Kolmogorov-Smirnov test  rejects the possibility that the low and high FWHM samples are drawn from the same parent distribution at the $99\%$ confidence level. 
A similar effect is also noticeable   for the $M_{20}$ index;  the Gini coefficient is instead less sensitive to PSF blurring. 
All such biases disappear  once our corrections are applied, as shown in the lower panels of the same figure. 
The inspection of the distributions of Figure \ref{fig:ZEST+_corrected} furthermore  shows that the corrections recover the   peak of high $C$ values for the early-type galaxies, which would otherwise be absent from the raw measurements; similarly, the  peak at ``more negative" values is recovered in the corrected $M_{20}$ distributions. 

\begin{figure*}
\begin{center}
\includegraphics[width=130mm,angle=90]{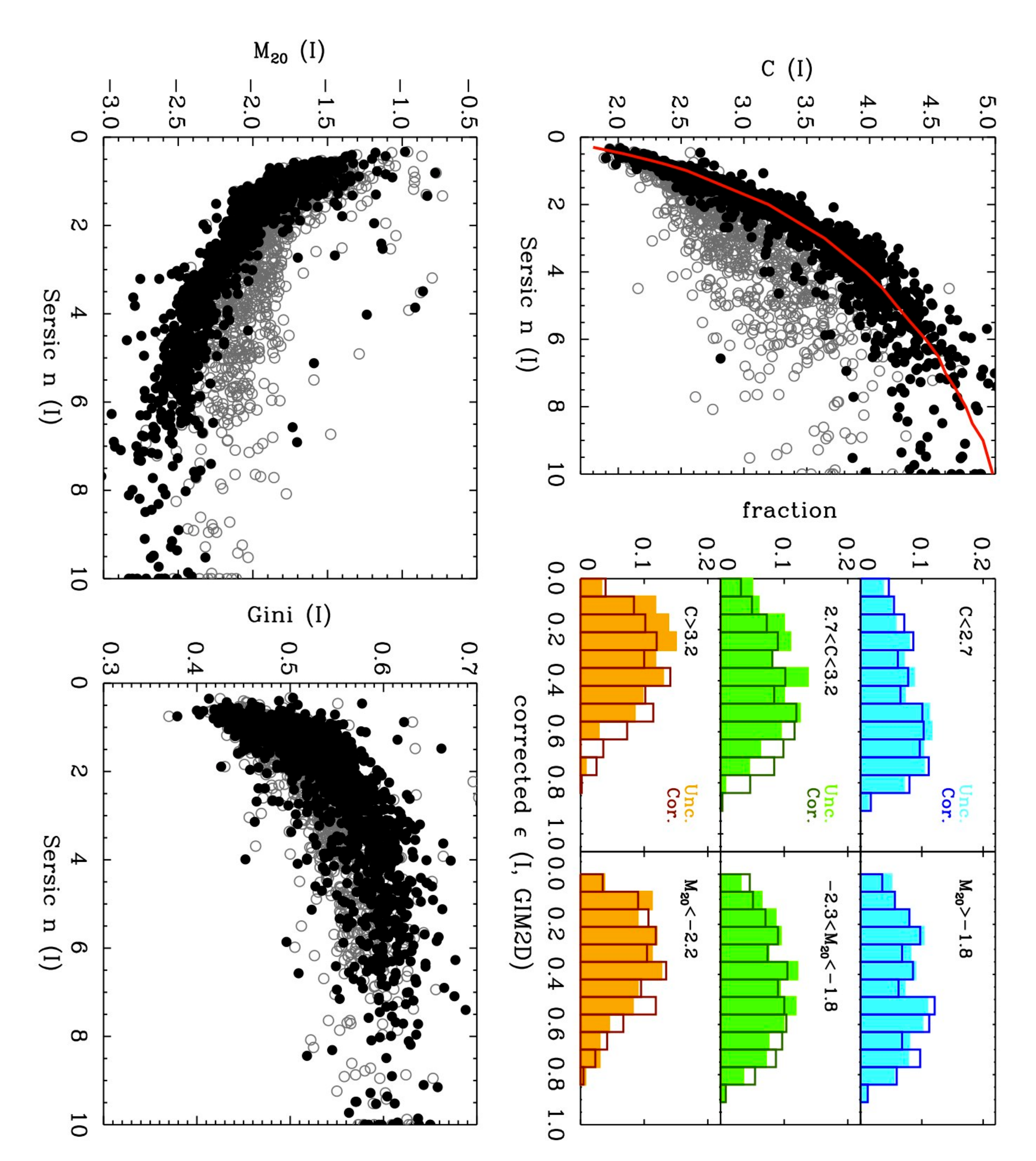}
\end{center}
\caption{\label{fig:ZEST_vs_Sersic} Relation between $I-$band S\'ersic index $n$ and non-parametric structural quantities, i.e., $C$, $M_{20}$ and Gini coefficients before and after applying our correction scheme, described in Sections \ref{sec:GiniCorr} and \ref{sec:correctionMaps} (top left panel and bottom panels). Gray points show the uncorrected measurements and   black   points   the  corrected measurements. In the upper left panel the solid grey (red in the online version) line marks the values of the concentration index which are theoretically expected for a pure S\'ersic profile of a given $n$.  Both $C$ and $M_{20}$ require substantial corrections, while the Gini coefficient is largely unaffected by observational biases. The upper central and  right panels finally show,  for galaxies classified as disks in Section \ref{sec:MorphClass} (S0 and later-types),  the dependence of uncorrected and corrected concentration and $M_{20}$ parameters on ellipticity. 
Specifically, we present the distribution of ellipticity for galaxies divided in three bins of  concentration and $M_{20}$, as indicated by the Figure legend.
The filled light histograms are for the uncorrected parameters, the dark empty ones for the corrected indices.
The uncorrected $C$ and $M_{20}$ values show a bias with $\epsilon$ which is removed by our correction procedure.
(A color version of this figure is available in the online journal.)}
\end{figure*}

 This is also highlighted in Figure \ref{fig:ZEST_vs_Sersic}, where we show the comparison between
 corrected and uncorrected non-parametric structural estimators and the corrected S\'ersic indices. 
The upper-left  panel shows the comparison for the concentration index:  the  solid red line marks the expected
 values of the concentration inside the Petrosian radius for a 
perfect S\'ersic profile of a given index $n$. 
It is clear that  before correction, the measured concentration flattens rapidly at values $\sim3-3.5$ for $n\geqslant3$, lying far away from the theoretical line. After applying our correction,  the measured points well match this line. 
A similar adjustment is  also observed for the $M_{20}$ index, while, as already mentioned, the Gini coefficient is measured quite robustly and needs virtually no corrections. The  upper right inset show the dependence on ellipticity of uncorrected concentration and $M_{20}$ parameters for galaxies classified as disks in Section \ref{sec:MorphClass}; the spurious lack of high concentration galaxies at high elongation is largely ameliorated by our correction scheme.

We stress in concluding that all our corrections are  clearly ``statistical". For example,  individual models can retain an underestimation of the radius of up to a factor $30\%$, especially   models close to the surface brightness limit of the ZENS observations.  Nevertheless, in a statistical sense, our correction schemes return  well-calibrated measurements that can be robustly compared with each other. 

\subsection{A comparison with previous work}

\begin{figure*}[htpb]
\begin{center}
\includegraphics[width=100mm,angle=90]{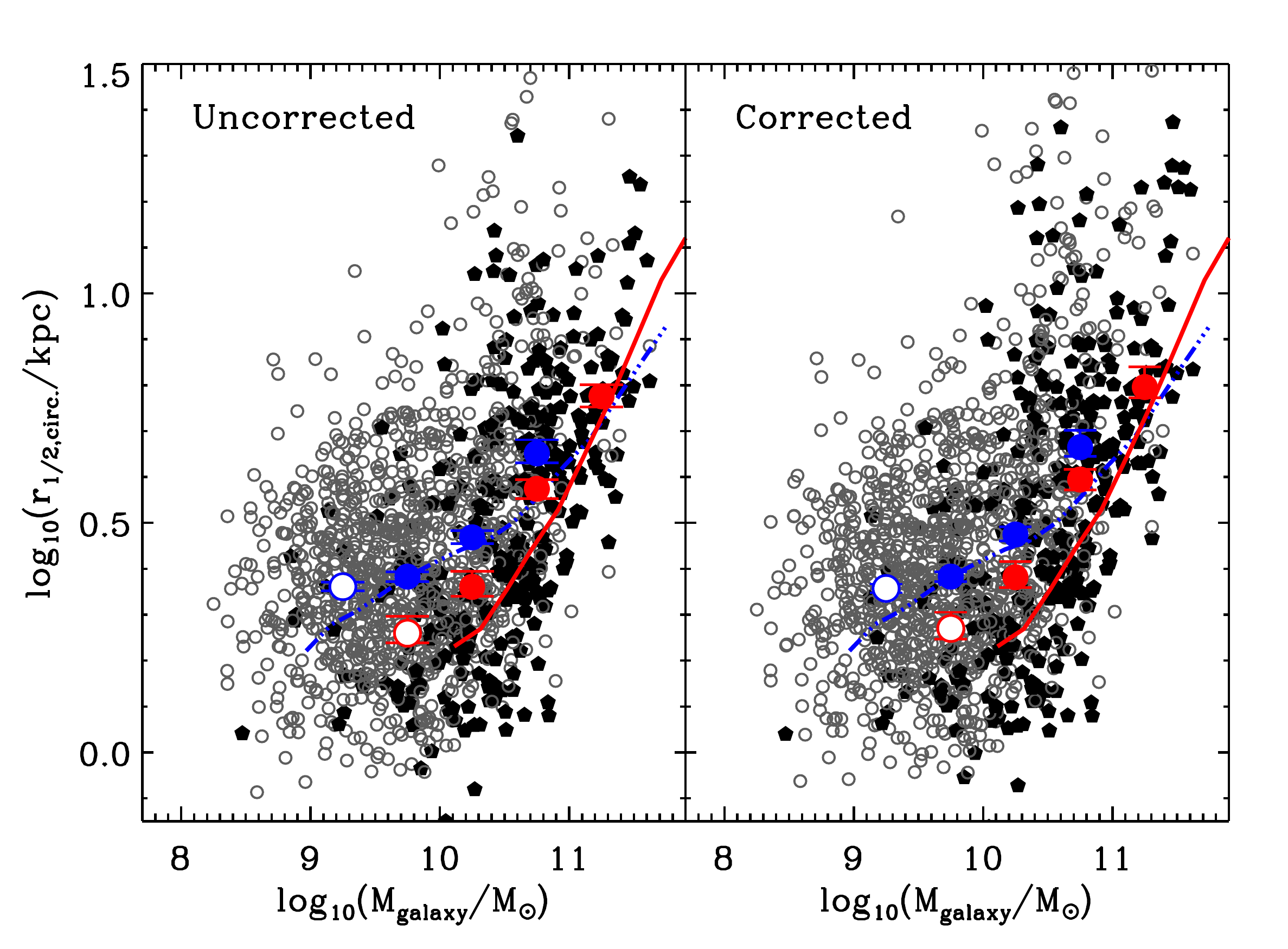}
\end{center}
\caption{\label{fig:ShenComp} Comparison of the mass-size relation between our \textsc{ZENS} sample and the early- (solid line, red in the online version)  and late-type (dashed-dotted line, blue in the online version)  samples of \citealt{Shen_et_al_2003}. The Shen's solid and dashed-dotted lines connect median values of sizes in the corresponding stellar mass bins (adapted from their Figure 11).
The left panel shows  results before applying to our data the corrections described in Section \ref{sec:correctionMaps}; our corrected sizes are shown in the right panel. Empty gray circles are \textsc{ZENS} galaxies classified as intermediate- and late-type type disks; filled pentagons are \textsc{ZENS} ellipticals, S0s and bulge-dominated spirals. The median values for
these two broad   \textsc{ZENS} morphological samples are shown  with large circles (E, S0 and bulge-dominated spirals grouped together. Red in the online version) and large triangles (intermediate-type and late-type disks group together. Blue circles in the online version).  White large symbols indicate \textsc{ZENS} data in mass bins below the \textsc{ZENS} mass completeness for the relevant morphological types.
 The major axis  \textsc{ZENS}  radii are circularized in post-processing mode in this figure, for a better comparison  with the SDSS sample.
 (A color version of this figure is available in the online journal.)}
\end{figure*}

We   show in Figure \ref{fig:ShenComp} a comparison between the mass-size relation obtained for the 
ZENS galaxies, both before and after applying the corrections outlined in Section \ref{sec:correctionMaps}, and the relation derived by \citealt{Shen_et_al_2003} in SDSS. The Shen's sizes are from the S\'ersic fits of \citet{Blanton_et_al_2003}, and are thus similar to the sizes of the VAGC catalogue  (see Appendix \ref{sec:testGIM2D} for a one-to-one comparison between our measurements and the Blanton et al data).  Circularized apertures are used in this comparison, to match the Shen et al.\, definition. Not surprisingly, given the global robustness of the structural parameters of disks, and more generally low S\'ersic index galaxies, the  \citet{Shen_et_al_2003} size-mass relation for the late-type galaxies is in excellent agreement with our relation. The Shen's sizes for the early-types are however systematically underestimated relative to our measurements, a difference which is further increased after the application of our corrections. In this respect, we note again that the SDSS data were typically acquired with
 a typical $r$-band PSF of $\sim$1.5$^{\prime \prime}$, which is comparable to the worst seeing in the $I-$band ZENS data, and no residual corrections as a function of size, magnitude, concentration and inclination effects (as done in this paper)  were  attempted in the SDSS analysis. Using our uncorrected sizes, the difference in size estimates is  $\sim 15\%$ at    $>10^{11}\Msol$, and increases towards lower galaxy masses,  up to $\sim20-25\%$ at $\sim10^{10} \Msol$. The median size of the Shen et al.\ early-type galaxy at this latter mass scale is $1.8$kpc, while  our  corresponding  \text{ZENS} estimate is 2.2 (2.4)  kpc   before (after) the application of our corrections.  

We  attribute the difference in size estimates between our analysis and Shen's to the combination of three factors: $(i)$  in ZENS the half-light radii are obtained using elliptical apertures, in contrast with the circular apertures used in the SDSS study. Although we have partly corrected for this difference in a post-processing mode,  by circularizing  our measurements for the comparison in Figure \ref{fig:ShenComp}, residual differences may remain; these differences are likely more relevant at relatively smaller sizes and relatively higher S\'ersic indices; $(ii)$ the use of different upper limits for the S\'ersic index in the SDSS and ZENS fits, see Appendix \ref{sec:AppNYVAGC}; $(iii)$  different morphological mixes in the two samples,  since Shen et al.\ use  a simple split in S\'ersic index to define their morphological classes;  and $(iv)$ finally, of course, to the fact that the Shen et al measurements were not corrected by observational biases as we do for our ZENS sample.
 
\begin{figure*}
\begin{center}
\includegraphics[width=130mm]{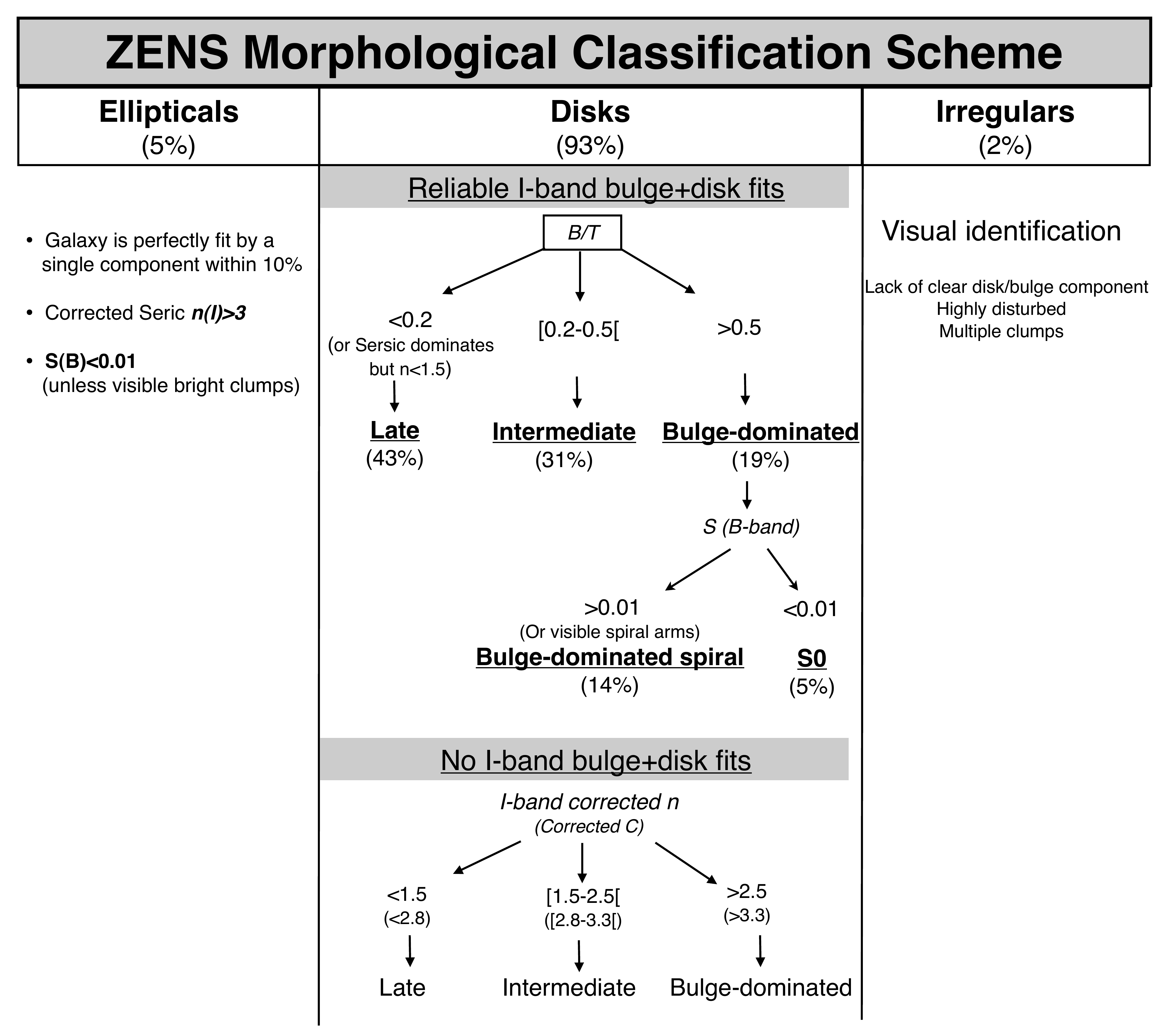}    
\end{center}
\caption{\label{pic:ClassFlowChart}Schematic description of the steps adopted for the \textsc{ZENS} morphological  classification. 
The numbers in parenthesis provide the fraction of
galaxies assigned to the different morphological classes with respect to the total ZENS sample.}
\end{figure*}

\section{A quantitative morphological classification for the ZENS galaxies} \label{sec:MorphClass}

\subsection{Classification Criteria}

The availability of a large suite of structural diagnostics for the \textsc{ZENS} galaxies enables us to perform an accurate and quantitative morphological classification, eliminating biases and dilution of signal in studies of galaxy properties as a function of morphology. 

We classified the  \textsc{ZENS} galaxies   into six morphological classes, primarily based on the prominence of the bulge component in the I-band. These   classes are, respectively, 
 ellipticals, bulge-dominated disks -- further divided into S0 and bulge-dominated spirals --
  intermediate-type disks, late-type disks and irregular galaxies.  
 The stamp images of the galaxies of different morphological types are shown
  in Figures \ref{fig:stampE}-\ref{fig:stampMergers} in Appendix \ref{app:ClassDist}.
   A flow-chart schematic description of the steps applied in our morphological classification scheme  is shown in Figure 
 \ref{pic:ClassFlowChart}, in which we also list the fraction of each morphological type in the  \textsc{ZENS} database.

In detail, our classification criteria are as  follows.
Ellipticals are required to be well fitted by a single S\'ersic profile with {\it corrected} S\'ersic index  $n>3$ and normalized residuals smaller than $10\%$ 
out to the sky level of the image (see e.g. \citealt{Kormendy_et_al_2009}).
We present  the $I$-band surface brightness profiles for all elliptical
 galaxies in Figure \ref{fig:EllipticalProfiles}: by definition, the light distribution of such galaxies is perfectly represented by a single (corrected) $n>3$  S\'ersic model.

 The distinction among different disk types is based on the following criteria:
galaxies with $B/T(I)<0.2$ are assigned to the late-type disk class, 
those with $0.2\leqslant B/T(I)<0.5$ are classified as intermediate-type disks and those with $B/T(I)\geqslant0.5$
 as bulge-dominated galaxies.
For many reasons, including a good degree of mixing of physical classes in the  visual classification of  the Third Reference Catalog \citep[RC3,][]{deVaucouleurs_1963}, we will refrain from using its popular nomenclature. We note however that, albeit with scatter,  our bulge-to-total separation for disk galaxies  roughly corresponds to a division in types S0 plus
S0a/Sa, Sab/Sbc and Sc/Sm, in order of decreasing B/T range.

Both S0 and spiral galaxies with large bulges fall in the bulge-dominated galaxies category.
The distinction among the two is based on the $B$-band smoothness parameter $(S_{B-band})$, which is set to $<$$0.01$ for S0 and larger than this value for early-type spirals, except in those case where faint spiral arms -- not prominent enough to cause a variation of the smoothness parameter -- are clearly 
 seen in the visual inspection of the images.
For nearly edge-on system it is obviously impossible to determine the presence or absence of
 spiral arms and the difference between S0 and bulge 
dominated spiral galaxies becomes more challenging.
 Given that we have no means to securely distinguish these  morphological types at these high inclinations,  
 we assigned to the bulge-dominated spiral class those galaxies which satisfy the B/T criterion above, and either present a dust lane in the disk plane or
have a high $B-$band smoothness parameter  ($S_{B-band}>0.01$). As seen in Figures \ref{fig:stampS0} and \ref{fig:stampSa},
 a large fraction of the highly-inclined 
galaxies are classified as bulge dominated spirals, and only those which clearly do not have a 
dusty disk are defined as S0.
The contamination of S0 galaxies into the bulge-dominated spiral class is hence likely non-negligible at the highest inclinations; a comparison between the relative fractions for face-on galaxies estimates this contamination at the 33\% level.

 The distinction between S0 and elliptical galaxies is also  non trivial, 
this time especially in face-on systems. To attempt to  distinguish  between these two classes we inspected all the residuals of the single component fits  and any sign of 
 a faint disk component was used as a discriminant diagnostic.
Finally the $B$-band smoothness $S$ parameter  was used to further validate the separation 
between the two types, by requiring  that for S0 $S_{B-band}>0.003$,
a condition which is however secondary to the fact of S0 not being perfectly fit by a single Sersic model.

\begin{figure*}
\begin{center}
\includegraphics[angle=90,width=165mm]{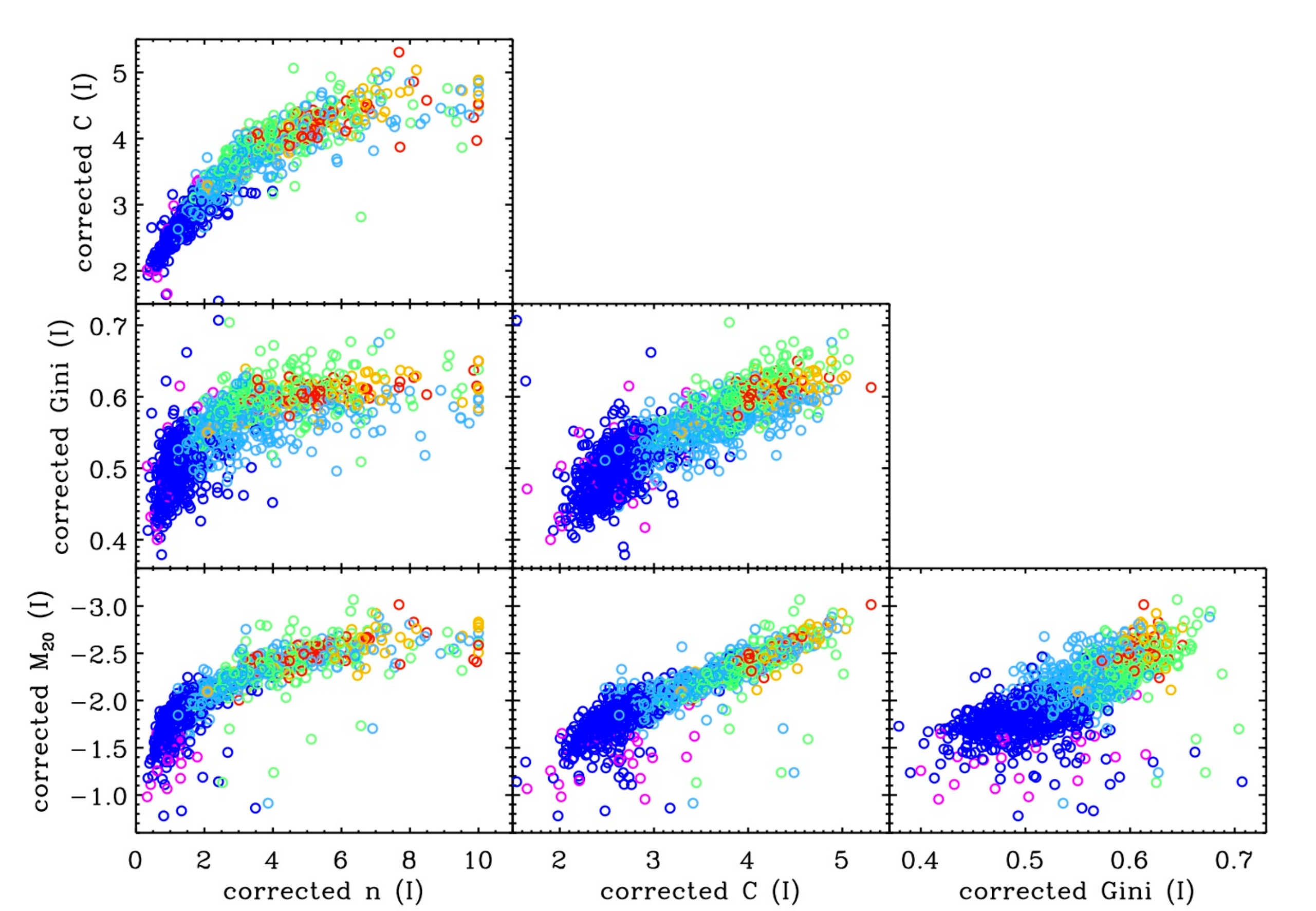}    
\end{center}
\caption{\label{fig:ZESTClasses}Location of the different morphological classes in the $C-Gini-M_{20}-n$ planes.
 The colors highlight the five morphological classes: red=ellipticals, orange=S0, green=bulge-dominated spirals, cyan=intermediate-type disks, blue=late-type disks and magenta=irregulars. }
\end{figure*}

We note that for $\sim15\%$ of all ZENS galaxies that were not classified as ellipticals or irregulars, and that would therefore be assigned a ``disk" classification,  it was not possible to obtain a reliable $I$-band bulge+disk decomposition
(see Appendix \ref{sec:testGIM2D}). In these cases the classification is based on the corrected S\'ersic indices from the single 
component fits, as shown in the right hand side of Figure \ref{pic:ClassFlowChart}. For the very few galaxies ($\sim1\%$) for which we could not obtain any reliable bulge+disk decomposition  or S\'ersic index,  we used the \emph{corrected} ZEST+ parameters (in particular the concentration) to define their morphological class. Specifically,  in such few cases, galaxies with $C<2.8$ were assigned to the late-type disk class, those with $2.8<C<3.3$ to the intermediate-type disks and galaxies with $C>3.3$ to the bulge-dominated spiral class. The precise values of these cuts were determined through the relation between $n$ and $C$ in Figure \ref{fig:ZESTClasses}.

Irregular (i.e., thus possibly disturbed by mergers/encounters) galaxies are identified in the ZENS sample by visual inspection;  in this category we 
included those galaxies which have no clear disk/bulge component or have a  highly disturbed morphology characterized by multiple clumps, as shown in Figure \ref{fig:stampMergers}.
 
 There are furthermore 26 galaxy-pairs or triplets in the \textsc{ZENS} sample which are also possible merger candidates, of which we show stamp images in Figure \ref{fig:stampMergers}.
  These galaxy pairs are identified as single objects in 2dFGRS,  and appear as single members in the group catalog, i.e., they have only one redshift measurements in the 2dFGRS catalogue. 
  For nine of such systems, we confirmed with SDSS \citep{York_et_al_2000} spectral or photometric redshifts, or with spectroscopic data available in the NED database, that the two galaxies are at the same redshift, strengthening the hypothesis that they are undergoing a merger event.
 There are 17 remaining  galaxies pairs, for which no such information is available. We note that five pairs/triplets show  tidal features and disturbed morphologies, supporting the merger scenario; the remaining 12 pairs display less clear signs of interactions, and hence for them a chance projection cannot   be excluded.
We will include nonetheless the latter in the sample of merger galaxies and, when necessary, test our results with and without these systems; these are flagged as `possible projections' in our ZENS catalogue.

For consistency with the original 2dFGRS and 2PIGG catalogs, we count the merging galaxies as a single galaxy pair system in the sample of 1455 galaxies, but we compute and provide photometric/structural properties for the individual members (thus making the total entries in the public ZENS catalog equal to 1484).
In the ``normal"  disk population, there are furthermore some galaxies which present clear tidal features and disk distortions; such galaxies are flagged has having plausibly undergone a recent galaxy-galaxy interaction or merger. 
Finally we also flag as candidate mergers those ZENS galaxies which, although identified as individual objects in the parent 2dFGRS/2PIGG catalog, are found at a distance from another  confirmed group member which is smaller or equal to the maximal separation observed in the above 26 galaxy pairs (i.e. 48$^{\prime\prime}$, roughly 50 kpc at the average ZENS redshift) and have a velocity difference from it $\Delta v \leqslant$ 500km/s. The properties of merger galaxies in the ZENS sample and the connection with the local and large scale environment are investigated in a forthcoming publication (Pipino et al.\ 2013, in preparation).

\begin{deluxetable*}{ccccccc}
\tablewidth{0pt}
\tabletypesize{\scriptsize}
\tablewidth{0pt}
\tablecaption{Median values of the corrected  structural diagnostics for the different morphological classes in bins of   ellipticity}
\tablehead{
\colhead{Type}  & \colhead{$<C>$}  & \colhead{$<$Gini$>$}  & \colhead{$<M_{20}>$}  & \colhead{$<S>$}  & \colhead{$<n>$} & \colhead{$<B/T>$}
}
\startdata \\
E                 &        4.11/-/-             & 0.60/-/-                     & -2.45/-/-                          &   0.005/-/-                   &           4.93/-/             &    -/-/-        \\   
S0               &       4.24/4.19/-       & 0.60/60/ -                 &  -2.48/-2.49/-                 &   0.006/0.0064/-          &         5.19/ 4.64        &    0.62/0.63/-  \\      
Bulge-dom. Spiral    & 3.88/3.94/3.92      &  0.59/0.59/0.61       & -2.30/-2.37/ -2.41         &  0.033/0.043/0.096       &  3.90/3.94/3.39       &  0.59/0.58/0.60 \\         
Interm.-type Disk         & 3.54/3.42/3.16      &  0.56/0.55/0.56       & -2.20/-2.15/-2.11          & 0.020/0.043/0.10      &  2.91/2.52/2.01       &  0.36/0.35/0.31\\       
Late-type Disk     &  2.58/2.60/2.56      &   0.48/0.49/0.51     & -1.80/-1.79/ -1.81        & 0.040/0.050/0.087   &  1.14/1.14/1.09       &  0.04/0.07/0.03\\            
\enddata
\tablecomments{\label{tab:ClassMedian}The median values of the corrected non-parametric diagnostics measured in the three ellipticity bins $\epsilon<0.33$, $0.33<\epsilon<0.55$
and $\epsilon>0.55$, from the lowest to the highest ellipticity. Note that Ellipticals and S0s have only one and two bins of ellipticity, respectively.  Note also that the dependence on ellipticity of  non-corrected  $C$ and $M_{20}$  values for disk galaxies shown in  Figure \ref{fig:ZEST_vs_Sersic} is largely eliminated by our correction scheme.}
\end{deluxetable*}

\subsection{Structural properties of the different morphological classes }

Our morphological classification scheme is primarily based on a bulge-to-total ratio separation. We inspect in Figure \ref{fig:ZESTClasses} the distributions of the \emph{corrected} $C$, $n$, Gini and $M_{20}$ for the various morphological classes, and we summarize in 
 Table \ref{tab:ClassMedian} the median values of these corrected non-parametric diagnostics for each morphological class, separately in three bins of ellipticity (i.e., inclination for disk galaxies).
Note that the six morphological classes largely segregate in specific region of the 
 structural planes. Again we stress that this consistency between B/T and non-parametric diagnostics is reached only after correcting the latter as described in Section \ref{sec:Simulations}: the raw measurements before the implementation of the corrections  would substantially mix different galaxy populations (e.g., through biased concentration and $M_{20}$ estimates, see Figures \ref{fig:ZEST+_corrected} and \ref{fig:ZEST_vs_Sersic}) -- a bias which could affect other published analyses based on   (uncorrected) non-parametric diagnostics. 
 
 There are of course some noticeable deviations which remain, even after correcting the non-parametric diagnostics. 
 A small number of galaxies classified as intermediate- or late-type disks have low $M_{20}$ indices for their concentration or Gini values, overlapping with the region of the plane which is mostly occupied by irregular galaxies. We visually inspected all these galaxies and verified that the $M_{20}$ deviations are mostly caused by the presence of bright clumps in the galaxies outskirts, but that a clear and ordered disk structure is present (we highlight these galaxies with a  ``*"  in Figures \ref{fig:stampSbc}-\ref{fig:stampSd}).
 
\begin{figure*}
\begin{center}
\includegraphics[width=120mm,angle=90]{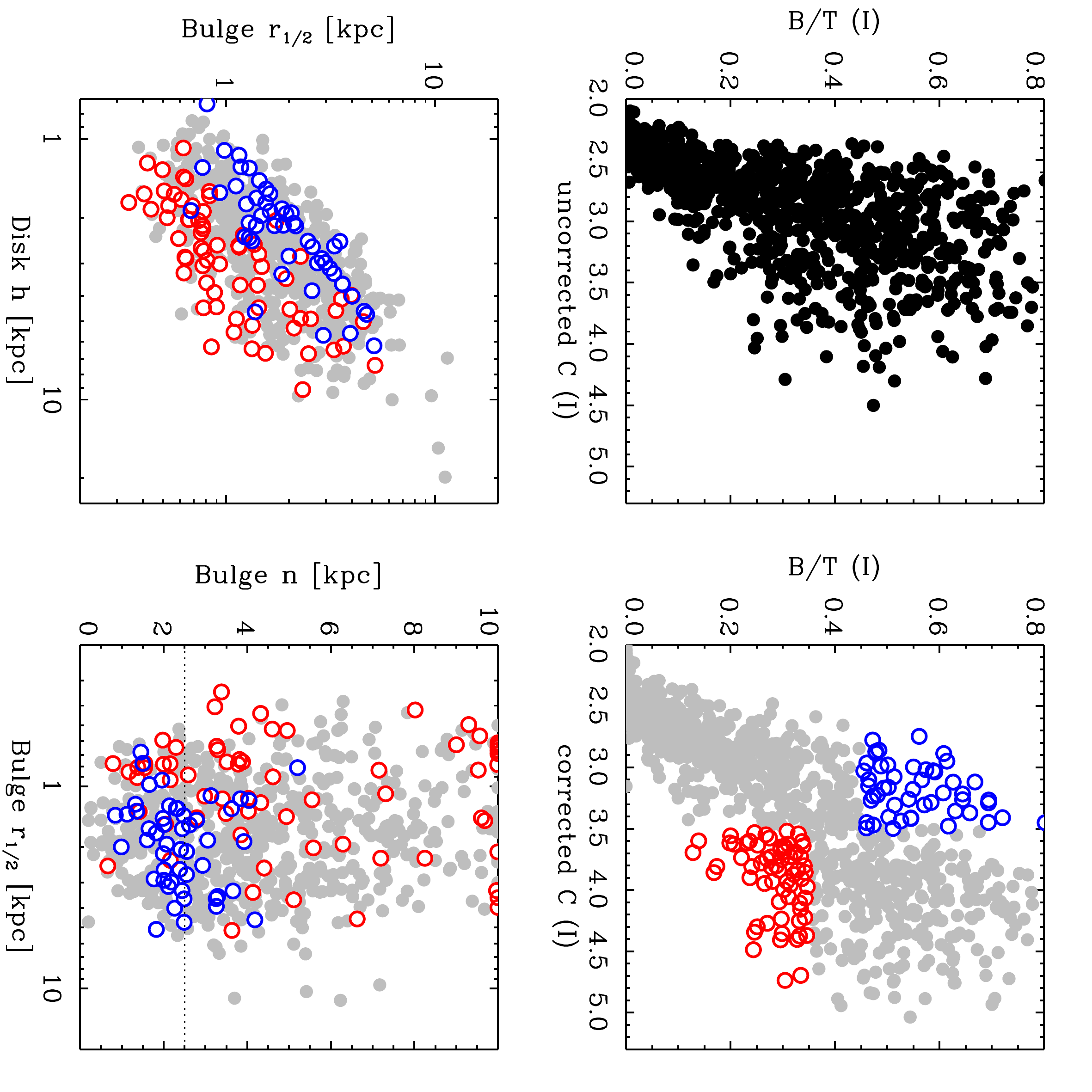} 
\end{center}
\caption{\label{fig:C_vs_BT}
The $I-$band relation between bulge-to-total ratio and uncorrected (top left panel) and corrected (top right panel) concentration for the  ZENS disk galaxies with bulge-to-disks analytical fits. In the plot with corrected-$C$ values,  with circles (red in the online version) are highlighted galaxies which are concentrated ($C>3.5$) but have a low bulge-to-total ratio  ($B/T<0.35$). With filled squares (blue in the online version)  are highlighted  galaxies which have low concentration $(C<3.5)$ and high $B/T>0.45$. Note the large scatter in the corrected-$C$ vs.\, B/T relationship, which makes $C$ a poor morphological indicator. Note also that uncorrected-$C$ values are, as expected, even worse representatives of the galaxy B/T (and thus physical morphology) properties. 
The two bottom panels show, from left to right, the  $I-$band bulge half-light radius vs.\, disk scalelength plane, and the $I-$band  bulge S\'ersic index vs.\, bulge half-light radius plane. The symbols (colors) in these plots represent the galaxy populations highlighted in the B/T vs.\, corrected-$C$ plane.
(A color version of this figure is available in the online journal.)}
\end{figure*}

Also, some  intermediate-type disk galaxies have relatively high concentration values ($C(I)>4$).
This demonstrate that concentration  is \emph{not} necessarily a proxy for the fractional contribution of the bulge component to the stellar light, as also pointed out by other authors \citep[e.g.][]{Scodeggio_et_al_2002}.
There is, of course, a correlation between $B/T$ and $C$ characterized by a  linear correlation coefficient of $\mathcal{R}=0.86$. However, the scatter is large and a given concentration value can be associated with a broad range of B/T ratios, as illustrated in the top right panel of Figure \ref{fig:C_vs_BT}.
We emphasize that the \emph{uncorrected} values of concentration  lead to a much larger scatter, which results in a weaker correlation between $B/T$ and uncorrected C with $\mathcal{R}=0.78$. 

We highlight with colors in Figure \ref{fig:C_vs_BT}  two regions of parameter space which are worth noticing, i.e., that of  galaxies with $B/T<0.35$ and $C>3.5$ (red points) and that of  $C<3.5$  and $B/T>0.45$ (blue points). As illustrated in the bottom  panels of the same figure, highly concentrated galaxies with low $B/T$ ratios are found below the bulk of the population in the relation between the bulge half-light radius and the disk scale length; the S\'ersic indices of their bulges are in most of the cases ($\sim70\%$) above $2.5$. Such objects hence host dense but small bulges, which are causing an increase in the concentration value but do not give a major contribution to the total galaxy light.
The blue points at  low $C$ and high $B/T$, in contrast,  are galaxies with typically  extended  bulges with low S\'ersic index (i.e.,  pseudo-bulge like structures, see e.g.,   \citealt{Carollo_1999,Carollo_et_al_1998,Carollo_et_al_2007} and \citealt{Kormendy_Kennicutt_2004} for a review and further references).


\begin{figure*}
\begin{center}
\includegraphics[width=120mm]{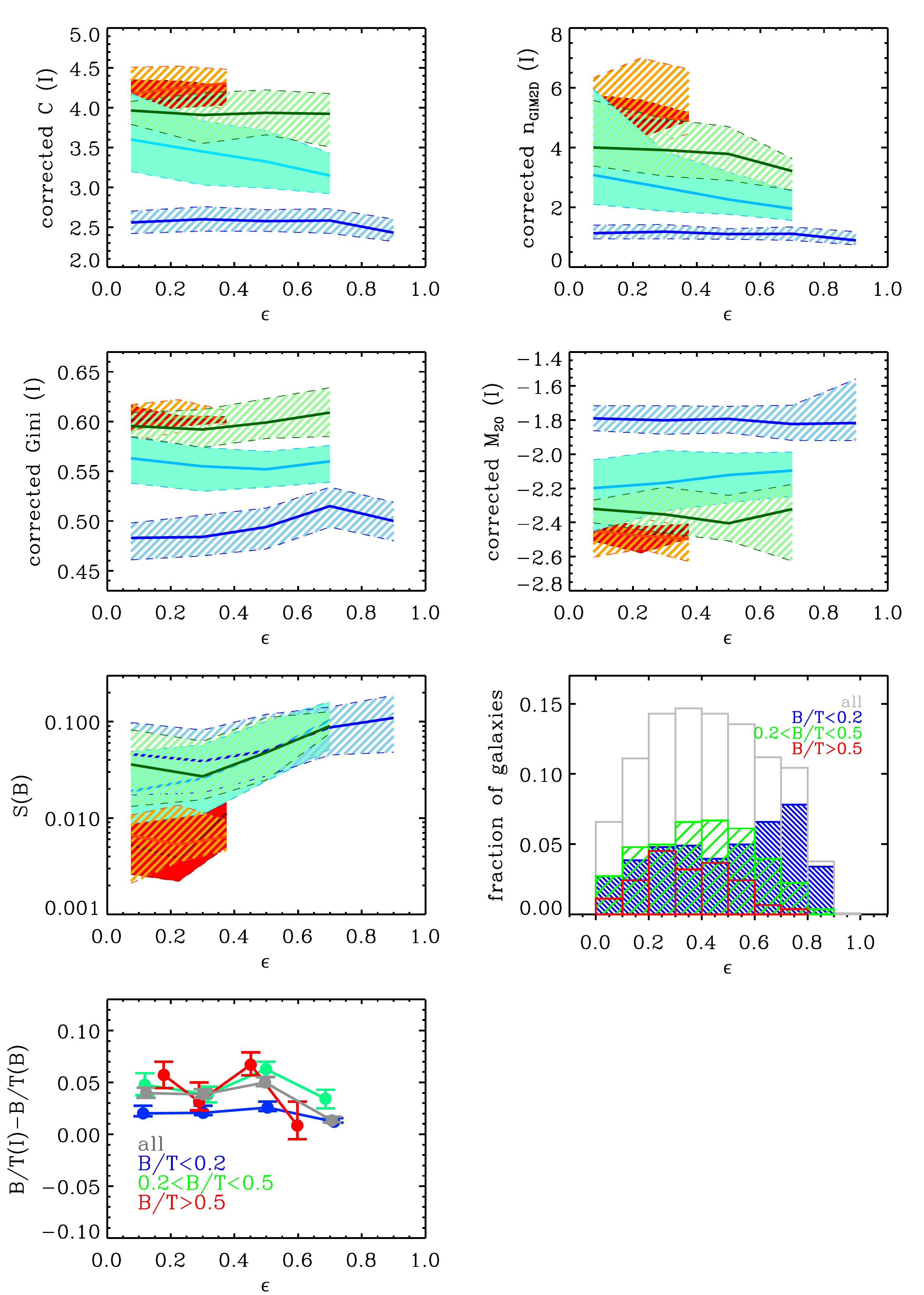}    
\end{center}
\caption{\label{fig:InclinationC} The shaded areas show the distribution of structural parameters (concentration, S\'ersic index, Gini coefficient, $M_{20}$ index and smoothness), as a function of measured ellipticity $\epsilon$, for galaxies of different morphological types. The color scheme reflects the morphological class: red=elliptical galaxies, orange=S0, green=bulge-dominated spirals, cyan=intermediate-type disks and blue=late-type disks. 
The distributions are calculated in bins of 0.15 in ellipticity for ellipticals and S0 -- which span a smaller range of ellipticity -- and in bins of 0.2 for all the other morphological types.
The thick solid lines mark the median values and the shaded areas the 25th and 75th percentiles.  Note the good segregation in  parameter space achieved by different morphological classes based on a B/T criterion; this is the result of our correction scheme, which eliminates observational biases (especially PSF-induced biases) from the measurements of the  non-parametric estimators of galaxy structure.  The  panel on the bottom-right presents the distribution of ellipticities for galaxies with $I-$band $B/T <0.2$ (blue histogram), $0.2<B/T<0.5$ (green histogram) and $0.5<B/T<0.8$ (red histogram) normalized to the total number of galaxies with available $I$-band B+D decompositions, for which the global distribution is shown in gray. 
Finally, the bottom-left panel shows the difference between the $I-$band and $B-$band B/T in four bins of ellipticity. Colors are as above; the points are located  on the x-axis at the median ellipticity of each given bin. The good agreement between the $I$- and $B$-band B/T measurements gives confidence that the B/T estimates are largely unaffected by inclination effects.}
\end{figure*}

In Figure \ref{fig:InclinationC} we furthermore show the different corrected structural indices for the morphological classes, as a function of the galaxy inclination. The distribution of corrected parameters for each individual class are fairly flat with ellipticity, as they should (in contrast with the non-corrected parameters, which we have shown to have a bias with ellipticity; see Section \ref{sec:ReliabilityCorrections} and Figure \ref{fig:ZEST_vs_Sersic}).
Note the galaxy smoothness maintains a substantial dependence on inclination by increasing towards higher inclinations; this parameter has not been corrected, since we  used only smooth galaxy models to determine our correction matrices. Part of the dependence of $S$ on ellipticity might thus be  a residual observational bias; part might however be a consequence of the prominence  of dust lanes in edge-on galaxies. 
 
Dust  absorption  in the disk could also in principle artificially lower the derived bulge-to-total ratios for edge-on galaxies  (see for example \citealt{Driver_et_al_2007}),  and hence introduce a bias in the B/T-based morphological  classification. 
We expect  this effect to be minimal in our analysis, since we use the relatively  unaffected $I-$band to derive  our fiducial $B/T$ estimates, 
but it is nonetheless worthwhile to test whether there are indication of  biases among the morphological classes given that some attenuation of the disk (and bulge) light is certainly occurring. 

In the lower right panel of Figure \ref{fig:InclinationC} we plot the distributions of ellipticity for  disk galaxies with bulge-to-total ratio $<0.2$ (blue histogram), $0.2<B/T<0.5$ (green histogram), and $0.5<B/T<0.8$ (red histogram).  
For $\epsilon\lesssim0.6$ we find a  reasonable constancy with ellipticity of the relative contribution of three disk types to the total disk population, indicating that there are no major biases in this regime.
At higher axis ratios, there is a depletion of $B/T>0.2$ galaxies and an increase of the ``bulgeless" disk galaxies, which we  interpret mostly as a 
consequence of a true change in the morphological mix rather than of strong dust absorption. The presence of a bulge component, even in highly-inclined disks, produces a rounding of the isophotes and thus of the measured axis ratio. Therefore, 
the decrease in the number of bulge-dominated galaxies and intermediate-type disks at $\epsilon \gtrsim 0.6$ likely reflects this correlation between ``morphology'' and  measured ellipticity.
Likewise,  the peak  of $B/T<0.2$ disk galaxies at $\epsilon \simeq 0.7$  can be explained by the ``rounding" of isophotes due to PSF effects in these galaxies and the intrinsic galaxy thickness. Indeed no galaxy is observed at $\epsilon>0.90$ and the fraction of $B/T<0.2$ disk galaxies with $0.6<\epsilon<0.90$ relative to the total sample of disk galaxies  is $\sim$25$\%$, i.e., if they were (infinitely thin disks and) equally
distributed up to $\epsilon=1$, each of the four 0.1-wide bins between $0.6<\epsilon<1$  would contain 6$\%$ of the sample, which is the average number observed
at lower axis ratios. 

To further test whether the estimates of galaxy bulge-to-total ratio are strongly affected by dust extinction, we compare in the bottom panel of Figure \ref{fig:InclinationC} the difference between the  B/T   values measured in the $B-$band and those measured in the $I-$band, as function of the galaxy ellipticity. In grey we show the full ZENS sample; other colors are the results for galaxy samples  split in the three B/T classes discussed as above.
The $I-$band B/T values are typically larger by about 5\% than those in the $B$ filter, which is consistent with larger dust absorption effects at the shorter wavelengths.
There is however no clear variation with $\epsilon$ of the difference  $B/T(I)-B/T(B)$ within each of the three morphological classes.
We thus conclude that dust effects are generally a modest contribution to the total error budget for the B/T measurements, which is dominated by other sources of random and systematic errors (see Figure \ref{fig:BTscatter} and the corresponding discussion in the text).

We note that other authors have combined empirical disk attenuation-inclination relations with  theoretical dust models in order to derive inclination-dependent corrections for dust absorption in bulge and disk fluxes \citep[e.g][]{Driver_et_al_2007,Driver_et_al_2008}. Using the $I$-band parameters in Table 1 of  \citealt{Driver_et_al_2008} together with  their equations 1 and 2, we estimate a correction of  order of $\Delta B/T\simeq0.1$ to the ZENS  $I$-band B/T values,  comparable with or even smaller than the typical uncertainty on the B/T measurements. While such corrections could be important in some contexts, they inherit the uncertainties on the adopted dust models and would not cause a significant shift of morphological class in the ZENS sample.  We hence choose not to apply any such (model-dependent) dust correction,  and to adopt the dust-uncorrected GIM2D measurements of B/T as our fiducial estimates for this parameter.

In summary, the above considerations, and  the constancy of the median B/T ratio for the different morphological classes in three ellipticity bins shown in Table  \ref{tab:ClassMedian}, lead us to conclude that our $B/T$ measurements are largely independent of axis ratio and that our morphological classification is thus largely unaffected by inclination effects.  

As a final remark, we note that the $n$ S\'ersic indices of intermediate-type disks also show a residual (but marginal) dependence on ellipticity: this is of course the morphological class for which the likely similar light contribution of bulge and disk components along  the line-of-sight is the most difficult to disentangle. Nevertheless, the   $n$ values spanned by these morphological class at the different ellipticities are consistently in the ``intermediate" range of $n\sim2-2.8$ (see table \ref{tab:ClassMedian}), providing a consistent characterization of these systems across the whole ellipticity range.  

\section{The concentration of satellites galaxies of different Hubble types in different environments} \label{sec:Results}

The main goal of ZENS is to study galaxy properties, at fixed stellar mass and Hubble type, as a function of several environmental parameters derived self-consistently for the same galaxy sample, i.e.,  the host halo mass, the large scale density $\log(1+\delta_{LSS})$ and the projected group-centric distance. We furthermore split the ZENS galaxy sample in central galaxies in their host group halos, and satellite galaxies, which orbit the central galaxies within those halos (see Paper I for the precise definition of central and satellite galaxies, and for the environmental parameters).  Several ZENS analyses are underway, which use the ZENS database, including the structural and morphological information derived in this paper,  to investigate how galaxy structural and morphological properties vary across the different environmental regimes. 

In this second paper in the ZENS series we want to show a first utilization of the structural measurements  presented above to answer, from a purely observational perspective, a simple question, namely how the concentration of \emph{satellite galaxies} depends, at fixed stellar mass, on $(i)$  Hubble type, and $(ii)$ on group mass, group-centric distance and large-scale structure density.

\subsection{The role of Hubble type at fixed stellar mass} 
 
Galaxy structure, as described by the concentration parameter, is a strong function of galaxy mass  \citep[e.g.][]{Kauffmann_et_al_2003,vanderWel_2008}.  As we have discussed above, concentration is however a relatively poor indicator of galaxy morphology (when this is defined according to a more physical parameter such as the bulge-to-total ratio, as we do for the ZENS galaxies).  Furthermore, morphology also is a function of stellar mass  (see Paper III and also, e.g., \citealt{Pannella_et_al_2006,Bamford_et_al_2009,Oesch_et_al_2010,Bernard_et_al_2010,Vulcani_et_al_2011}).  With available robust (corrected) concentration values, and a B/T-based morphological classification, we thus investigate how much of the concentration vs.\, galaxy stellar mass correlation is to be ascribed to a variation in the Hubble type of galaxies, and how much is actually driven by a real change in concentration within each individual morphological types. Stellar masses are taken from Paper III.
The results of this analysis will also help us interpret  the analysis of satellites' concentration in the different  environmental parameters, for which, in order to gain statistics, we will use broader morphological classes.

Figure \ref{fig:Cmass} shows the corrected concentration as a function of galaxy stellar mass.
The points connected with lines are the median concentrations for satellite  galaxies, divided 
according to the morphological types (note that S0 and bulge-dominated spiral galaxies are joined together into a single broad morphological bin of  ``bulge-dominated" galaxies).
The medians were calculated using a running box above the mass completeness of each morphological type (the results are noisier but consistent when using independent mass bins).
The inspection of this figure shows an increase in concentration with increasing galaxy stellar mass  at fixed morphological type for disk galaxies. Specifically, our  elliptical satellites' sample covers a small  range of stellar mass ($\sim10^{10-10.5}M_\odot$), over which the concentration is  observed to remain constant. For all types of disk galaxies where the range of masses covered is larger, a trend of concentration with stellar mass is clearly observable, separately for each morphological disk type (the latter defined according to the galaxy bulge-to-total ratios as described in Section \ref{sec:MorphClass}).  We note that residual bulge-to-disk variations with stellar mass {\it within} each morphological disk class are possible, since our morphological classification was done independent of galaxy stellar mass, and each morphological bin covers a relatively broad range of bulge-to-total 
ratio. Quantitatively, however, the a posteriori computation of the median bulge-to-total ratios  vary, between the mass bin centered at  $10^{10} M_\odot$ and the mass bin centered at $10^{10.7} M_\odot$, only from 59$\%$ to 60$\%$, 34$\%$ to $36\%$, and $7\%$ to $12\%$, respectively for bulge-dominated, intermediate-type and late-type disk galaxies. Thus, we conclude that the known concentration-mass correlation is  not solely driven by a correlation between morphological type and stellar mass, but is, at least  partially, driven also by a genuine increase in concentration of galaxies of similar  Hubble type, i.e., of similar bulge-to-total ratio.

We fit the observed trends of corrected concentration versus galaxy stellar mass with a linear relation, $C=\alpha+\beta \log(M/\Msol)$, in the mass range which is probed by all morphological type, i.e., for those bins which are above $10^{10}\Msol$ for their full width.
The global $C-M$ relation for the late-type disks shows a break around $\sim 10^{10} M_\odot$, below which mass the relation flattens; a linear fit is however a good approximation at $M>10^{10}\Msol$ where we can make a meaningful comparison with the other types.
The slopes $\beta$ for the four morphological classes above this threshold mass are, respectively,  
$\beta_{bulge-dom}=0.18\pm0.04$, $\beta_{interm.-type}=0.21\pm0.07$ and $\beta_{late-type}=0.34\pm0.15$; satellite elliptical galaxies are consistent with no dependence of concentration on the galaxy mass, i.e $\beta_{ellipticals}=0$. 
Although, as commented above, the lack of dependence of $C$ on stellar mass for the ellipticals is established only across a relatively limited mass range, generally speaking this finding follows the global trend, established on the disk galaxies, of a flattening of the $C$ vs.\, galaxy stellar mass relation from the later- to the earlier morphological type.

\subsection{The dependence of (disk) satellite concentration on  LSS density, group-centric distance and group halo mass} 

At the masses of our study, there are essentially no irregular satellite galaxies; furthermore, elliptical galaxies contribute only 5\% to the total number  of satellites in the ZENS sample at these masses. 
The small number of elliptical satellites does not enable us to perform a statistically significant analysis of their  dependence on environment.
Since the vast majority ($\sim95\%$) of satellites in the sample are disk galaxies, we choose to maximize morphological ``purity''  in our analysis of the dependence of satellite concentration on the different environmental parameters by  limiting it to  to the  sample of {\it disk} satellites only.

\begin{figure*}
\begin{center}
\includegraphics[width=89mm,angle=90]{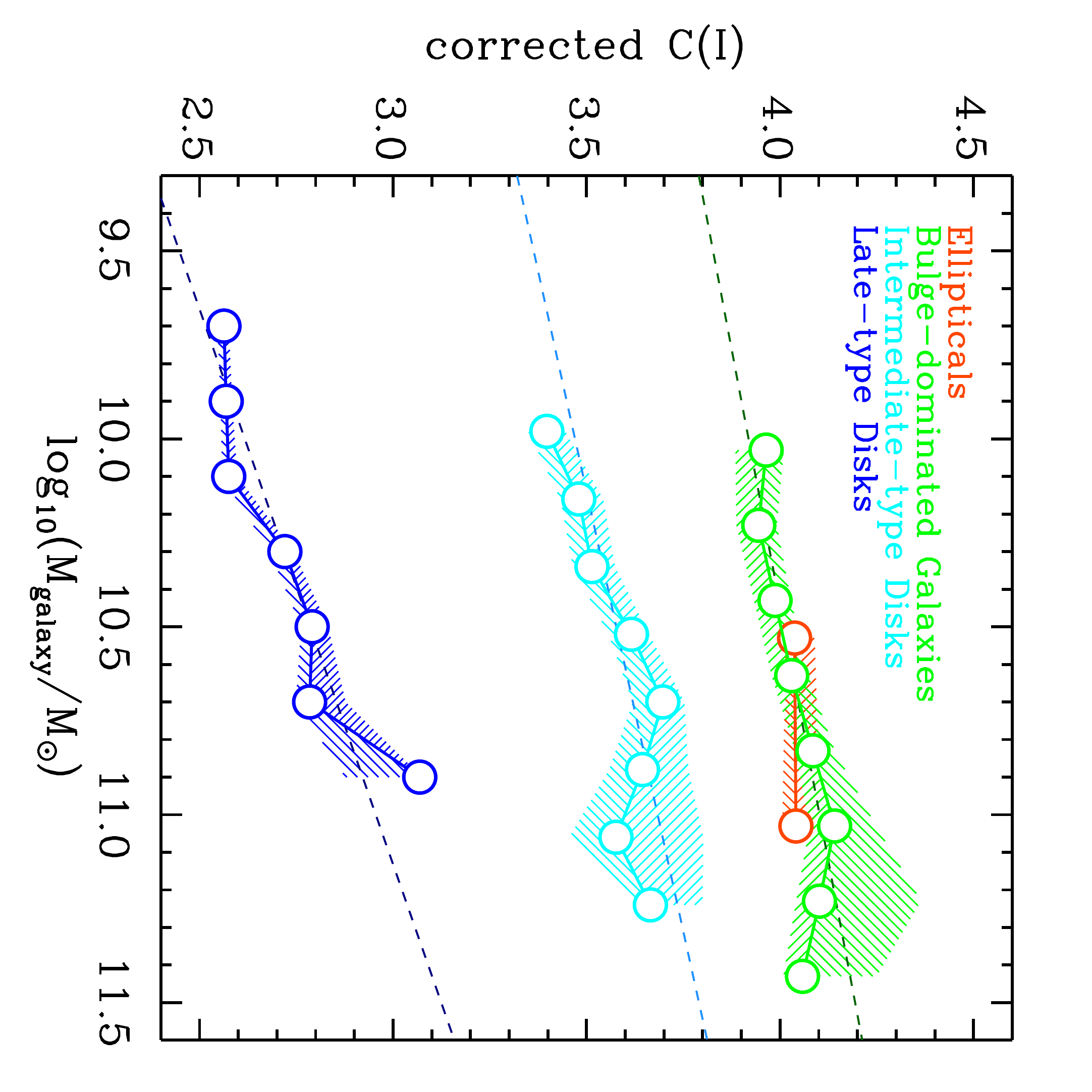}    
\end{center}
\caption{\label{fig:Cmass}Corrected concentration $C$ versus galaxy stellar mass (from Paper III), for satellite galaxies and  different Hubble types defined in Section 
\ref{sec:MorphClass} (S0 and bulge-dominated spirals are joined together in the broad bulge-dominated galaxies class). The points connected with lines show the   median galaxy concentration,  calculated over a running box of width 0.3 dex in mass for intermediate-type and late-type disks, and of width 0.4 dex for the earlier types. The dashed lines are the best linear fits to the observed relations over the mass range probed by all morphological types, i.e., for masse above $10^{10}$ \Msol. The shaded areas correspond to the 1$\sigma$ error bars on the median values.
(A color version of this figure is available in the online journal.)}
\end{figure*}

\begin{figure*}
\begin{center}
\includegraphics[width=120mm]{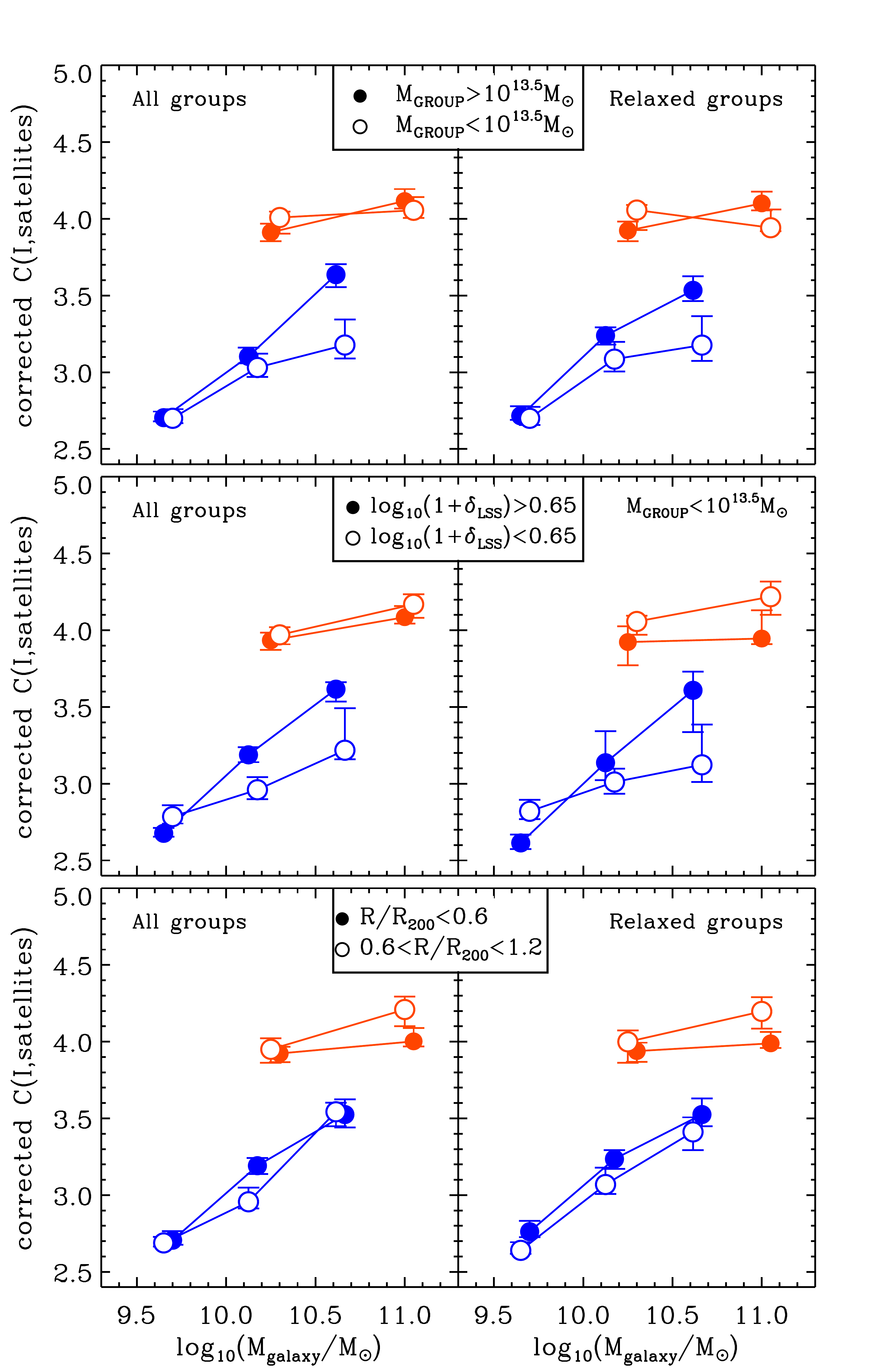}    
\end{center}
\caption{\label{fig:CmassGrp}The panels show the corrected concentration parameter $C$ as a function of galaxy stellar mass, split in bins of environments as follows: top panels show the dependence on group mass $M_{GROUP}$; central panels on LSS (over)density $\delta_{LSS}$; bottom panels on group-centric distance ($R/R_{200}$). 
Only disk satellites are shown in this analysis. These are split in two broad bins of morphologies, specifically bulge-dominated galaxies (which include bulge-dominated spirals and S0 galaxies; grey circles, red in the online version) and disk-dominated galaxies (which include intermediate-type and late-type disks; black squares, blue in the online version). 
For all three environmental indicators, filled symbols show results for the ``denser" bin, and empty symbols for the ``lighter" bin. In the top and bottom panels, left plots are for all ZENS groups, and right plots  for relaxed groups only (see text).  In the central panels, the left plot shows again results for all groups, but this time the right plot shows results for groups with $M<10^{13.5}\Msol$ only, to avoid spurious effects with group mass when studying the effects of the LSS density field.
Galaxies located at a radial position $R>1.2R_{200}$ are excluded from the analysis.
Symbols are positioned at the center of any given mass bins, with a small offset  between denser/lighter environments applied for visual clarity.
(A color version of this figure is available in the online journal.)}
\end{figure*}

In  Figure \ref{fig:CmassGrp} we present the  (corrected) concentration vs.\, galaxy stellar mass relation for disk satellite galaxies, split  into two broad morphological bins of ``bulge-dominated" galaxies and `` disk-dominated" galaxies by adding together  S0s and bulge-dominated spirals in one bin, and intermediate-type and late-type disks in the other bin. The relation is plotted separately as function of group mass $M_{GROUP}$, LSS (over)density $\delta_{LSS}$, and group-centric distance $R_{R200}$ (from top to bottom). 
The broad bulge-dominated galaxies bin is split in three bins of galaxy stellar mass, i.e., $\log (M/\Msol) \in $[9.5,9.9[, [9.9,10.45[,  [10.45,10.88];  we use only  two bins of galaxy stellar mass for the broad bulge-dominated galaxies bin, i.e.,  $\log (M/\Msol) \in $[10.,10.6[,$>10.6$. 

Before discussing the results we present some considerations on the  biases and uncertainties which may affect the analysis.  First, we remind that trends with group mass that we measure are, intrinsically,  possibly even stronger. As discussed in Paper I,   the uncertainty on our group mass estimates  tends to ``wash  out"    trends  with this environmental quantity; quantitatively,  real trends  with group mass have slopes $\approx 1.3-1.4$ steeper on average than what we can measure in the ZENS sample. 

In Paper I we furthermore discuss a classification of the ZENS groups into ``relaxed" and ``unrelaxed" groups, according to whether a self-consistent solution for a central galaxy can be found (in relaxed groups) or not (in unrelaxed groups). The latter are expected to be combination of genuinely dynamically young, merging structures, contaminated by physically relaxed groups for which several observational biases prevent us to properly identify their central galaxy. In the spirit of checking that our results are not affected by the inclusion of the unrelaxed groups in our studies, in Figure \ref{fig:CmassGrp} we show the results for all  ZENS groups (left panels) and for the subset of relaxed groups only (right panels) when studying the dependence of the $C$-mass relation on either $M_{GROUP}$ or $R_{R200}$.

We furthermore exclude from the following analysis those galaxies which are located at group-centric distances  $R>1.2R_{200}$. 
 In some of the ZENS groups relatively massive galaxies are found to lie in the outskirts of the group. As discussed above, this is partially a consequence of observational limitations but it may also be the signature of  on-going accretion of sub-haloes. 
 The sample selection at  $R<1.2 R_{200}$ thus helps  minimizing the contamination at these large radii of in-falling substructures and maximizing the ``purity"  of galaxies that are satellites within a single common halo. 

In Paper I we also comment on the fact that, at group masses above $\sim10^{13.5} M_\odot$, group mass $M_{GROUP}$ and LSS (over)density $\delta_{LSS}$ are correlated by nature, since massive groups are found by definition only in dense regions of the LSS. In contrast, groups of  masses $<10^{13.5} M\odot$ are found at any value of $\delta_{LSS}$. It is thus possible to disentangle the effects of $M_{GROUP}$ and $\delta_{LSS}$ by limiting the studies as a function of $\delta_{LSS}$ to group masses below this threshold.  We adopt this strategy here as well, and, in order to search for a dependence of the $C$-mass relation for satellites on $\delta_{LSS}$, we test that the results obtained when including all  groups in the analysis with $\delta_{LSS}$ (left central panel) hold also when   restricting the group  sample to   groups  with $M_{GROUP}<10^{13.5}\Msol$ (right central panel).

Having established  that the concentration parameter is a  function of both galaxy stellar mass and morphological type (see Figure \ref{fig:Cmass}),  it is important  to keep dependencies of both factors with the environment under control in order to proceed in this analysis. In Paper III we discuss the variation of the stellar mass of each broad morphological type with our environmental indicators and show that in our sample galaxy mass is largely independent of environment at fixed morphological type.

Here we note that, for the broad disk-dominated galaxies bin, biases in the median mass with any of the environments are not an issue, as for this class we can afford a fine galaxy mass bin splitting, and compare the $C$-mass relation for different environments within relatively small galaxy mass bins. For the broad bulge-dominated galaxies bin, however, given  the fairly broad ranges covered by each of the two galaxy stellar mass bins,  a difference  in the median galaxy stellar mass  (and, more generally, of the galaxy stellar mass distributions) between some of the different environments could in principle  be present. This  could  induce a spurious dependence on such environments  of the $C$ vs.\ stellar mass relations plotted in Figure \ref{fig:CmassGrp}.
To keep this potential bias under control, we computed the median galaxy stellar masses within each environmental bin,   and found that, for all   three environments under study, the median stellar mass of galaxies in the broad bulge-dominated galaxies morphological bin changes by less than 0.1 dex between the environmental bins under comparison (i.e., low/high $M_{GROUP}$, $\delta_{LSS}$ and inner/outer group regions). 

We also verified if the morphological mix  changes substantially with environment. 
The fractional contribution of  intermediate-type  disk galaxies to the total intermediate- plus late-type disk populations which  compose  the broad disk-dominated galaxies morphological bin defined above  varies at most by $5\%$ between the low/high $M_{GROUP}$ samples. Precisely, in the three bins of galaxy stellar mass defined for the broad disk-dominated galaxies morphological bin this fractional contribution, i.e.,  
 $\frac{N_{interm.}}{N_{interm.}+N_{late-type}}$,  equals, respectively,  35\%, 55\%, 85\% for $M_{GROUP}<10^{13.5} M_\odot$  and 36\%, 56\%,  82\% for $M_{GROUP}>10^{13.5} M_\odot$.
The  fractional contribution  of  S0s to the broad bulge-dominated galaxies morphological bin is also almost identical between the samples of $M_{GROUP}<10^{13.5} M_\odot$  and $M_{GROUP}>10^{13.5} M_\odot$, and equal to  
$\sim15\%$  at $M_{galaxy}\sim10^{10} \Msol$ and   $\sim30\%$ at $M_{galaxy}>10^{10.5} \Msol$. 
Likewise at $M\gtrsim10^{10}\Msol$, the maximal variation in the fraction of intermediate galaxies within the disk-dominated galaxies class is 5\% across the two $\delta_{LSS}$ bins.
On the other hand, variations of $\sim20\%$ are detected in the contributions of S0s  to the broad bulge-dominated  morphological bin between samples at small/large group-centric distances  and low/high LSS (over)densities at  $M\sim10^{11}\Msol$;  hence we use caution in  interpreting, for this broad morphological type, possible differences in the $C$ vs.\, galaxy stellar mass relation at these mass scales. 

The first thing to realize from   Figure \ref{fig:CmassGrp} is that  there are at most weak trends with the different environments of the concentration vs.\, stellar mass relation for disk satellite galaxies.  Two environmental  trends are however potentially  important:

$(1.)$ There is a global trend at and above galaxy masses of order $10^{10} M_\odot$, for satellites with a bulge-dominated morphology to be, albeit at the $1-2\sigma$-level, systematically more concentrated, by about 5-10\%,   at low  environmental densities (large group-centric distances and low LSS densities) than at high environmental densities (high mass groups, small group-centric distances, high LSS densities). 
This difference could be at least partially driven by a change in the mix between S0 and bulge-dominated spirals within the broad bulge dominated morphological bin.  

$(2.)$ In contrast, the disk-dominated population shows a systematic increase in  median concentration, at fixed galaxy stellar mass above $10^{10}\Msol$, with increasing  group mass and LSS density. In this case the similar contributions of intermediate-type disks and late-type disks to the broad disk-dominated morphological bin suggest that there is a genuine environmental effect at work. The difference in concentration between low and high mass groups is $\Delta C \simeq 0.3$; furthermore, the effect is robust towards the inclusion/exclusion of the sample of unrelaxed groups.   The significance of this concentration difference, as assed through a $t$-test on the two samples, gives  a probability $>91\%$ that the two populations have truly different median concentrations (or a Kolmogorov-Smirnov test give a 95\% probability of a common parent distribution).  
On the other hand, the trend with $\delta_{LSS}$ becomes weaker when only groups with $M_{GROUP}<10^{13.5} M_\odot$ are considered for the LSS analysis. As discussed in Paper I, the effect observed with $\delta_{LSS}$ when also the most massive groups are included is likely a spurious reflection of the dependence of concentration on halo mass.  

As indicated above, a full investigation of the implications of  either of these trends is beyond the scope of this paper. We will study these effects in more detail, also within the framework set by other independent analyses of the satellite population, in a future ZENS analysis.

\section{Summary and concluding remarks}\label{sec:Conclusions}     
      
We have presented  detailed structural analyses   performed on the 1455 galaxies in the 141 ZENS groups introduced in  Paper I. We remark that the corresponding ZENS catalogue has however 1484 galaxy with valid entries, since we measure parameters   for individual galaxies that are members of 29 galaxy pairs, which are given as one single entry in the parent 2dFGRS galaxy catalogue.

The parametric and non-parametric structural measurements  presented here, together with the detailed environmental parameters from Paper I and the photometric (including stellar masses) measurements presented in Paper III, 
set the basis for a number of forthcoming publications which use the \textsc{ZENS} data to 
explore which  environmental scales are  relevant for the evolution of morphologically different  galaxy populations at different mass scales. The measurements are published in the global ZENS catalog with Paper I. 
In detail, the measurements that we have presented here are:

\begin{itemize}

\item  Strength of bars in disks, quantified through  an isophotal analysis (Section \ref{sec:IsophotalAnalysis});

\item  Single- and double-component (bulge+disk) S\'ersic fits parameters both in the $B$ and $I$ bands, including model-based galaxy sizes  (Sections \ref{sec:SersicFits} and  \ref{sec:GIM2D_decomp}), as well as bulge+disk+bar fits for disks with a noticeable bar component (Appendix \ref{app:BulgeBarDisk});
      
\item Non-parametric structural indices of concentration, asymmetry, smoothness of the light distribution, and  Gini and $M_{20}$ coefficients, and ``aperture photometric" size estimates (Section \ref{sec:ZEST});

\item Morphological classes, based on a quantitative bulge+disk criterion, augmented (or supported) by quantitative criteria regarding the non-parametric diagnostics, after   correction  for observational biases (Sections \ref{sec:GIM2D_decomp} and \ref{sec:MorphClass}).

\end{itemize}

Crucially, we do indeed derive  correction matrices,  which we apply to  the relevant structural estimates, to minimize biases which, depending on PSF size as well as galaxy magnitude, size, shape of light profile and ellipticity, would otherwise prevent a reliable comparison of the structural properties of  galaxies observed in different seeing conditions, and lying in different  regions of this four-dimensional galaxy parameter space (see Section \ref{sec:Simulations}). 

As expected, biases in the model-fit parameters are substantially reduced thanks to the  treatment of the PSF-blurring effects in these algorithms; still, some are present even in the model-fit parameters, which may have an impact in some analyses if left uncorrected. Disk properties are well measured and require very modest or no further corrections; so are bulge-to-total ratios, which therefore offer an excellent parameter to base a quantitative morphological/structural classification. In contrast, bulge $n$ S\'ersic indices and half-light radii are degenerate in some circumstances. This implies that, on global galactic scales,  concentration parameter or n-S\'ersic index, alone, are not a good proxy for bulge-to-disk ratio and thus morphology:  bulges can give large contributions to the total light budget and have large half-light radii, leading to low galaxy concentrations, or, vice-versa, bulges can be very compact and lead to high  galaxy concentrations despite a modest contribution to the total light. Finally, we particularly warn against using uncorrected non-parametric estimators as galaxy classifiers since, understandably,  they suffer from severe observational  biases  and introduce severe  errors in the classifications (see Section \ref{sec:Simulations}).

As a first application of our corrected structural measurements, we have studied
 the variation of concentration in \emph{satellite} galaxies of fixed  stellar mass (from Paper III) with morphological type, and with the three environments detailed in Paper I, i.e., the mass of the host group halo, the projected group-centric distance, and the density of the large-scale structure  cosmic web (Section \ref{sec:Results}).

We find that the known correlation of satellite concentration with galaxy stellar mass  \citep[e.g.][]{Kauffmann_et_al_2003} holds at a fixed morphological type. Specifically,  there is a genuine increase in concentration with increasing stellar mass for disk satellite galaxies within each separate bin of bulge-dominated galaxies, intermediate-type and late-type disk morphologies. The slope of the concentration vs.\, galaxy stellar mass relationship flattens from the later to the earlier types (and becomes $\sim0$ for satellites with an elliptical morphology, which however cover a limited range in galaxy stellar mass in our sample).  It is not trivial to disentangle, in a physically meaningful way, the contributions of  an increasing bulge-to-total ratio with increasing mass, and an increasing bulge concentration at fixed bulge-to-total ratio with increasing mass, to the increase in concentration of satellite disk galaxies with increasing stellar mass. This is true even in our study, in which we did  indeed base our morphological classification on  the bulge-to-total ratio, and thus  bulge-to-total  variations within each Hubble type should be minimized. There are nevertheless some residual effects, as the median
$B/T$  is found to vary in our sample from $\sim7\%$ at $10^{10}\Msol$ to  $\sim12\%$ at $\sim10^{10.5}\Msol$ for late-type disks, and from 59$\%$ to 64$\%$  between $10^{10}\Msol$ and $10^{11}\Msol$  for bulge-dominated disks. Still, for intermediate-type disk satellites, in which the bulge-to-total ratio is tightly constrained by   firm lower an upper boundaries by definition, the increase in satellite density with stellar mass can be genuinely ascribed to a $\sim 30\%$ decrease in the bulge-to-disk  size  ratio, i.e., to a genuine increase in  the concentration of the bulge component at fixed bulge-to-total  light  ratio. We tentatively assume this as the explanation for the increase of disk satellite concentration with stellar mass at fixed  Hubble type, an hypothesis which we will test with further analyses.

When considering the galaxies environment,  we find that, at galaxy stellar mass $\sim10^{10} \Msol$ and above,  $(i)$ bulge-dominated satellites tend to be marginally more concentrated at low LSS densities and high group-centric distances, and $(ii)$ in contrast, disk-dominated  satellites are significantly more concentrated in high group masses.

The interpretation of the first  weak trend as a hint for an environmental effect is further hampered, at $M\sim10^{11}\Msol$, by a higher fraction of S0s relative to bulge-dominated spiral galaxies in low $\delta_{LSS}$ relative to high densities (60\% and 0\%, respectively in $M_{GROUP}<10^{13.5} \Msol$ groups and 58\% and 35\% over all groups) and in the outskirts relative to the cores of groups (50\% and 31\%, over all groups). This change in morphological mix at the high-end of the B/T sequence is consistent with other observational works which have also reported an increase in the S0 fraction in the outskirts of groups and clusters, although with no distinction between central and satellite galaxies.

The early study of \cite{Whitmore_et_al_1993} on local cluster reveals a  drop in the S0 fraction close to the cluster centers, which they interpret as the outcome of disk galaxies destruction happening at the cluster cores.  The more recent analysis of the morphology-density relation at redshift $0.05<z<0.1$  by \citealt{Goto_et_al_2003} shows a depletion of S0 galaxies within 0.3 virial radii, and  an increase in the early spirals population at the same distances, consistent  with the variation we see in our ZENS  sample. Also in $z\sim0.4$ groups \citealt{Wilman_et_al_2009}  find hints  for an excess of S0 galaxies at $r\gtrsim0.3$ Mpc with respect to the group centers, which instead host a higher fraction of late-type disks. These authors furthermore  find that the fraction of S0 galaxies in groups is comparable to the one in clusters at the same redshift, suggesting that galaxy pre-processing and S0 formation is effective already at these low densities.
The increase of S0 in the outer regions of the groups is  particularly interesting, as it counter to the intuitive idea that gas-rich bulge-dominated spirals may become S0 galaxies as they fall deeper into their group potential wells (see e.g. \citealt{Bekki_et_al_2002} who find  that gas stripping in the group environment is only effective close to the group centers).

Speculating on  the above, a possible scenario is that $(i)$ bulge-dominated galaxies are transformed into  S0s  as soon as disk galaxies enter the group potential, and $(ii)$ further stellar evolution within the groups replenished the dried-out disks of S0 galaxies of fresh gas, establishing/restoring in them a bulge-dominated morphology. Note that an increase in the fraction of dusty star-forming galaxies at  high densities and close to cluster centers has been observed at low redshift, e.g. \citealt{Gallazzi_et_al_2009,Mahajan_Raychaudhury_2009}. Furthermore, the evidence of polar and extended HI disks  around S0s -- e.g. \citealt{vanGorkom_1987,Noordermeer_et_al_2005,Sage_Welch_2006} --  may also support this "disk-regrowth" scenario.  

The second, statistically more significant trend for $M>10^{10} M_\odot$ disk-dominated galaxies   to have a higher  concentration in high-mass ($M>10^{13.5} M_\odot$)   than in lower-mass groups appears to be a genuine  environmental effect which should suffer from no morphological complications.  

\citet{Weinmann_et_al_2009,Guo_et_al_2009}   found evidence in their SDSS sample that, among galaxies with  $C<3$, satellites are more concentrated than centrals with identical stellar mass.
These authors interpret the variation in concentration within the framework of gradual
 stripping and subsequent quenching of satellite galaxies  during infall into the group potential. This is also believed to cause a reddening of the satellite galaxies and a shrinking of their typical sizes.  
While we postpone an analysis on the subject  to a forthcoming paper, here we wish to note that the SDSS studies divide the early- and late-type morphological classes on the basis of a non-corrected concentration criterion 
($C<3$ for late-type and $C>3$ for early-type galaxies). Comparing their Figure 1 with our Figure \ref{fig:C_vs_BT}, we see that $C<3$ in the SDSS system roughly corresponds to a value of $C=3.5-4$ in ZENS, as our concentrations show a small offset due to the corrections we applied. As illustrated in Figure \ref{fig:Cmass},  a cut at constant concentration sub-divides the \textsc{ZENS} galaxy sample into  two broad bins in which, however,  individual Hubble types are mixed together below the chosen $C$ threshold.  

It thus remains an open  question whether
the difference in concentration between centrals and satellites reported by those authors  is the result of 
 a variation in the morphological mix of the central vs.\, satellite populations,  or rather a
change in the structure of central and satellite galaxies at a fixed morphological type. We will address this question in a dedicated  ZENS analysis.

\section*{acknowledgements}
A.C., E.C.  and C.R. acknowledge support from the Swiss National Science Foundation.
This publication makes use of data from ESO Large Program 177.A-0680, and data products from the Two Micron All Sky Survey, 
which is a joint project of the University of Massachusetts and the Infrared Processing 
and Analysis Center/California Institute of Technology, funded by the National Aeronautics and
 Space Administration and the National Science Foundation.
GALEX (\emph{Galaxy Evolution Explorer}) is a NASA Small Explorer, launched in April 2003. 
We gratefully acknowledge NASA's support for construction, operation, and science analysis for the GALEX mission.


\clearpage


\appendix


 \section{Properties, reduction and photometric calibration of the WFI $B$ and $I$ data}  \label{app:DataReduction} 
 
\begin{figure}
\begin{center}
\includegraphics[width=100mm,angle=90]{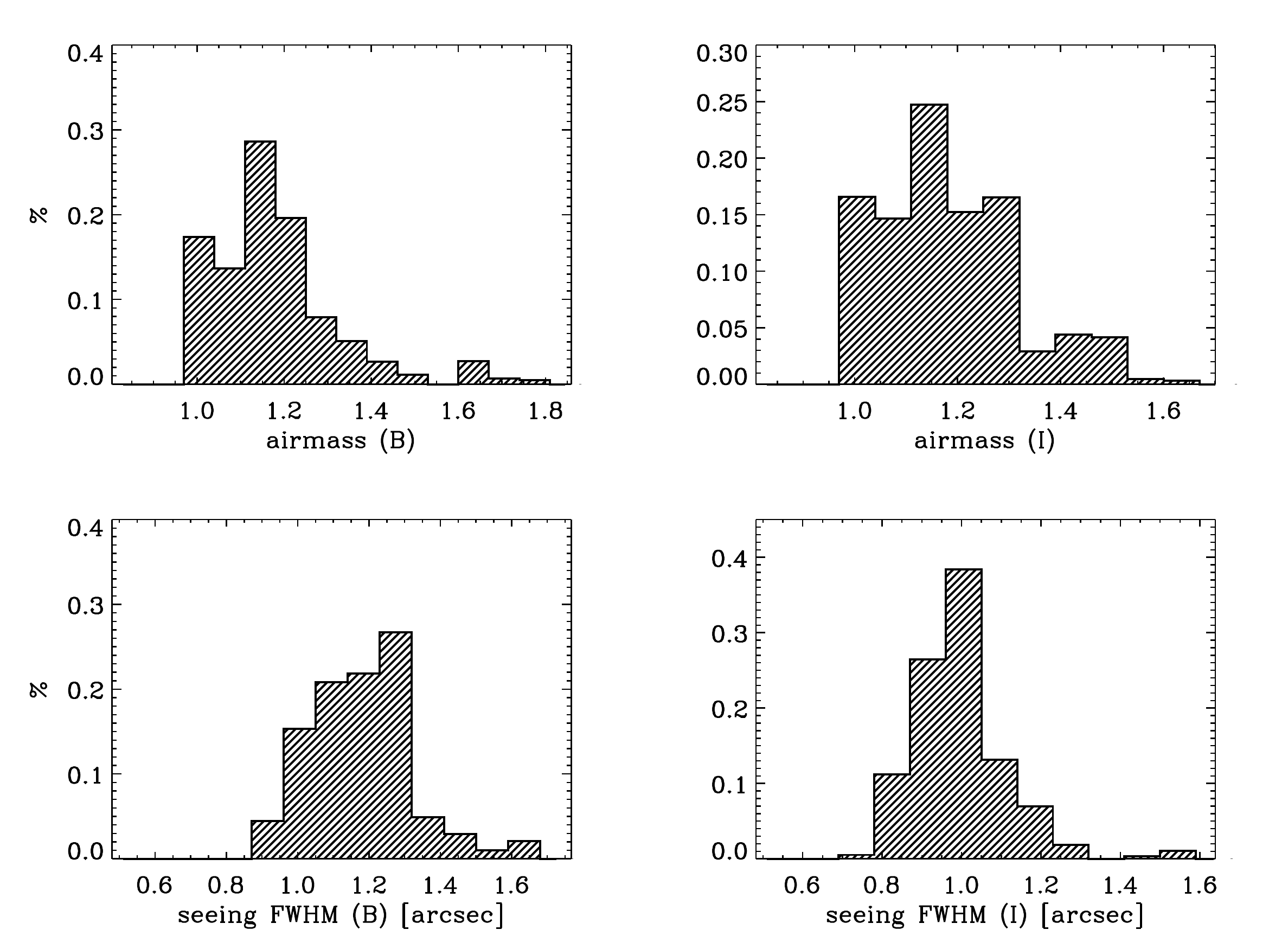}
\end{center}
\caption{\label{fig:skyConditions}Summary of the observing conditions for the \emph{ZENS} WFI data. Shown are the distribution of airmass and (instrumental plus atmospheric, i.e., total PSF) seeing values over the targeted fields, for the $B$- (left panels) and $I$-band (right panels), respectively.}
\end{figure}

Table \ref{tab:obsruns}  gives a  log of the observing runs for the new ESO/2.2m $B$ and $I$ WFI images.
	These  images were taken with the $4\times2$ mosaic of 2k$\times$4k coated CCDs, which provides a scale of 0.238$^{\prime\prime}$/pix and covers a total field of view of 34$^{\prime}\times 33^{\prime}$, well suited to the typical sizes of the \emph{ZENS} groups.  Each group was observed with five dithered 
	exposures to remove inter-chip gap effects and cosmic rays events; the single exposure times were 132$s$ for the 2005 runs and 144$s$ for all other runs. All observations were carried out under clear night conditions. 
The final seeing after image stacking is typically $\sim 1^{\prime\prime}$, reaching down for the best runs to $0.7^{\prime\prime}$ and $0.9^{\prime\prime}$ in the $I$- and $B$-band, respectively, and, in a few worst cases, degrading to $1.5^{\prime\prime}$ and $1.6^{\prime\prime}$. Figure \ref{fig:skyConditions} gives a summary of the airmass and seeing conditions of the WFI observations. 

The \emph{ZENS} detection limits are given by the $1\sigma$ background fluctuations  in a uniform area of 1 arcsec$^2$(18 pixels)  and correspond to  $\mu(B)=27.2$ mag arcsec$^2$ and $\mu(I)=25.5$ mag arcsec$^2$ (AB magnitudes). The depth of the WFI $B$ and $I$ data allow us to robustly quantify the structure of the \emph{ZENS} galaxies out to at least 2 half-light radii, and to detect low surface  brightness tidal tails and merger-induced features. At the given depth limits, and with a total sky coverage of 45 deg$^2$, \emph{ZENS}  is a relatively  deep survey on a  relatively large area, and thus a valuable dataset also for legacy science. 

\begin{deluxetable}{clcccc}
 \tablewidth{0pt}
\tabletypesize{\scriptsize}
\tablewidth{0pt}
\tablecaption{Log of Observing Runs\label{tab:obsruns}}
\tablehead{
\colhead{Run} & \colhead{Date} & \colhead{Exposure Time (sec)} & \colhead{Number of Groups} &\colhead{Program ID} &\colhead{Mode} 
 }
\startdata
1 & Jan. 5 - Feb. 18, 2005                      &   131.917      & 20    &  074.B-0570          & Service\\
2 &  May 1 - Dec. 23, 2006                     &   143.917      & 50    & 177.A-0680(A/B)  & Service\\
3 & Feb. 18 - July 14, 2007                     &   143.917      &  21   & 177.A-0680(A/B)  & Service\\
4 &  Jan. 14 - July 31, 2008                &   143.917      &  12   & 177.A-0680(A/B)  & Service\\
5 & Oct. 30 - Nov. 5, 2008                         &   143.917      & 25    & 177.A-0680(C)     & Visitor \\
6 & Mar. 25 - Mar. 28, 2009                      &   143.917      &  13   & 177.A-0680(D)     &Visitor \\
\hline \\ 
\enddata
 \tablecomments{For each set of observations we list the exposure times, the number of observed groups and the program ID and mode. Sequential numbers in the first columns indicate the run ID.}
\end{deluxetable}

\subsection{Data Reduction}

The WFI $B$ and $I$ images were reduced using standard IRAF
\footnote{IRAF is distributed by the National Optical Astronomy Observatories, which
 are operated by the Association of Universities for Research in Astronomy, Inc., under cooperative agreement with the National Science Foundation.} routines and in particular the NOAO mosaic software $MSCRED$ was used for most of the processing. 

 The bias subtraction involved two steps: each frame was initially subtracted of a bias level using, for each pixels row, the median value of the overscan pixels in the same row. Remaining large scale features in the bias were further eliminated by subtracting a combined bias image, which was obtained separately for each night by stacking together at least 10 (overscan subtracted) bias frames. Dark current was not subtracted since its value is negligible ((0.3-0.6)$\times 10^{-3}$ADU s$^{-1}$ pix$^{-1}$). 
The images were corrected for pixel-to-pixel variation of the quantum efficiency by dividing them 
 with a twilight flat-field, obtained combining several tens of flat field exposures acquired 
 throughout the different nights of each observing run. The presence of residual large-scale illumination inhomogeneities was removed with a super-flatfield obtained from the science exposures themselves, stacked together with a $3\sigma$-clipping algorithm to remove all bright sources.
 
The $I$-band images were affected by fringing and required intermediate processing before deriving the super-flat images. All the bias-corrected and flat-fielded $I$-frames were combined to obtain an image of the fringing pattern. 
 The large-scale features of the fringing image were removed by subtracting, from it, 
 a smoothed version of itself; this produced the final fringe template, containing only the signal arising from the fringing itself. 
 The fringing template was then subtracted from the science exposures with an interactive task, allowing us to verify the quality of the correction. These fringing-removed images were finally used to create the super-flatfield for the red filter.

Further reduction steps were made to correct for detector cosmetic effects, such as bad pixels, hot or dead columns. 
 The five individual reduced images of each group in each filter were stacked together to obtain the final science images that we used in our analysis. From the large group frames we extracted postage stamp images for each of the galaxies,
of size equal to 3 Petrosian radii. Sky subtraction was performed locally on the cleaned stamps,
so as to account for small inhomogeneities in the sky level. 
Cleaning of the postage stamps involved removing most of the light coming 
from other galaxies (or from spurious stellar spikes and other contaminations), 
and substituting to the relevant pixels an iterated value of the average sky level in the stamp.
Companion galaxies were identified using \textsc{SExtractor} as sources above 1.5$\sigma$ 
of the sky level and the pixels belonging to these objects were replaced with random blank sky regions.
All stamps were furthermore visually inspected to verify the quality of the cleaning process.

\subsection{Photometric Calibration}

Photometric calibration was performed using Landolt (\citealt{Landolt_1992}) standard stars. 
Several Landolt fields, with at least 5 standard stars, were observed for each night. 
Aperture magnitudes for these stars were measured with the IRAF task \emph{qphot} and used to derive the photometric zero point (ZP) for the two pass bands.

We used the following calibration equations between the Landolt B and I and the 
WFI instrumental magnitudes \emph{b},
 \emph{i}: $B=ZP_B+b+\alpha_B(B-V) -k_BX$ and  $I=ZP_I+i+\alpha_I(V-I)-k_IX$, respectively.
  In these calibration equations  $X$ is the airmass value at the time of observation and $k$ is the extinction coefficient. Zero points, color terms and extinction coefficients are presented in Table \ref{tab:phot_calib}. Since our standard stars were not observed at sufficiently differing airmasses to allow for a sensible extinction correction, magnitudes were airmass-corrected by using the extinction coefficients given on the WFI website \footnote{http://www.ls.eso.org/lasilla/sciops/2p2/E2p2M/WFI/zeropoints/} as a reference. All magnitudes are calibrated onto an extra-atmospherical airmass of zero. 
  
While there is no clear color dependence of the calibration for the $I$ filter ($\alpha_I=0$),
   it is necessary to correct for a color term ($\alpha_B=0.215$) in the case of the $B$ band. 
    We also found that the ZP points themselves show a shift over the different observing runs, 
    even though the slope of the calibration remains constant (most likely be due to different 
    CCD/ambient conditions during the observations). For this reason, we calibrated each night separately,
     fixing the slope of the color contribution to the well-established common value for all the nights,
  while using the night-by-night determined ZPs for each set of observations.

 \begin{deluxetable}{cccccccc}
\tabletypesize{\small}
\tablewidth{0pt}
\tablecaption{Zero points, color terms and extinction coefficients for the WFI B and I filters. \label{tab:phot_calib}}
\tablehead{
\colhead{Run} & \colhead{Date} & \colhead{$ZP_B (AB)$} & \colhead{$ZP_I (AB)$} & \colhead{$\alpha_B$} & \colhead{$\alpha_I$} & \colhead{$k_B$} & \colhead{$k_I$} }
\startdata
1 &  Jan. 5 - Feb. 18, 2005         & 24.75                 &  23.74              &  0.215  & 0  &  0.22  &   0  \\   
2 &   May 1 - Dec. 23, 2006      &  24.75-25.08      &  23.74-24.0     &  0.215  & 0  &  0.22  &   0  \\    
3 &  Feb. 18 - July 14, 2007      & 24.85-24.96       &  23.81-23.90   &  0.215  & 0  &  0.22  &   0  \\        
4 &  Jan. 14 - July 31, 2008      &  24.53 -24.99     &   23.74-23.92  &  0.215   & 0  &  0.22    &   0  \\        
5 &  Oct. 30 - Nov. 5, 2008         & 24.93-24.99       &  23.89-23.92 &  0.215  & 0  &  0.22  &   0  \\        
6 &  Mar. 25 - Mar 28, 2009     &   25.00-24.94       &   23.88-23.98 &  0.215  & 0  &  0.22  &   0                                                                                                                                  
\enddata
 \tablecomments{The zero points (in the AB system and for magnitudes in ADUs, a term +0.75 must be added to convert to electrons) and other parameters for the photometric calibrations
  of the observing runs of Table \ref{tab:obsruns}. For each run we list the range of measured ZPs. The coefficients $\alpha$
  are the color terms in the calibration equations. The airmass extinction coefficients $K_B$ and $K_I$ could not be
  derived from our own data as the acquired standard stars did not sample a sufficient range of zenith distances.
  For these quantities we used the latest measurements available on the WFI calibration web page.} 
\end{deluxetable}
  
The Landolt system is tied to the standard Johnson-Cousin system which uses Vega for reference; 
  we thus converted our magnitudes to the AB system (\citealt{Oke_1974}) through the following procedure. 
  For the B-band, we found the ZP corresponding to an object with zero color in AB in the $ZP-(B-V)$ relation, and used this value as our final calibration factor for the blue filter. As mentioned before, the color term was negligible for the $I$ band and thus it was not possible to apply such method. In this case the following transformation was applied: $m_{AB}=m+m_{AB}(Vega)$. We computed the conversion factor using the SYNPHOT IRAF package 
  (see also \citealt{Fukugita_et_al_1995}) and found it to be $I_{AB}(Vega)=0.45$.  
  The values for the ZP we find are in good agreement with those reported on the WFI calibration web-page.
  
We assessed the robustness of our  calibration by comparing the magnitudes 
  obtained for (the stars in the \emph{ZENS} fields, as well as for) the \emph{ZENS} galaxies with the others available from the literature. 
  For the \emph{ZENS} galaxies we used the magnitudes obtained with the SExtractor (Source Extractor, \citealt{Bertin_et_Arnouts_1996}) software and determined from the flux curve of growth inside a Kron-like elliptical aperture (MAG$\_$AUTO in the code). 
  The Kron (1980) aperture radius correspond to 2.5 times the radius of the first image moment. 
  All magnitudes are corrected for galactic absorption using the dust maps of \citet{Schlegel_et_al_1998} 
  and the \citet{Cardelli_et_al_1989} dust law. 
  As reference values we consider the 2dFGRS $b_j$, the SuperCosmos Survey  $b_j$ and $r_F$ \citep{Hambly_et_al_2001} 
  and the SDSS $i$ and $b_j$ magnitudes \citep{Abazajian_et_al_2009}.  
  The latter is derived as in \citet{Norberg_et_al_2002} using the relation $b_j=g+0.155+0.152\times(g-r)$.
  
\begin{figure*}
\begin{center}
\includegraphics[width=140mm,height=130mm]{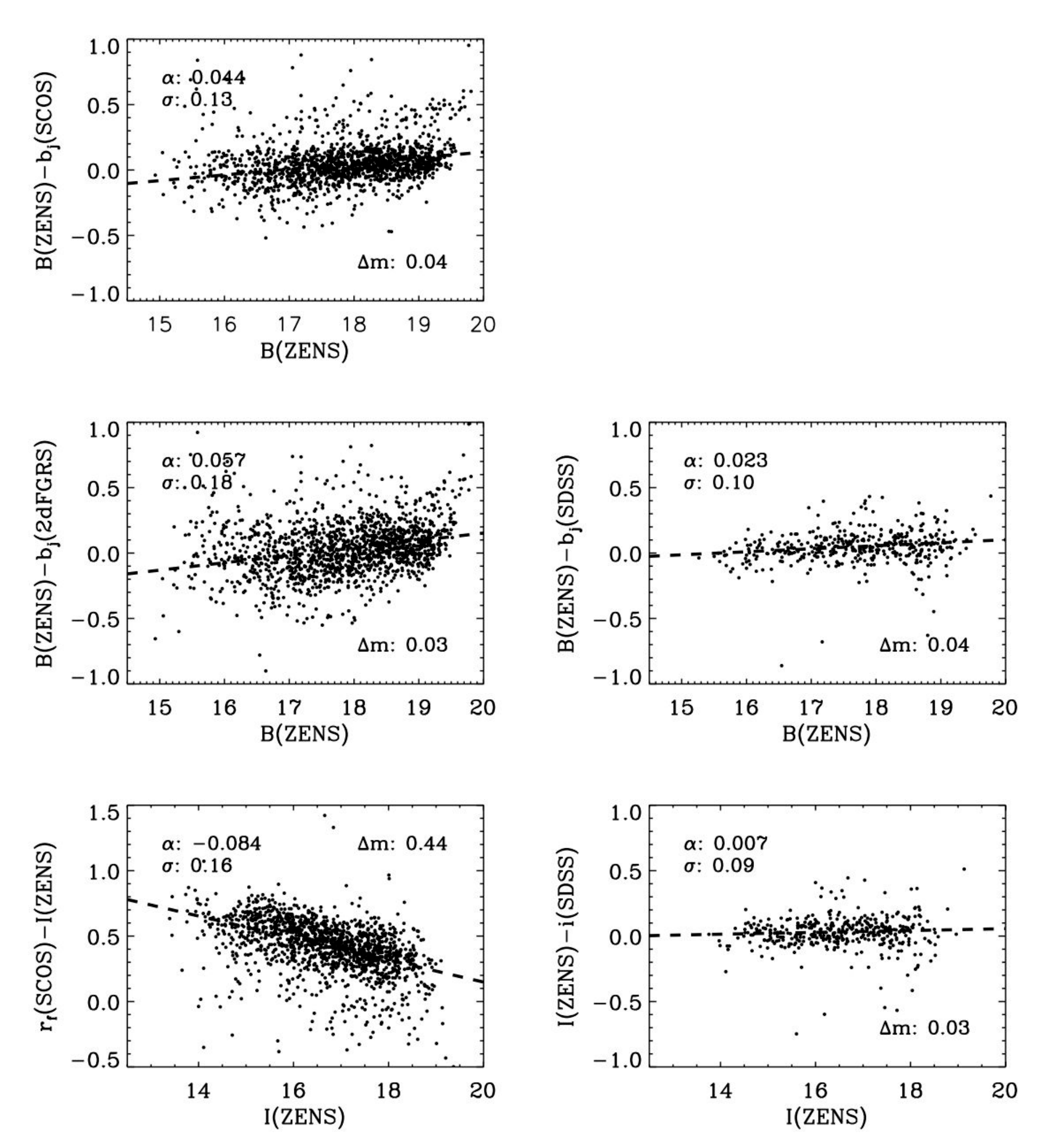}
\end{center}
\caption{\label{fig:photAccuracy} Accuracy of the \emph{ZENS} WFI photometric calibration. Each panel shows the difference between the \emph{ZENS} apparent magnitude in either the $B$- or $I$-band, and relevant magnitudes available from the literature: SuperCosmos Survey $b_j$, 2dFGRS $b_j$, SDSS-based $b_j$, SDSS $i$, SuperCosmos Survey $r_F$. The quantities $\alpha$ and $\sigma$ are the slope of the best linear robust fit to the data and the dispersion around this relation, respectively. $ \Delta m$ is the median difference between each couple of measurements. All magnitudes are corrected for galactic extinction.}
\end{figure*}

We show in Figure \ref{fig:photAccuracy}  the comparison between the derived \emph{ZENS} magnitudes and the published ones. We have a good agreement with the SuperCosmos values (within a standard deviation of $\sigma=0.1$ mag), which is better than with the 2dFGRS one  ($\sigma=0.16$ mag). This is very consistent with the quoted photometric errors of the 2dFGRS ($b_j$ magnitudes, $0.15$ mag) and SuperCosmos magnitudes ($0.1$ mag). We mention that our $B$ measurements show a small non-linearity with respect to both the SuperCosmos data and 2dFGRS $b_j$ magnitudes.  This non-linearity is however absent when we compare with SDSS photometry, which is  available  for a subset of our galaxies. With the exception of the expected shift between the SuperCosmos $r$ passband ($r_F(SCOS)$) and the \emph{ZENS} $I$ passband, the median magnitude offsets in the relevant comparisons are  small  at the level of  $\sim 2\%$. 

We remark that in deriving galaxy stellar masses and other photometry-inferred parameters in Paper III, 
the  ZEBRA+ software \citep{Feldmann_et_al_2006,Oesch_et_al_2010} was first run in ``photometry-check mode", to detect residual offsets from the individual passbands; 
no offsets were found for the WFI $B$ and $I$ passbands.

\section{Quality-Control checks for the GIM2D  model fits} \label{sec:testGIM2D}

\subsection{Comparison with previous works}\label{sec:AppNYVAGC}

\begin{figure*}[htpb]
\begin{center}
\includegraphics[width=60mm,angle=90]{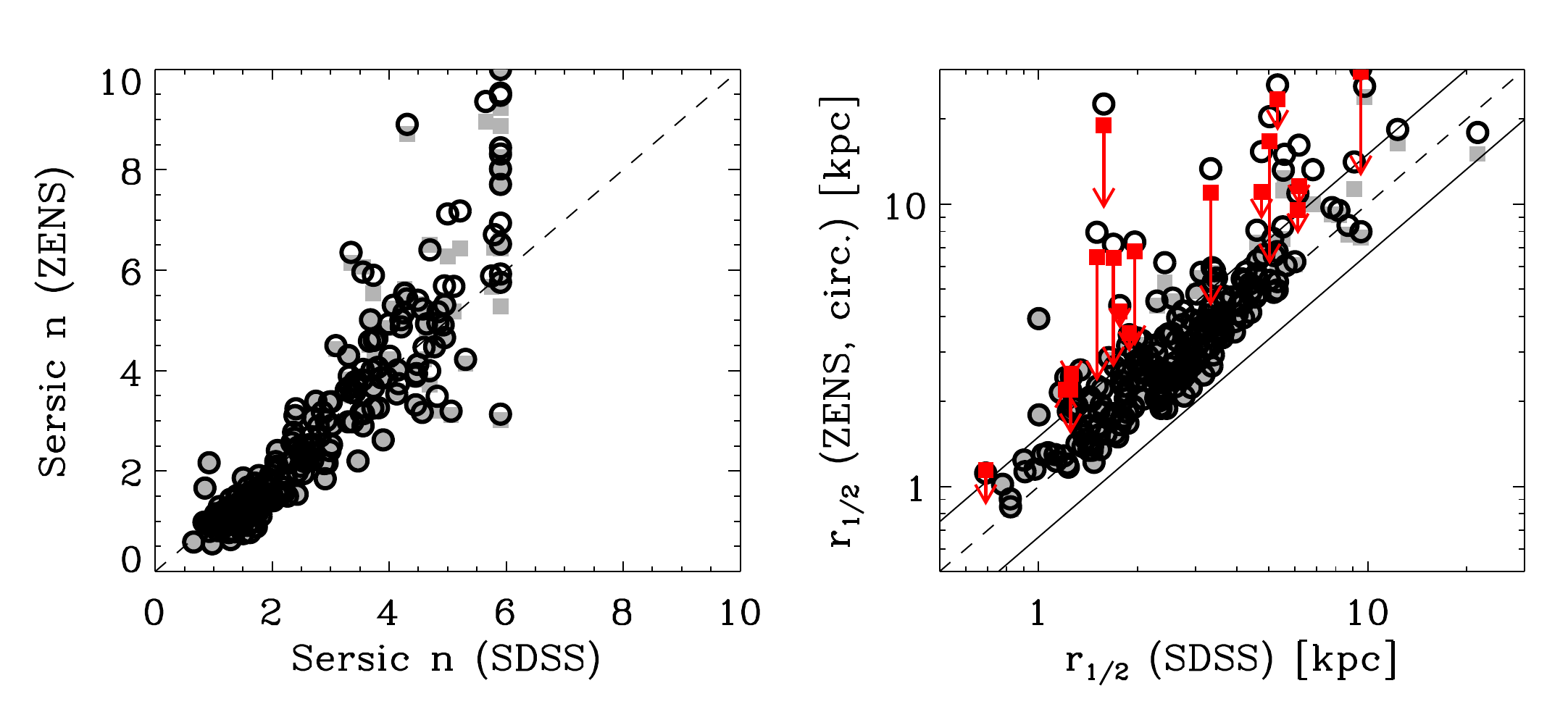}
\end{center}
\caption{\label{fig:ZENS_SDSS_sersic} Comparison of  S\'ersic indices  (left) and half-light radii (right)  between our single-component  $I$-band S\'ersic fits and corresponding  $i$-band single-component S\'ersic parameters  from the  SDSS NYU-VAGC. To match these previous  measurements, which use circular aperture, we plot here post-processing circularized (i.e., $\sqrt{ab}$) half-light radii also for  \textsc{ZENS}.
Empty circles present the comparison using the corrected ZENS structural parameters (see Section \ref{sec:correctionMaps}) while the light gray or dark grey (red in the online version) squares are the original GIM2D measurements prior to such corrections. 
The dashed line in the right plot is the identity relation; the solid lines indicate a variation of a factor of 1.5.
Dark gray squares (red in the online version) highlight  galaxies  with ZENS S\'ersic indices $n>6$, i.e.,  the maximum allowed values in the Blanton et al.\ SDSS fits. The arrows connect the original (uncorrected) ZENS measurement, performed allowing $n$ to vary  freely up to a value of 10, and  the half-light radius that is measured imposing $n<6$, as done in the SDSS fits.
(A color version of this figure is available in the online journal.)}
\end{figure*}

For a number of  galaxies in our sample, the SDSS New York University Value-Added Galaxy 
Catalog (NYU-VAGC) value added catalogue \citep{Blanton_et_al_2005} provides  half-light radii and
 indices $n$ from single-component S\'ersic fits to the azimuthally averaged surface brightness profiles.
We thus compare in Figure \ref{fig:ZENS_SDSS_sersic} our own best fit parameters for the single component fits, both {\it prior to and after }  the application of our corrections described in Section \ref{sec:correctionMaps},   with these previously published data. To match the  VAGC definition, we \emph{circularize} in post-processing mode, using the galaxies' ellipticities,  our fiducial major-axis size estimates based on elliptical apertures.
As illustrated in the Figure, the agreement between the two sets of measurement is generally good.
Only about 5\% of the galaxies show in ZENS larger half-light radii and S\'ersic indices than those measured in SDSS. Regarding the differences in S\'ersic index, we note that the SDSS fits have a maximum $n$=6 whereas for the ZENS sample we allowed $n$ to vary up to $n$=10. Indeed, the most discrepant galaxies in the $n$-comparison plot are those which have reached the saturation value of $n\sim6$ in the SDSS analysis (left panel in the figure). 
These same galaxies are actually also 
many of those which show a large difference in the half-light radius comparison plot (red filled squares in the right panel of Figure \ref{fig:ZENS_SDSS_sersic}). Part of the observed discrepancy in the radii is thus likely caused by the different $n$ values in the ZENS and SDSS fits. 
To test this directly, we refitted these galaxies (including other galaxies which have  $n>6$ in ZENS but  $n<6$ in the SDSS fits) imposing a  maximum $n=6$ value for the S\'ersic index in their  single-component surface brightness fits, as done in the SDSS analysis. The red arrows in the right panel of the figure show the differences in $r_{1/2}$ that result from constraining the fits to a maximum $n=6$ value. These differences highlight the interdependence of the derived parameters in such kind of fits, and the systematic errors that such interdependence might introduce.
Finally we note that many of the ZENS galaxies whose nominal single-component size estimates are substantially larger than the SDSS size estimates are those galaxies which were already flagged in our ZENS catalog as having a problematic single-component size estimate in the analysis of Section \ref{sec:LargeRadiiDiff} (filled red squares in the right panel of Figure \ref{fig:ZENS_SDSS_sersic}).  These galaxies were found to have  substantially larger half-light radii from their
single-component fits    than from their double-component fits. This evidence, and a visual inspection of  the data and models, motivated our choice to adopt the double-component half-light radii  as our fiducial size estimates in ZENS (Section \ref{sec:LargeRadiiDiff}).

\subsection{Validation of the \textsc{GIM2D} bulge+disk models} \label{app:FilterBD}

Caution should be exercised in using blindly  the  {\ttfamily GIM2D} double component results, 
as a number of factors can produce profiles which do not correspond to a meaningful bulge+disk 
decomposition (see also \citealt{Allen_et_al_2006} for an extensive discussion on this issue).
For this reason, all residuals images, models and profiles obtained from {\ttfamily GIM2D} were 
visually inspected to look for possible failure of the fitting algorithm and to identify physically-reasonable double-component models.  
As in \cite{Allen_et_al_2006}, we   analyze the radial surface brightness profiles of bulges and disks   to identify problematic models which would not be recognized as unphysical by simply looking at the fit $\chi^2$ or at residual images.
To this purpose we analyzed the surface-brightness profiles of the bulge and disk along both the semi-major 
and semi-minor axis; given that the two components can be twisted by several degrees, this allow us to make a consistent comparison.

For all galaxies for which we attempted a (S\'ersic-profile  bulge plus exponential-profile disk) decomposition, i.e., all galaxies which are not classified as ellipticals or irregulars, we classify  the double-component model fits into the following categories:
 
\begin{deluxetable}{lcc}
\tabletypesize{\small}
\tablewidth{0pt}
\tablecaption{Fraction of reliable single S\'ersic fits and  bulge+disk decompositions for galaxies of different morphological types}
\tablehead{
\colhead{Type}   & \colhead{Reliable Single S\'ersic Fits}  & \colhead{Reliable Bulge+Disk Fits}    }
\startdata
  \cline{1-3}\\    
 $B$-Band  & Total (96$\%$) & Total (81$\%$) \\
 \cline{1-3}\\         
 Elliptical                               &  98$\%$  &         -     \\
  S0                                        &  95$\%$  &       $81\%$       \\                     
Bulge-dom. Spiral               &  97$\%$  &       $76\%$\tablenotemark{a}        \\                                                                              
Intermediate Spiral     	 & 94$\%$  &       $92\%$      \\    
Late Spiral 	                   & 96$\%$   &       $74\%$\tablenotemark{b}     \\    
  \cline{1-3}\\    
  $I$-Band  & Total (96$\%$) & Total (77$\%$) \\
  \cline{1-3}\\    
Elliptical                                &  100$\%$  &           -    \\
S0                                          &  95$\%$   &      $74\%$      \\        
Bulge-dom. Spiral                & 97$\%$ &        $69\%$\tablenotemark{a}     \\                                                                    
Intermediate Spiral              &  94$\%$ &        $90\%$      \\   
Late Spiral  	                  &   97$\%$ &         $70\%$\tablenotemark{b}         \\    
\enddata
\tablecomments{The fraction 
of reliable single-S\'ersic and bulge+disk decompositions (the latter obtained either with \textsc{GIM2D} or with \textsc{GALFIT}) for each  morphological type, separately 
 for the $B$- and $I$-bands. Elliptical  are by definition galaxies very well fit by  a single-Sersic component only, whose $n$ index is $>3$.
\label{tab:GIM2DRejected}}
\tablenotetext{a}{If excluding the merging galaxies, for which the decomposition is made difficult by the strong contamination from the companion and the generally disturbed morphology, the fractions increase to 86\% and 81\% for the $B-$ and $I-$band respectively.}
\tablenotetext{b}{In the fraction of failed bulge+disk decompositions we include  those galaxies in which the formal S\'ersic-bulge profiles dominates at all  radii  with an  $n<1.5$. These failed disk+bulge decompositions identify however good disk-dominated cases;  these are 6$\%$ and $10\%$ of the late-type disks in the $B$- and $I$-band, respectively.}
\end{deluxetable}

\begin{enumerate}

\item bona-fide bulge+disk decompositions. 
These models have to satisfy the following criteria: the bulge dominates the light in the inner regions, and the disk component  dominates at large radii; furthermore,  the bulge half-light radius has to be smaller than the disk  half-light radius. 
 In the $B$- and $I$-bands, precisely 51\%  and 61\%  of the galaxies in our sample fall into this class.

\item Another $20\%$ and $10\%$  $B$ and $I$ models are classified as ``pure" (bulge-less)  disks ($B/T=0$). 

\item For the remaining $\sim 30\%$  of disk galaxies,  the bulge+disk decompositions failed to match either of the above categories. We separate a few cases within this class.  
In about  $7\%$ of such galaxies, the bulge component always dominates the surface brightness profile. 
These models are not  necessarily unphysical decompositions: some galaxies which fall in this category have a S\'ersic indices $n<1.5$ and are simply disk-dominated galaxies. GIM2D  hence correctly recovers the information that the 
structure is dominated by the disk component. Nonetheless the $B/T$ ratio are not considered as valid.

We consider instead as truly unphysical those fits ($\sim23\%$) which result in a disk half-light radius that is smaller than the bulge's half-light radius and/or in which the disk and bulge profiles are ``inverted",  i.e., the disk dominates in the inner regions and the bulge dominates in the outer regions. Also unphysical were judged those fits in which the bulge and disk surface brightness radial profiles cross twice (again, see also \citealt{Allen_et_al_2006}).

We re-fitted all these problematic cases with the software GALFIT \citep{Peng_et_al_2002}, which allows for more stringent constraints on the input parameter. As initial guesses for the galaxy sizes, position angles, ellipticity and magnitudes we used the values obtained from SExtractor.

 We were able to recover 30\% of the previously-classified 'unphysical' fits. For a total of  $19\%$ ($B-$band) and $23\%$ ($I-$band) of  \textsc{ZENS}  disk galaxies, however,  also GALFIT did not converge  to a  physical  bulge+disk decomposition. The fraction of successful bulge+disk decompositions for each morphological class are summarized in Table \ref{tab:GIM2DRejected}.
The largest formal failure rate is observed for the late-type disks, as in this case many fits results in a dominating S\'ersic-bulge profile with $n<1.5$, as mentioned before. These are meaningful fits, but not meaningful bulge+disk decompositions; these systems are considered to be single-component, disk-dominated galaxies. 
 Excluding these late-type disks, galaxies which do not have bulge+disk decompositions in either one or both filters sum up to $\sim$20$\%$ of the S0- to intermediate-type disk sample.  

\end{enumerate}

 \begin{figure*}[htpb]
\begin{center}
\includegraphics[width=130mm,angle=90]{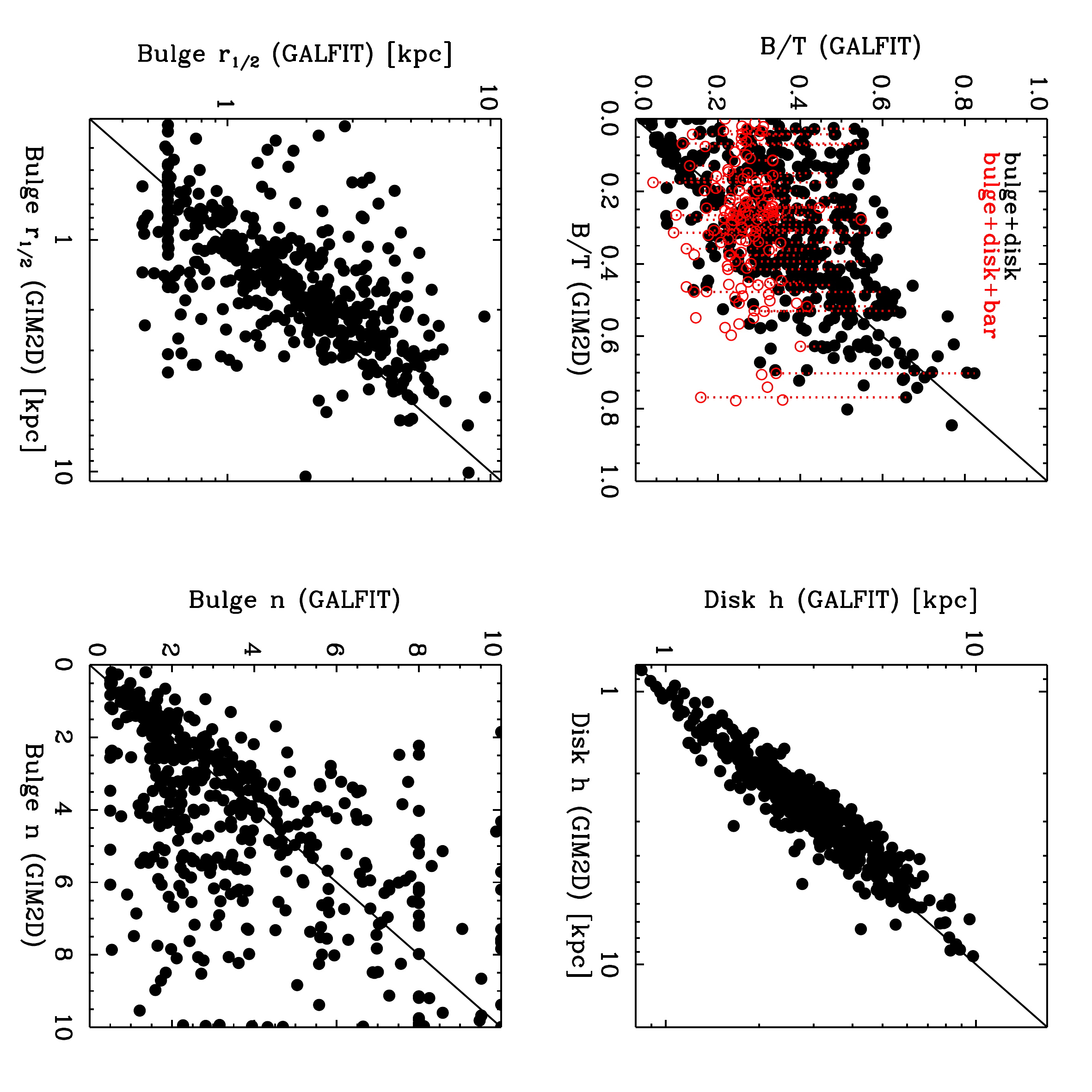}
\end{center}
\caption{\label{fig:GALFIT_GIM2D} Comparison between the bulge and disk structural parameters obtained with GALFIT and GIM2D.
From top to bottom and from left to right: bulge-to-total ratio, disk scale-length, bulge effective radius and bulge S\'ersic index.
Shown are only galaxies for which both the GIM2D and GALFIT models were judged to give  reliable fits according to the criteria that we describe in  Appendix \ref{app:FilterBD}. The empty gray symbols (red in the online version) in the top
left panel show the comparison between the B/T obtained with the simple GIM2D bulge+disk decomposition (x-axis) and the GALFIT bulge+disk+bar fits (y-axis). The vertical gray (red in the online version) lines connect the B/T obtained by GALFIT with the bulge+disk+bar fits or the bulge+disk only fits.
(A color version of this figure is available in the online journal.)}  
\end{figure*}

 In $\sim 17\%$ of galaxies belonging to genuine bulge+disk decompositions, the GIM2D fits result  in rather elongated bulges ($0.6<\epsilon<0.7$). 
 A closer inspection  showed hat a good fraction of these models are associated to galaxies classified as barred disks.  
 In a few other cases,  a resolved bar is not observed and hence these central structures 
 may either be  unresolved bars or truly flattened bulges (although we ourselves wonder whether such distinction is meaningful, not only from a descriptive but also from a physical perspective).
 
 \subsection{The impact on the bulge+disk fits of a bar component: Comparisons with bulge+disk+bar fits}\label{app:BulgeBarDisk}

It has been argued that neglecting the bar component  when modeling the light profile  can  
affect the best fit parameters and cause an over-estimation of the bulge-to-total ratio of a factor of two to four, and 
an artificial increase of bulge effective radii (\citealt{Laurikanen_et_al_2005}, \citealt{Gadotti_2008}, \citealt{Weinzirl_et_al_2009}).

To test the importance of such an effect  in our sample, we 
 carried out on those galaxies classified as barred in Section \ref{sec:bars} a bulge+disk+bar decomposition with the GALFIT software by modeling the bar with a S\'ersic profile having 
 $n<1$. The addition of a third component can introduce further degrees of degeneracy: applying only broad constraints on the bar structural properties resulted in a high fraction of unphysical models. For this reason we  used the results of the  {\ttfamily{ELLIPSE}} isophotal analysis to constrain the bar structural parameters, when available, and to stick to bulge+disk only fits (without a bar component) for galaxies without this a priori information. Given  possible degeneracies  between bulge and bar components, when appropriate we will discuss our results with and without those GIM2D fits which result in very elongated bulge components.  
 
 Specifically, for galaxies with an {\ttfamily{ELLIPSE}} isophotal  fits, we used
 the bar size, ellipticity and position angle measured in Section \ref{sec:bars}  as initial guesses for the fits. Furthermore we imposed  the bar half-light radius to be within $\pm50\%$ of the size measured with the {\ttfamily{ELLIPSE}} method, and at the same time, to be larger than the bulge $r_{1/2}$ and smaller than the disk scale-length. This avoids the inversion of the disk and bar components or  fitting a spurious ``nuclear" bar. Likewise, we limited isophotal  twists between the bulge and the bar to be within $15^{\deg}$ and also imposed a minimum S\'ersic index , $n>1.5$, and a maximum ellipticity, $\epsilon_{bulge}<0.4$, to the bulge. Extensive tests validated the effectiveness of these constraints in leading to reliable fits when the extra degrees of freedom introduced by the bar parameters are allowed in the fits.

The results   are presented in the top  right panels of Figure \ref{fig:GALFIT_GIM2D} (red points). Consistently with previous work, comparing the GIM2D B/T ratios with those obtained from the bulge+disk+bar GALFIT fits, we find that the latter tend to be typically smaller, although exceptions are present. Only in $\sim10\%$ of the cases the difference is of a factor of $\sim2$ or larger. For most of the galaxies the difference in the final parameters are quite modest, with an average difference of $\Delta B/T=0.05$.  In several cases,  fitting also a bar  brings the GALFIT B/T values in better agreement with those derived  with GIM2D, indicating that the algorithms involved in the derivation of the B/T value in  GIM2D are less sensitive to  the presence of a bar component. 
 We thus conclude that $(i)$ adding the bar component may no doubt be relevant in some   analyses of our barred sample, for  which we will utilize the bulge+disk+bar decompositions; $(ii)$  the bulge+disk GIM2D fits return generally robust descriptions of these galactic subcomponents, and we thus assume these fits as our fiducial two-component description of the galaxy light distributions.
 
\subsection{A code versus code comparison}

 We furthermore tested the general consistency between the parameters obtained with bulge+disk fits performed respectively with GALFIT and GIM2D. Specifically, we ran   bulge+disk GALFIT decompositions   on all disk galaxies with available GIM2D models (independent of the availability of the {\ttfamily{ELLIPSE}} isophotal profiles), and compared with each other the best fit parameters obtained with the two independent codes. This is illustrated in the four panels of Figure \ref{fig:GALFIT_GIM2D} (black points), for galaxies with reliable GALFIT and GIM2D fits (see above). The relations shown in the Figure are consistent with  what we expect from the discussion presented in Section \ref{sec:BulgeDiskBias}: the disk scale length is the most robustly determined parameter, with little scatter among the two measurements, whereas  the bulge structural properties are subject to the largest discrepancies.
The B/T ratios obtained by the two codes agree on average  reasonably well.
The discrepancies observed in B/T in a minority of cases is mostly caused by the  differences in the bulge structural parameters. 
We note   that  we found a general tendency to associate low S\'ersic indices ($n<1.5$) to the bulge component in the GALFIT fits. This is particularly true for disk dominated galaxies (these galaxies have single S\'ersic indices $n<1.5$) for which  GALFIT often converges  to a double component solution  with $B/T>0.4$ but $n<1.5$. Such bulges have furthermore elongations and sizes similar to that of the disk. In these cases, GIM2D produces instead a model which is either a pure disk ($B/T=0.0$) or very close to it ($B/T<0.2$).  Such GALFIT fits were flagged as not reliable, and are not plotted in Figure \ref{fig:GALFIT_GIM2D}.  In our studies we will adopt as our fiducial bulge+disk decompositions those derived with GIM2D; we flag however in our catalog individual galaxies with substantially different (and yet both formally valid) GIM2D and GALFIT fits,  so to be able to test against these uncertainties our results concerning the properties of bulges and disks.

\subsection{Properties of galaxies with no reliable GIM2D fits} 

\begin{figure*}[htpb]
\begin{center}
\includegraphics[width=100mm]{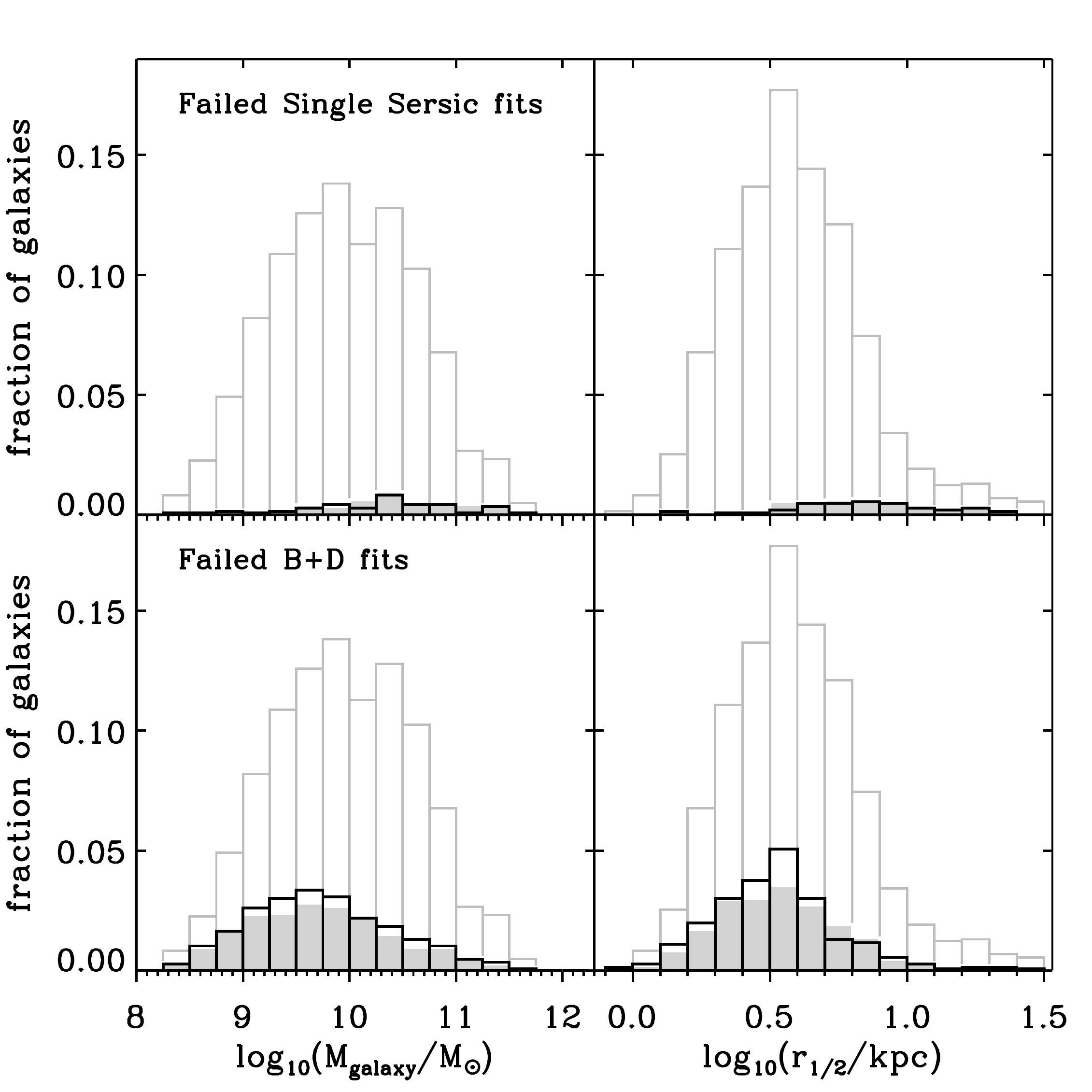} 
\end{center}
\caption{\label{fig:failedGIM2D}Distribution of galaxy masses and $I$-band corrected sizes (see Section \ref{sec:Simulations}) for the total sample of galaxies
 (empty, light gray histograms) and for  
 galaxies with no reliable GIM2D S\'ersic fit (top) or bulge+disk decompositions (bottom). 
 The black histogram is for galaxies with no fit in the 
 $I$-band and the filled dark-gray histogram for failed fits in the $B$-band. 
 All histograms are normalized to the total number of galaxies in our sample.
 For galaxies which do not have half-light radii derived from the single S\'ersic fits we use in this figure the corrected \textsc{ZEST+} radii.}
\end{figure*}

As discussed in Section \ref{sec:GIM2Dmeasurements}, reliable S\'ersic fits and bulge+disk decompositions are not available
for some galaxies in our sample (see also Appendix \ref{app:FilterBD}). 
We inspected whether the galaxies that remained without either or both analytical fits were somehow a biased component of our \textsc{ZENS} sample (apart from having on average, as expected, a more complex light distribution).  Figure \ref{fig:failedGIM2D} shows the distributions of galaxy 
masses and $I-$band half-light radii for such galaxies with no GIM2D models. Galaxies without any (single- and double-component) size measurements, we use the  \textsc{ZEST+} radii, {\it corrected} as described in Section \ref{sec:correctionMaps}.
To establish the presence/absence of possible biases, for comparison we also show the global distributions of masses and sizes of galaxies  with  reliable analytical fits.
A summary of the fraction of failed models for each morphological type is given in Table \ref{tab:GIM2DRejected}.

Galaxies with no single S\'ersic fits tend to populate the high mass/radius tail of the distributions and their incidence 
is highest for intermediate-type disks.
 For these galaxies we will use in our \textsc{ZENS}  analyses the corrected \textsc{ZEST+} sizes, which, on the remaining galaxies, we have shown to be in excellent agreement with the (corrected) sizes derived from the analytical fits; only 4\% of galaxies are in this category. We hence reckon that the impact on the final results  will be in any case negligible.
Although the overall success of the bulge+disk decompositions is lower than the success rate of the single-component fits  (see Table \ref{tab:GIM2DRejected}),
galaxies with no bulge and disk parameters  are more uniformly distributed in mass and size relative  to the
global \textsc{ZENS} population. It is thus reasonable to assume that   the  exclusion of these galaxies from analysis based on bulge/disk parameters will lower the statistical significance of the results, but should not introduce strong biases.

\section{Correction maps for sizes and magnitudes: Variations with PSF size and robustness toward contaminants}\label{app:corrections_PSF}

\subsection{Correction maps obtained for the best and worst PSF}

Correction maps for sizes and magnitudes, similar to those shown in the main text in Section \ref{sec:Simulations} for the median PSF-FWHM of the \textsc{ZENS} WFI $I$  images, are shown here for the  two ``bracketing" -- i.e., the best and the worst -- PSF FWHM values our dataset.   A comparison of the corresponding correction maps derived for the three different PSF sizes clearly shows the major impact of PSF-blurring in the (uncorrected) structural measurements. The implementation of our correction maps rends  all measurements comparable onto a consistently calibrated grid.

\subsection{The effect of contaminants on the derivation of the corrections}

As commented in Section \ref{sec:Simulations}  for the ZEST+ measurements, there are regions of the observed $C$-$\epsilon$-$mag$-$r_{1/2}$ hyper-plane which are highly degenerate: they can be populated by models with intrinsic parameters which originate  in that same region of parameter space, as well as by models with very different  ``intrinsic" ellipticity or concentration, which are scattered into that given bin by observational errors.
It is thus important to verify that the  size and magnitude corrections that we infer  in these regions of parameter space are not strongly dependent on the precise way in which we populate the input grid of  simulated models.
 If this were the case and without an a priori knowledge of the relative fraction of the two populations in the real Universe, the derivation of the corrections in the parameters space would be affected by biases introduced by the precise choice of the simulation grid. 
To assess whether our correction suffer from this problem, we calculated the corrections in these ``troublesome" grid points by using only those models   with original, intrinsic  $\epsilon-C$ values in that same bin, and, for comparison, 
using only models which were originated in a different $\epsilon-C$ bin and were scattered into that bin  by   measurement errors. In many cases we found that both model samples gave consistent results for the corrections to the magnitude and sizes, which we could then use for our real galaxies.  This is illustrated in Figure \ref{fig:ZESTCorrections_withnoContaminers} for the lowest ellipticity bin, which is also the one with the highest contamination, and for models convolved with the typical $I$-band PSF. The corrections derived with the two sub-samples are reassuringly very similar over a wide are of size-magnitude space (and thus very similar  to the global corrections  presented in Figure \ref{fig:ZESTCorrections}). We therefore argue that our implementation of the correction scheme of Section \ref{sec:correctionMaps} is free from biases introduced by the choice of the simulation grid.

\begin{figure*}[htbp]
\begin{center}
\includegraphics[width=120mm]{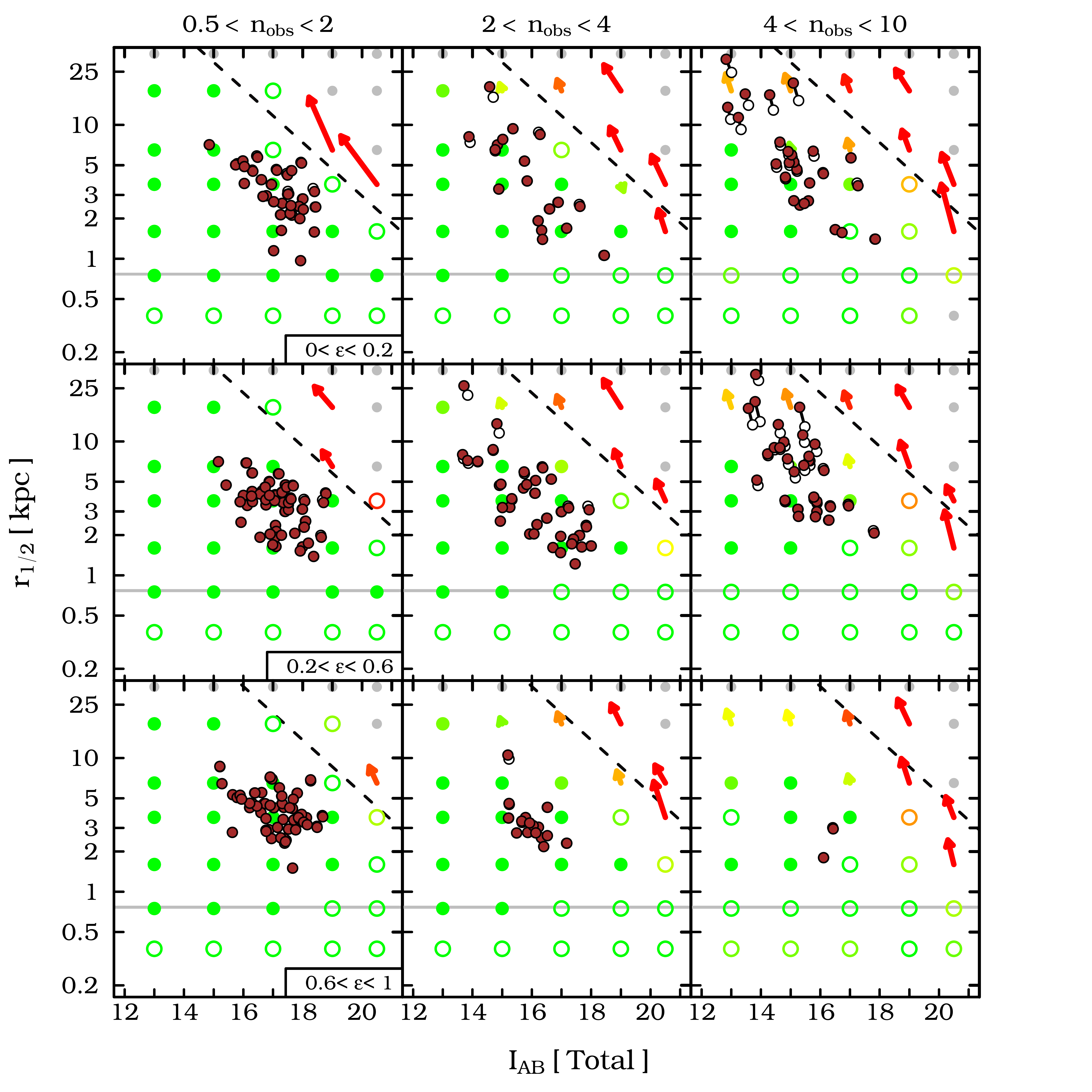}
\end{center}
\caption{\label{fig:GIM2DCorrections_best}As in Figure \ref{fig:GIM2DCorrections} but for galaxy models convolved with the best PSF ($0.7^{\prime \prime}$) measured in the $I$-band WFI ZENS images}
\end{figure*}

\begin{figure*}[htbp]
\begin{center}
\includegraphics[width=120mm]{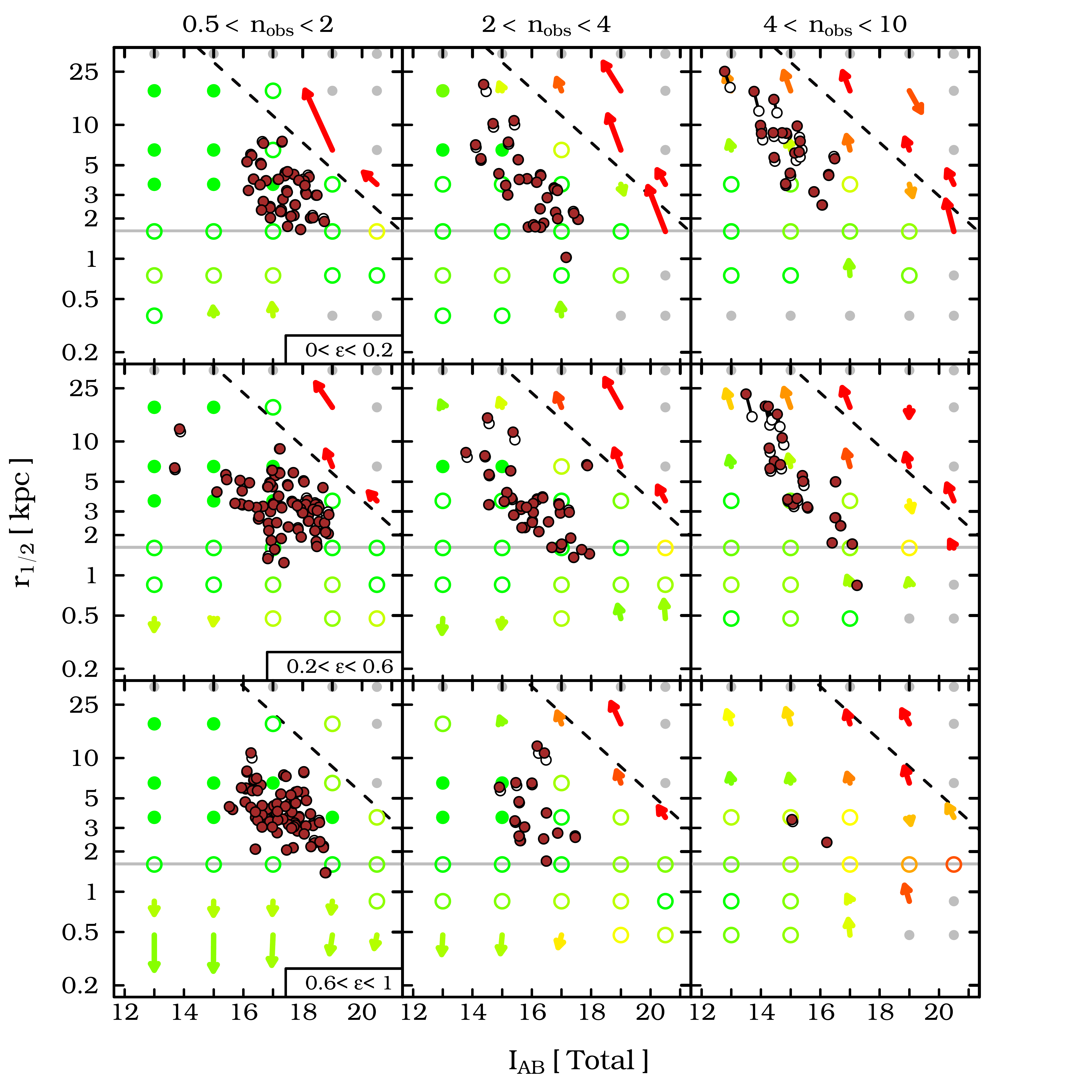}
\end{center}
\caption{\label{fig:GIM2DCorrections_worst}As in Figure \ref{fig:GIM2DCorrections} but for galaxy models convolved with the worst PSF ($1.5^{\prime \prime}$) measured in the $I$-band WFI  ZENS images}
\end{figure*}

\begin{figure*}[htbp]
\begin{center}
\includegraphics[width=120mm]{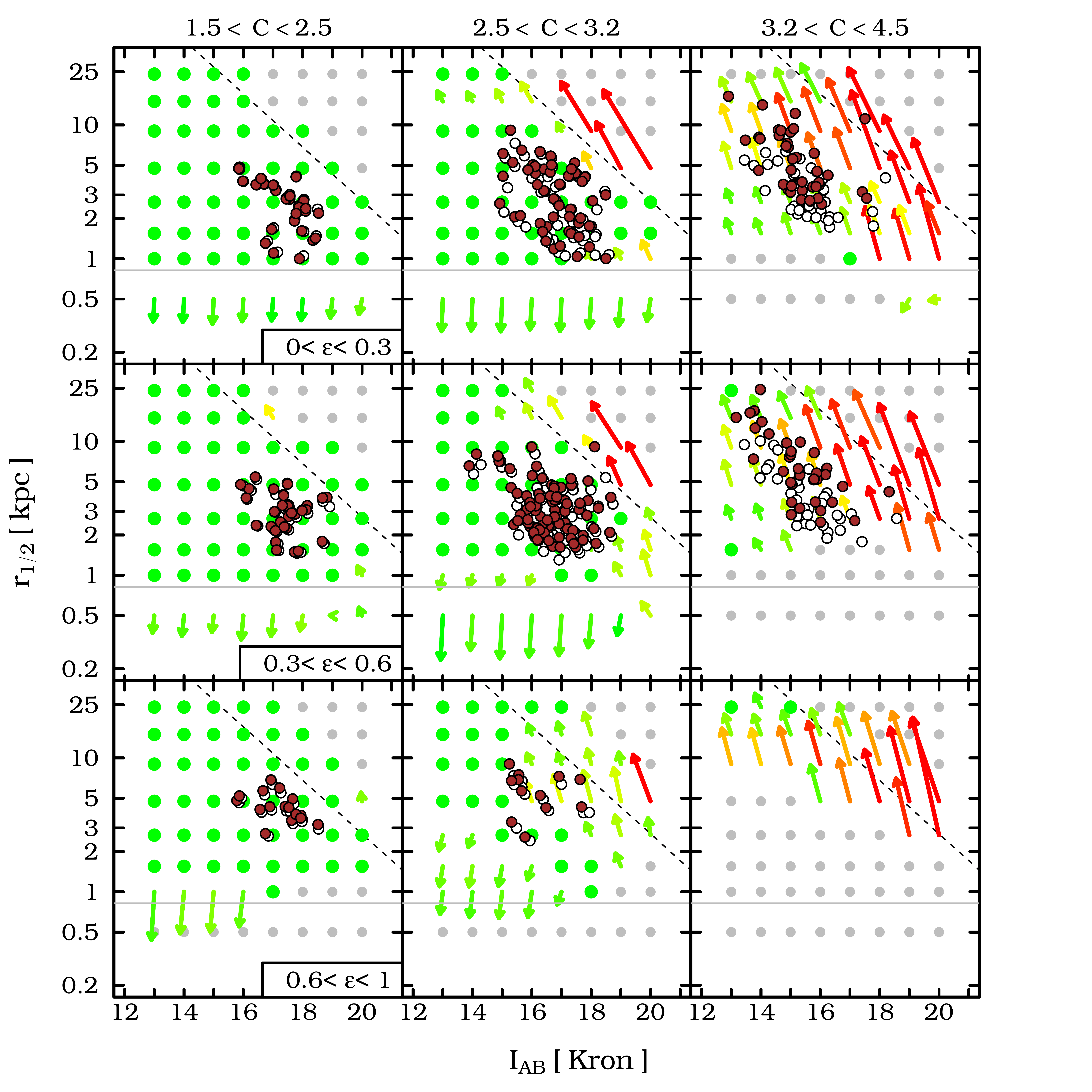}
\end{center}
\caption{\label{fig:ZESTCorrections_best}As in Figure \ref{fig:ZESTCorrections} but for galaxy models convolved with the best PSF ($0.7^{\prime \prime}$) measured in the $I$-band WFI ZENS images}
\end{figure*}

\begin{figure*}[htbp]
\begin{center}
\includegraphics[width=120mm]{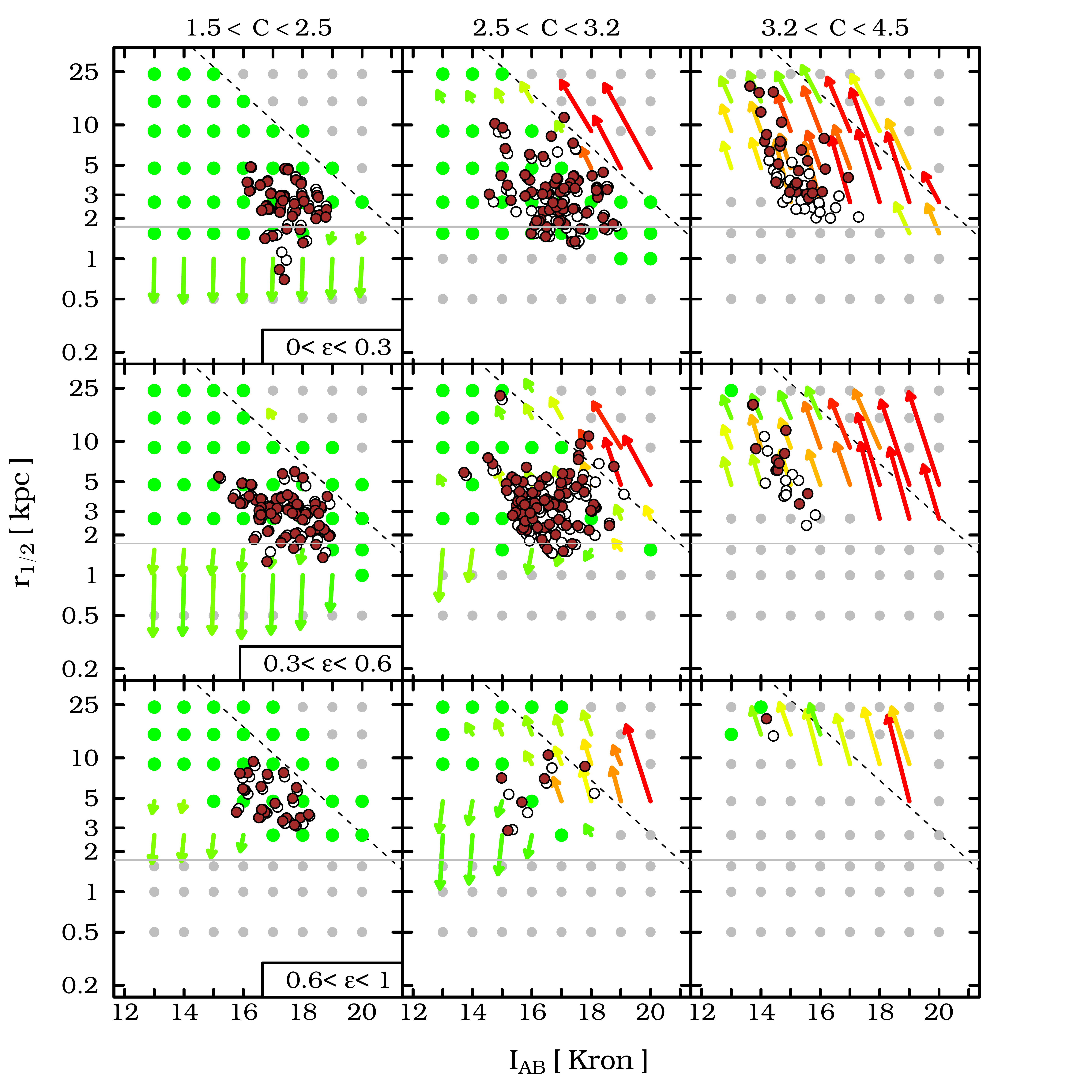}
\end{center}
\caption{\label{fig:ZESTCorrections_worst}As in Figure \ref{fig:ZESTCorrections} but for galaxy models convolved with the worst PSF ($1.5^{\prime \prime}$) measured in the $I$-band WFI ZENS images}
\end{figure*}

\begin{figure*}[htbp]
\begin{center}
\begin{tabular}{c}
\includegraphics[width=120mm]{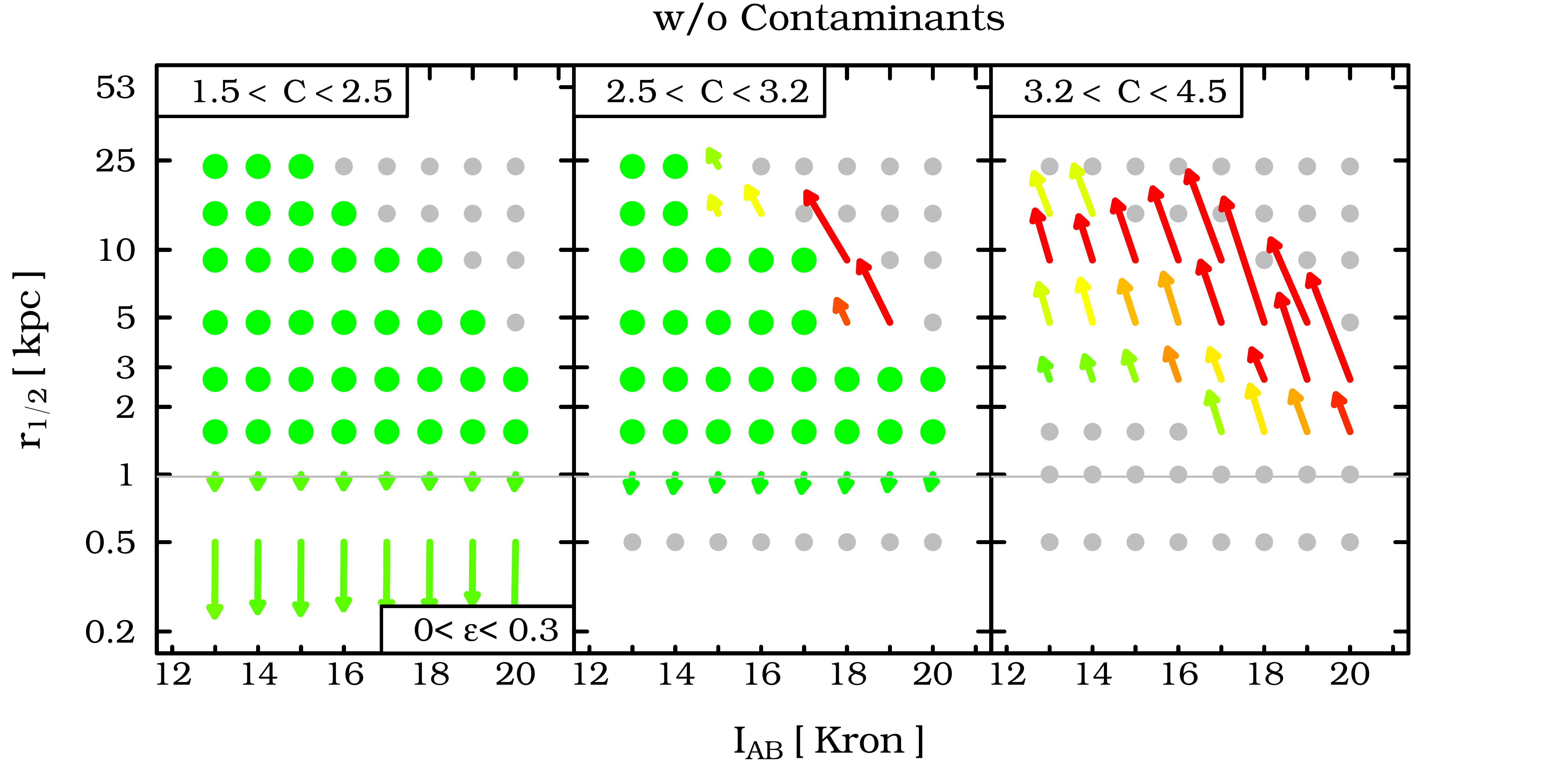}  \\
\includegraphics[width=120mm]{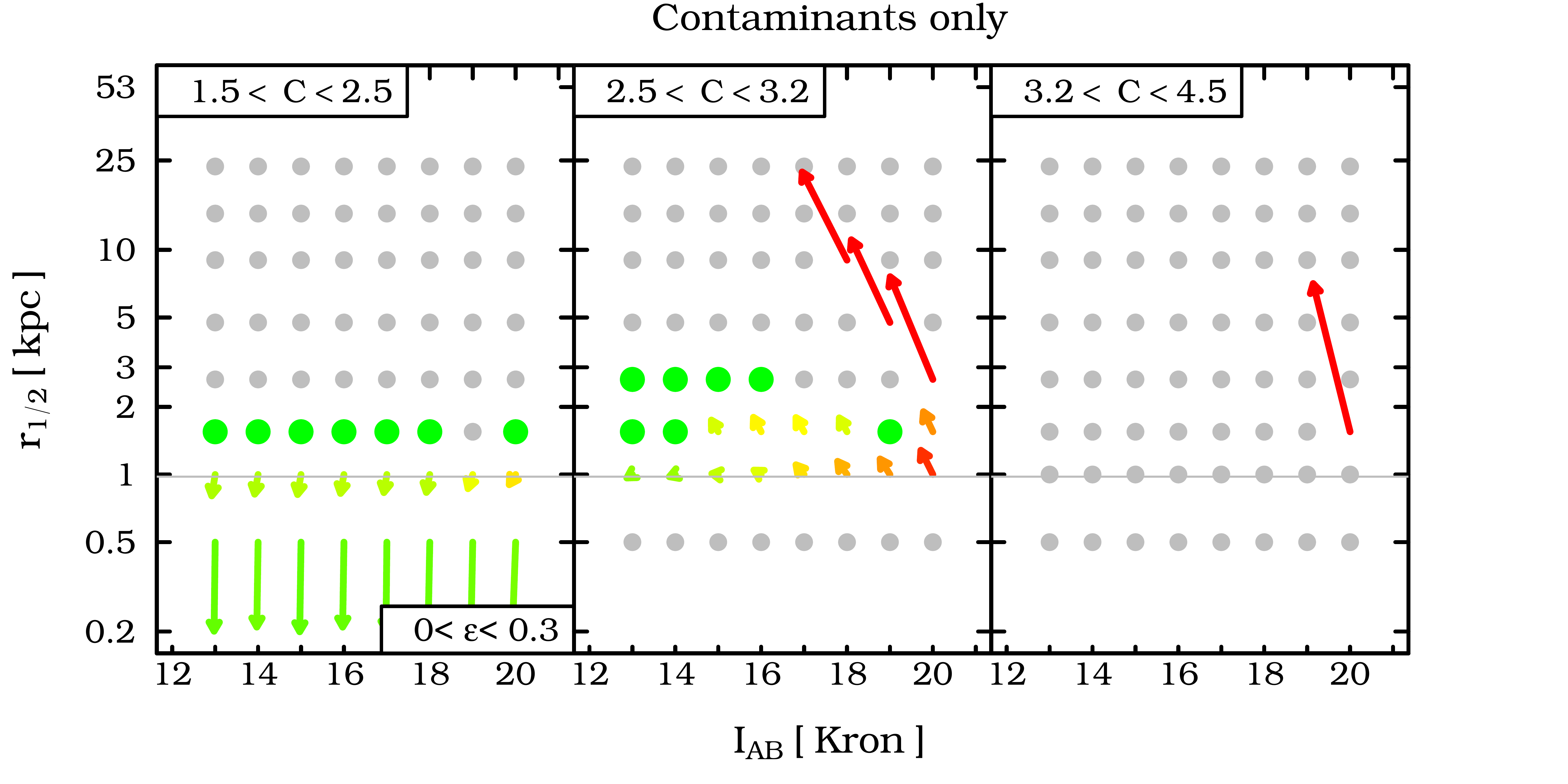}  
\end{tabular}
\end{center}
\caption{\label{fig:ZESTCorrections_withnoContaminers} Comparison between the size-magnitude correction grid obtained from galaxies whose intrinsic parameters lie within the same $\epsilon$-$C$ bin (upper panels, ``without contaminants"), and from galaxies with $\epsilon$  and/or $C$ values outside of the measured bin, which are scattered into the latter by measurements errors (lower panels; ``contaminants only"). 
The comparison is shown for  the lowest ellipticity bin, which suffers from the highest contamination of scattered galaxies into low concentration bins, and for the median PSF FWHM in the WFI \textsc{ZENS} data; similar results hold for all PSF sizes.  
Corrections for sizes and magnitudes  obtained from ``indigenous" and ``scattered" models are well in agreement with each other, indicating that our correction maps do not depend significantly on the precise way we populate our simulation grid. }
\end{figure*}

\clearpage


\section{\textsc{ZENS} galaxies split in their morphological classes} \label{app:ClassDist}
Finally we show the postage-stamp images of the \textsc{ZENS} galaxies, split according to their morphological type. 
The relevant structural parameters characterizing each type are indicated on top of each stamp image. The $n>3$ single-component profiles that {\it define} our elliptical galaxy sample are also shown.
 
 \begin{figure*}[htbp]
\begin{center}
\includegraphics[height=150mm]{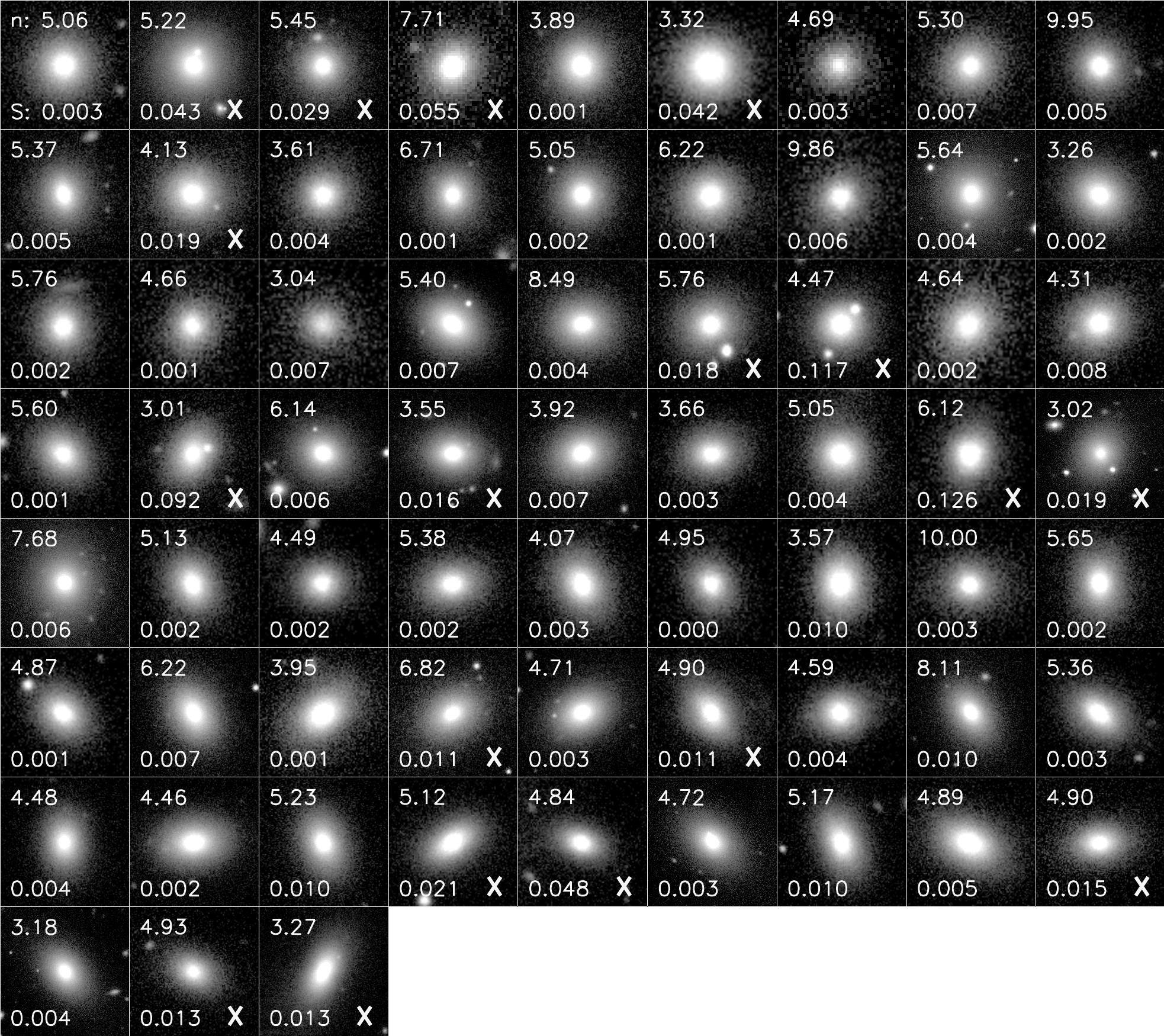}
\end{center}
\caption{\label{fig:stampE}$I$-band stamp images of \textsc{ZENS} galaxies classified as ellipticals. Galaxies are ordered from top-left to bottom-right by increasing ellipticity. The criteria defining galaxies falling into the elliptical class are given  in Figure \ref{pic:ClassFlowChart}. Specifically, elliptical galaxies have surface brightness profiles perfectly fitted by a single-component S\'ersic profile with $n>3$,  as shown in Figure \ref{fig:EllipticalProfiles}, and $B$-band smoothness $S_{B-band}<0.01$.  
The value of the corrected S\'ersic index $n$ (top left) and $B-$band smoothness (bottom left) is indicated in each stamp.
There are some exceptions, in which we maintain an elliptical classification despite the galaxy fails the $S_{B-band}$ criterion. These galaxies are marked with a white cross on the image; in a large fraction of them, the increased smoothness is associated with a bright star cluster or small galaxy clearly visible in the stamps.  For the other such galaxies,  our visual inspection did not detect  any clear substructure which would suggest a different classification.
The surface brightness profiles obtained from the GIM2D fits of all the elliptical galaxies are shown in Figure \ref{fig:EllipticalProfiles}, and show that these systems are genuine single-S\'ersic ($n>3$) galaxies.
This condition, together with the absence of a faint disk in the residual images, was given priority in the classification relative to the smoothness criterion.}
\end{figure*}

\begin{figure*}
\begin{center}
\includegraphics[height=180mm]{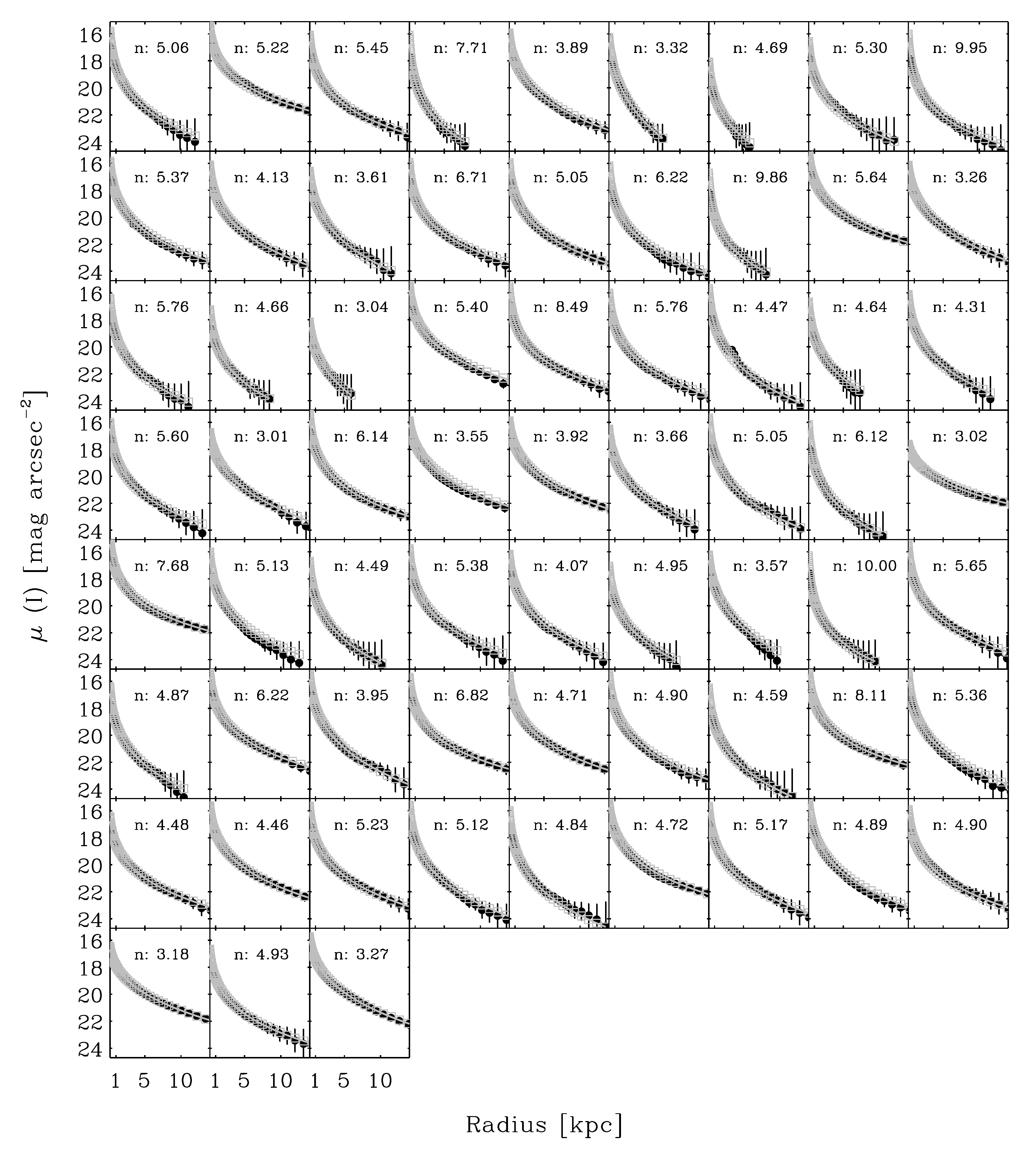}
\end{center}
\caption{\label{fig:EllipticalProfiles}$I$-band  surface brightness profiles and best fit single-S\'ersic profiles for galaxies classified as ellipticals. The filled black points with error-bars are the surface brightness profiles from the {\ttfamily{ELLIPSE}} isophote fits,   the gray empty squares show the single-S\'ersic GIM2D models. Galaxies classified as ellipticals are required to be well fitted by a single S\'ersic component  with $n>3$ from the inner galaxy regions out to the outermost measured points. Each panel shows the value of the Sersic index of these single-component fits.}
\end{figure*}

\begin{figure*}[htbp]
\begin{center}
\includegraphics[height=140mm]{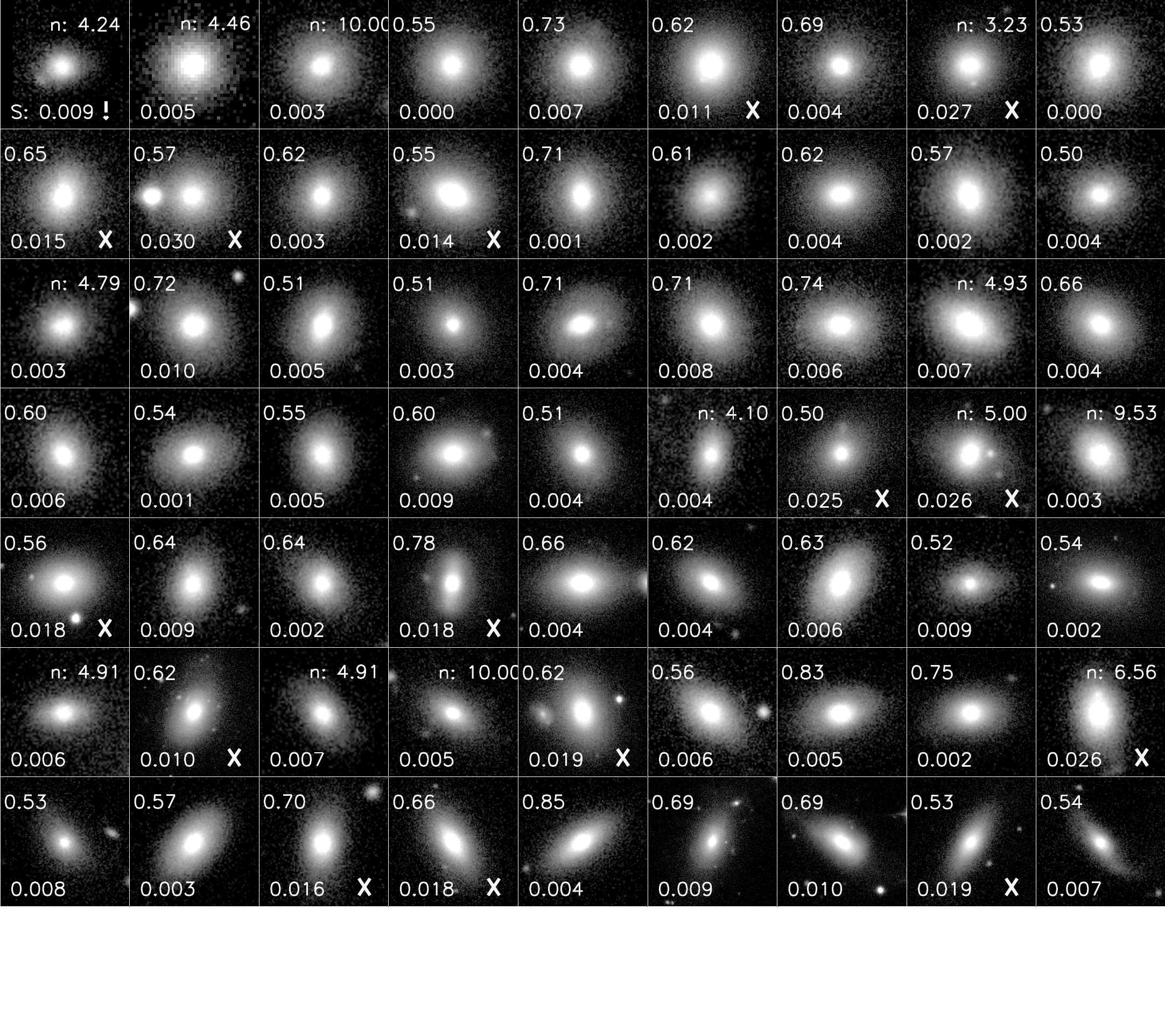}
\end{center}
\caption{\label{fig:stampS0}$I$-band postage-stamp images of \textsc{ZENS} galaxies classified as S0s. Galaxies are ordered from top-left to bottom-right by increasing ellipticity.   These galaxies are defined to have $I-$band $B/T  >0.5$ (or either  $I-$band $n >2.5$ or  $C>3.3$, when no B/T decomposition  is available) and $B-$band smoothness index $S_{B-band}<0.01$ (unless faint spiral arms are visible, in which case the galaxies are classified as a bulge-dominated spiral, or, in contrast, the smoothness is evidently boosted by e.g., bright star-clusters or small companion galaxies etc., in which case the galaxies are kept in the S0 class). The values of $B/T$ (top left) and smoothness (bottom left) are indicated in the stamps. 
For those S0 galaxies with no reliable bulge+disk decomposition we report the value of the corrected S\'ersic index   or   corrected concentration (if also no single-component index $n$ is  available; top left).
Also in this class there are a few exceptions, which are highlighted with   white cross, i.e., galaxies  with  $S(B)>0.01$, as for the ellipticals, in 10 out of 13 of which  a bright star-cluster/companion is again the reason for the increase in $S_{B-band}$;  in one other case, the boosted value of $B$ smoothness is   associated with residuals due to the gaps in the WFI detectors.  Two remaining   galaxies are  moderately inclined systems but do not show  any visible dust lane; for this reason they are included in the S0 sample. One galaxy marked with a `!' lie close to the edges of the WFI camera, thus $B$-band observations are available but not the $I$-band ones. The classification for this galaxy is based on the $B$-band parameters. }
\end{figure*}

\begin{figure*}[htbp]
\begin{center}
\includegraphics[height=175mm]{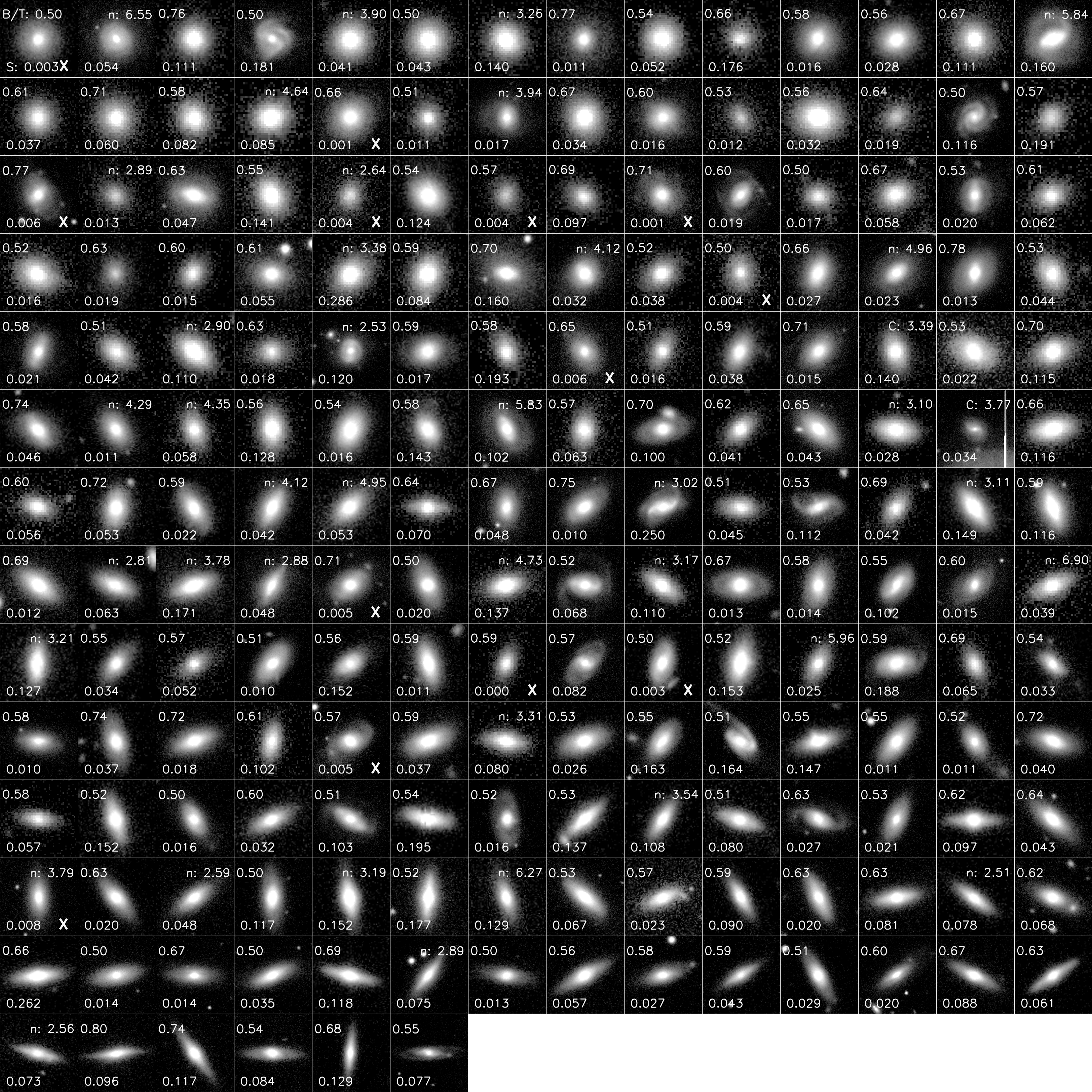}
\end{center}
\caption{\label{fig:stampSa}$I$-band postage-stamp images of \textsc{ZENS} galaxies classified as bulge-dominated spiral galaxies. Galaxies are ordered from top-left to bottom-right by increasing ellipticity.  These galaxies have $I-$band  $B/T>0.5$ (or  $I$-band $n >2.5$, when no B/T decompositions are available,  or $C>3.3$ if also the latter is missing) and either $B-$band smoothness parameter $S_{B-band}>0.01$ or visible  spiral arms.
Numbers on the stamp are as in Figure \ref{fig:stampS0}. The crosses highlight those galaxies which have $S_{B-band}<0.01$ but visual inspection reveals faint spiral arms.}
\end{figure*}

\begin{figure*}[htbp]
\begin{center}
\includegraphics[height=190mm]{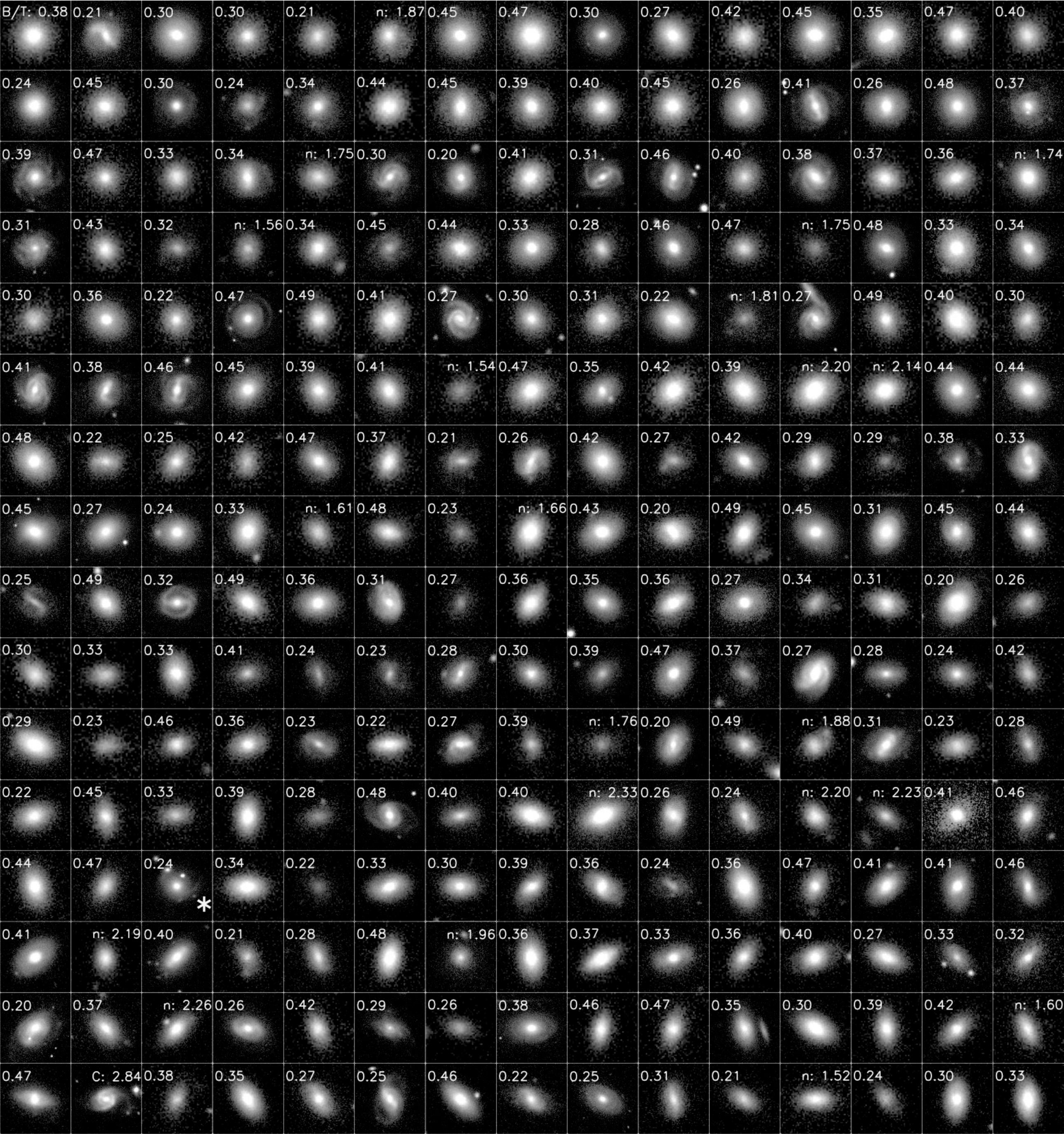}
\end{center}
\caption{\label{fig:stampSbc}$I$-band postage-stamp images  of \textsc{ZENS} galaxies classified as intermediate-type disks. Galaxies are ordered from top-left to bottom-right by increasing ellipticity.  To fall into this class galaxies must have $I-$band $0.2<B/T<0.5$ or, alternatively, $1.5<n <2.5$, when no valid bulge+disk decomposition is available (or $2.8<C<3.3$ if also the S\'ersic index is missing). Numbers on the images are the $B/T$ ratios of each galaxy or the S\'ersic index (or concentration parameter) when no $B/T$ (and also no $n$) is available.
Stamp images marked with an asterisk are for those galaxies which have low $M_{20}$ indices, and thus lie outside the global relations in the bottom panels of Figure \ref{fig:ZESTClasses}.
One galaxy marked with a `!' lie close to the edges of the WFI camera, thus $B$-band observations are available but not the $I$-band ones. The classification for this galaxy is based on the $B$-band parameters.}
\end{figure*}

\addtocounter{figure}{-1}

\begin{figure*}[htbp]
\begin{center}
\includegraphics[height=190mm]{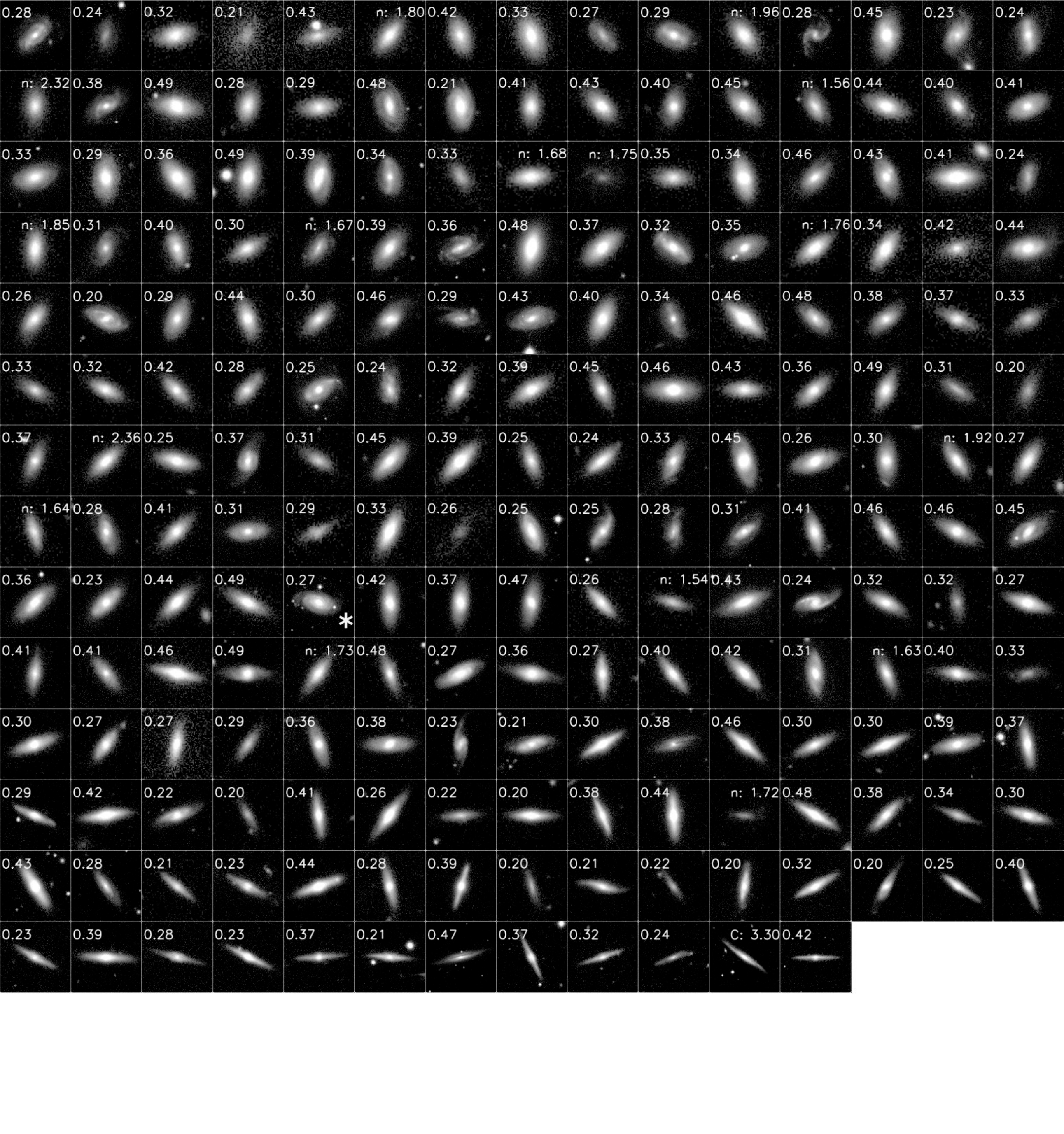}
\end{center}
\caption{ -- \emph{Continued}}
\end{figure*}

\begin{figure*}[htbp]
\begin{center}
\includegraphics[height=200mm]{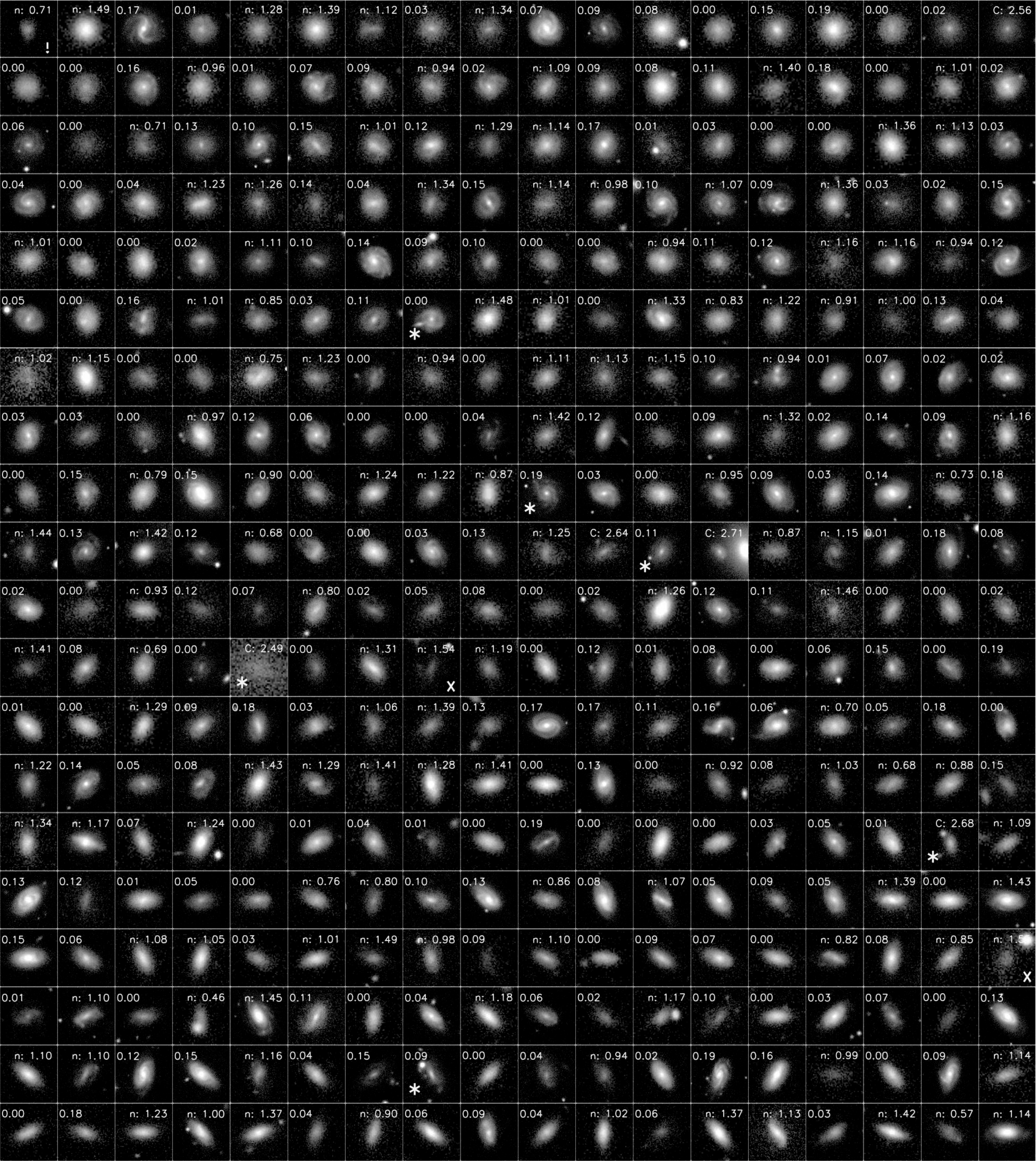}
\end{center}
\caption{\label{fig:stampSd}$I$-band postage-stamp images  of \textsc{ZENS} galaxies classified as late-type galaxies. Galaxies are ordered from top-left to bottom-right by increasing ellipticity.  To fall into this class galaxies must have $I-$band $B/T <0.2$ or $n <1.5$  (or $C<2.8$ if also the S\'ersic index is missing), when no valid bulge+disk decomposition is available. As in Figure \ref{fig:stampSbc}, outliers   in the bottom panels of Figure \ref{fig:ZESTClasses}   are marked with an asterisk. Two galaxies with high concentration, highlighted with a 'X' sign, were kept in this class after further visual inspection.}
\end{figure*}

\addtocounter{figure}{-1}

\begin{figure*}[htbp]
\begin{center}
\includegraphics[height=200mm]{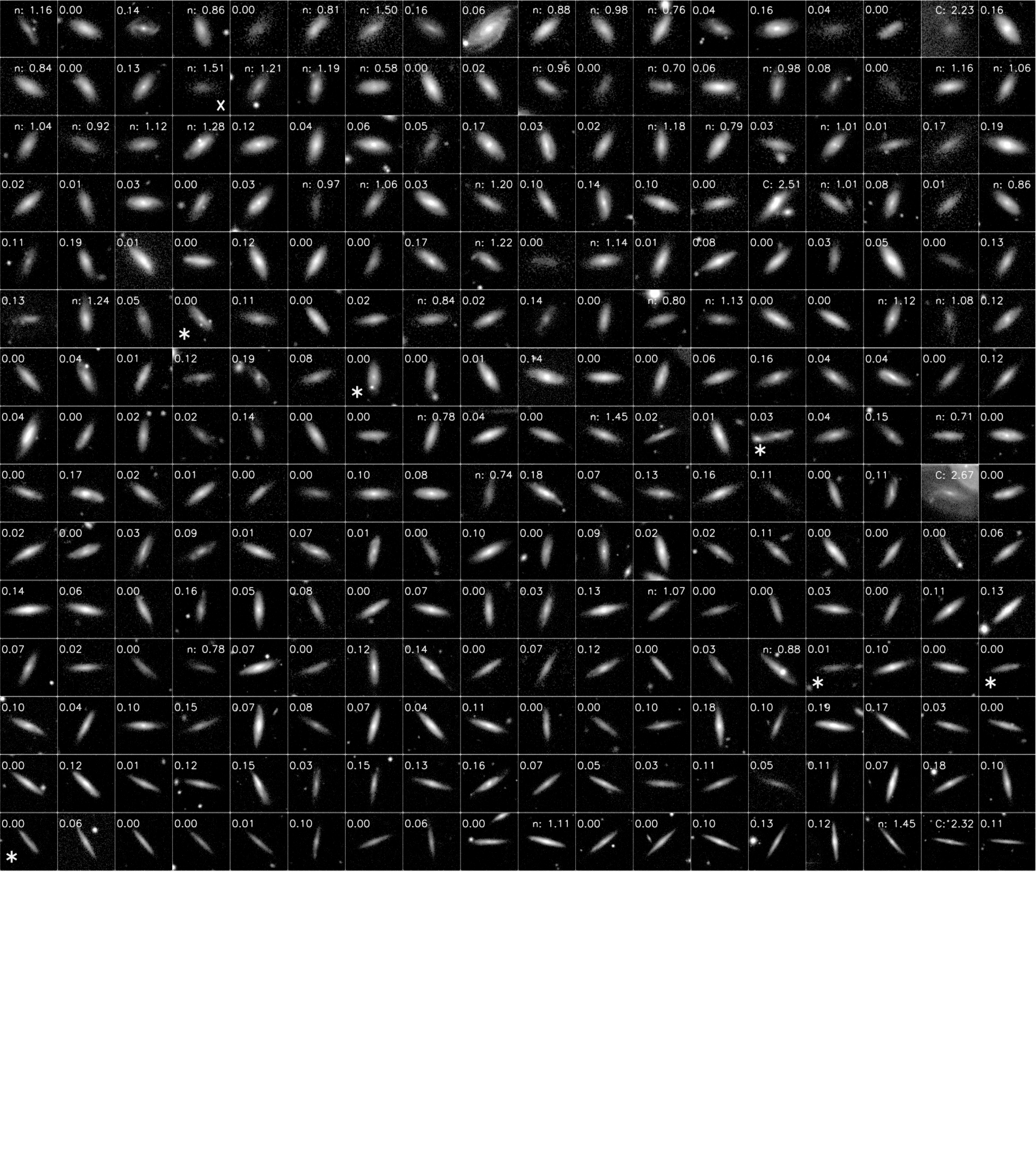}
\end{center}
\caption{ -- \emph{Continued}}
\end{figure*}

\begin{figure*}[htbp]
\begin{center}
\includegraphics[width=140mm]{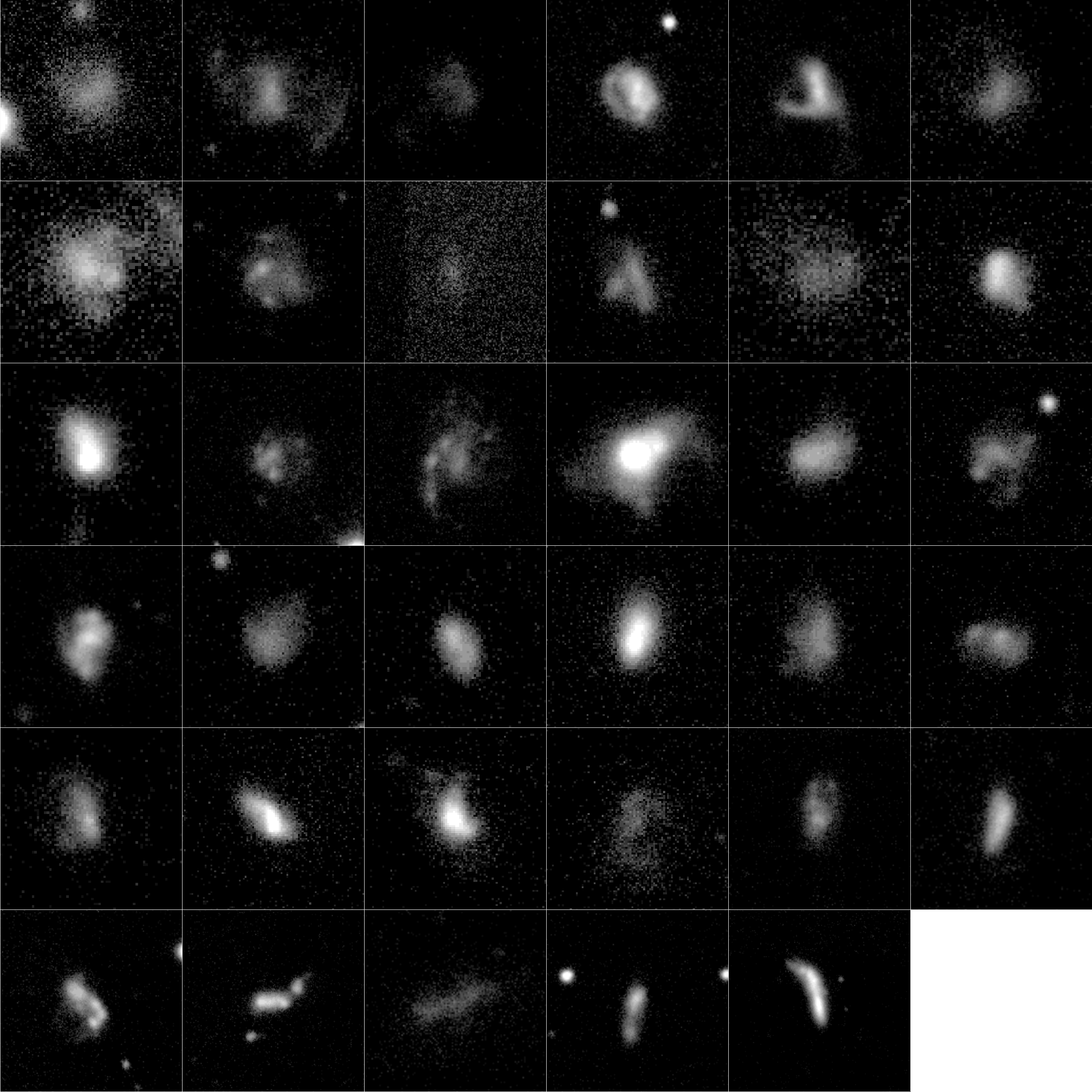}
\end{center}
\caption{\label{fig:stampIrr}$B$-band postage-stamp images of \textsc{ZENS} galaxies classified as irregular/interacting systems. These galaxies were visually identified as systems with disturbed morphologies.}
\end{figure*}

\begin{figure*}[htbp]
\begin{center}
\includegraphics[width=160mm]{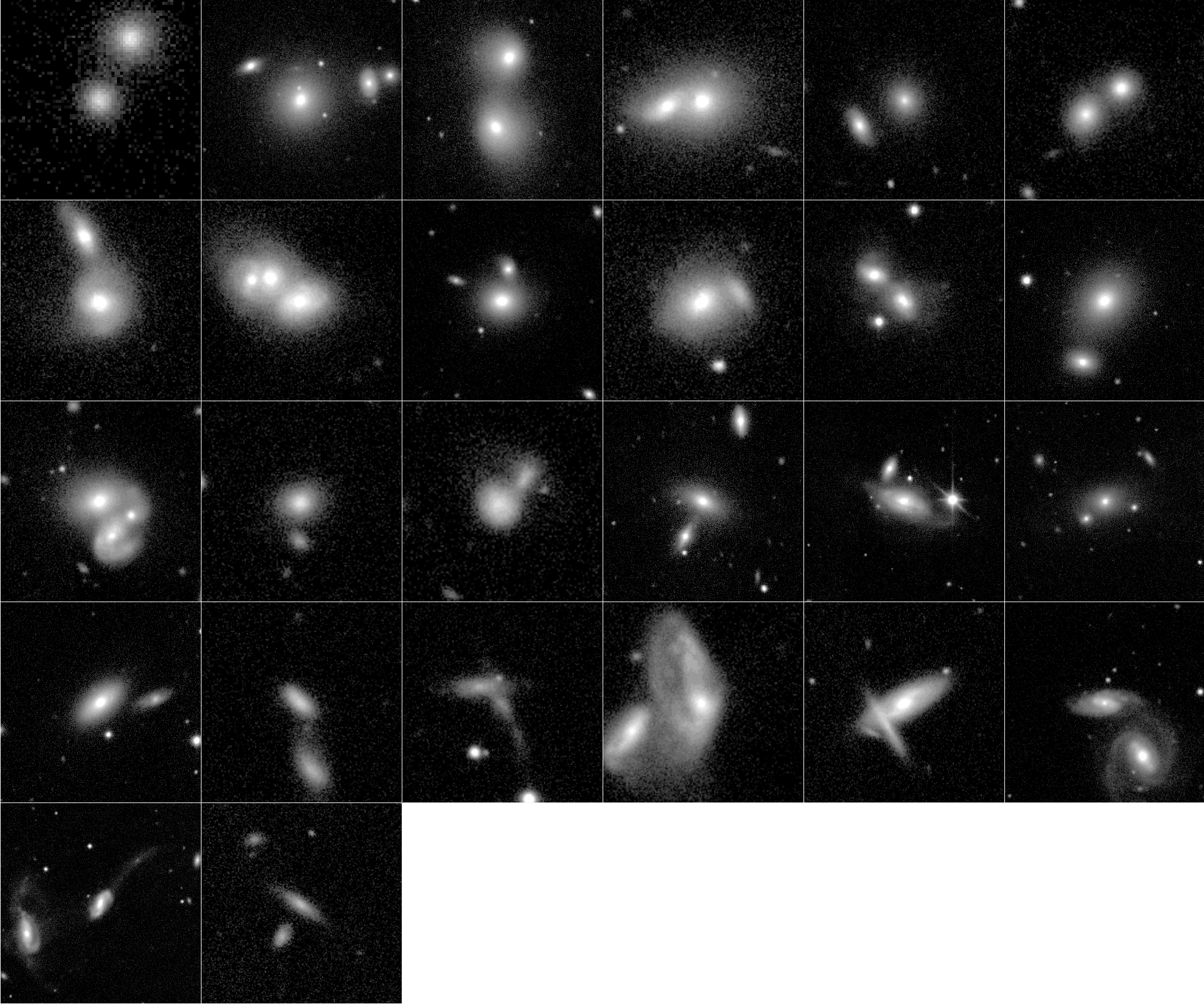}
\end{center}
\caption{\label{fig:stampMergers}$I$-band postage-stamp images of the \textsc{ZENS} galaxy-pairs which are plausibly  undergoing a merger.}
\end{figure*}

 \end{document}